\documentclass{aa}  

\usepackage{hyperref}

\hypersetup{
    colorlinks=true,
    citecolor=blue,   
    linkcolor=blue,   
    urlcolor=blue     
}
\usepackage{graphicx}
\usepackage{float}
\usepackage{txfonts}
\usepackage{siunitx}
\usepackage{pdflscape}
\usepackage{multirow}
\usepackage{booktabs}
\usepackage{array}
\usepackage{tabularx}
\usepackage{ragged2e}
\usepackage{graphicx}
\usepackage{pgffor} 
\usepackage{orcidlink} 
\usepackage{academicons}
\usepackage{afterpage}
\usepackage{subcaption}
\usepackage{sidecap}
\usepackage{multirow}
\usepackage{orcidlink}
\usepackage{mathtools}

\usepackage{txfonts}
\usepackage{xcolor}
\usepackage{ragged2e}
\usepackage[utf8]{inputenc}

\definecolor{gr}{rgb}{0.2, 0.20, 0.80}

\newcommand{\gsim}{\ \raise -2.truept\hbox{\rlap{\hbox{$\sim$}}\raise 5.truept\hbox{$>$}\ }}

\definecolor{emerald}{rgb}{0.4,0.66,0.31}

\newcommand{\customcitep}[3]{(#1 \citeauthor{#3} \citeyear{#3} #2)}
\setcitestyle{citesep={,}}

\begin{document}

   \title{Looking into the faintEst WIth MUSE (LEWIS): Exploring the nature of ultra-diffuse galaxies in the Hydra-I cluster}

   \subtitle{IV. A study of the Globular Cluster population in four UDGs.}

\author{Marco Mirabile\inst{1,2,3}\orcidlink{0009-0007-6055-3933}
       \and
        Michele Cantiello\inst{1}\orcidlink{0000-0003-2072-384X}
        \and
        Marina Rejkuba\inst{2}\orcidlink{0000-0002-6577-2787}
        \and
        Steffen Mieske\inst{4}\orcidlink{0000-0003-4197-4621}
          \and
        Enrichetta Iodice\inst{5}
        \and 
         Chiara Buttitta\inst{5}\orcidlink{0000-0001-9787-3067}
         \and
         Maria Luisa Buzzo\inst{2,7}\orcidlink{0000-0003-3153-8543}
        \and         
         Johanna Hartke\inst{9,10,11}\orcidlink{0000-0002-8745-689X}
         \and 
         Goran Doll\inst{5,6}
         \and 
         Luca Rossi\inst{5,6}
         \and 
         Magda Arnaboldi\inst{2}
         \and 
         Marica Branchesi\inst{3}
         \and
         Giuseppe D'Ago\inst{12}\orcidlink{0000-0001-9697-7331}
         \and
         Jesus Falcon-Barroso\inst{13,14}
         \and
         Katja Fahrion\inst{15}
         \and
         Duncan A. Forbes\inst{7}\orcidlink{0000-0001-5590-5518}
         \and
         Marco Gullieuszik\inst{8}\orcidlink{0000-0002-7296-9780}
         \and
         Michael Hilker\inst{2}\orcidlink{0000-0002-2363-5522}
         \and
         Felipe S. Lohmann\inst{2}\orcidlink{0000-0003-4896-5103}
         \and
         Maurizio Paolillo\inst{5,6}
         \and
         Gabriele Riccio\inst{1}\orcidlink{0000-0002-6399-2129}
         \and
         Tom Richtler\inst{16}\orcidlink{0000-0002-4887-5661}
         \and
         Marilena Spavone\inst{5}
        }

      

\institute{\scriptsize 
        INAF Osservatorio Astr. d’Abruzzo, Via Maggini, 64100 Teramo, Italy\\
              \email{marco.mirabile@inaf.it}
         \and
        European Southern Observatory, Karl-Schwarzschild-Strasse 2, 85748 Garching bei München, Germany
        \and
             Gran Sasso Science Institute,  Viale Francesco Crispi 7, 67100 L’Aquila, Italy
        \and
        European Southern Observatory, Alonso de Cordova 3107, Vitacura, Santiago, Chile  
        \and
        INAF -- Astronomical Observatory of Capodimonte, Salita Moiariello 16, I-80131, Naples, Italy 
        \and
        University of Naples “Federico II”, C.U. Monte Sant’Angelo, Via Cinthia, 80126 Naples, Italy
        \and
        Centre for Astrophysics and Supercomputing, Swinburne University, John Street, Hawthorn VIC 3122, Australia
        \and
        INAF - Osservatorio Astronomico di Padova, Vicolo dell'Osservatorio 5 I, 35122 Padova, Italy
        \and
        Finnish Centre for Astronomy with ESO, (FINCA), University of Turku, FI-20014 Turku, Finland
        \and
        Tuorla Observatory, Department of Physics and Astronomy, University of Turku, FI-20014 Turku, Finland
        \and
        Turku Collegium for Science, Medicine and Technology (TCSMT), University of Turku, FI-20014 Turku, Finland
        \and
        Institute of Astronomy, University of Cambridge, Madingley Road, Cambridge CB3 0HA, United Kingdom
        \and
        Instituto de Astrofísica de Canarias, Calle Vía Laáctea s/n, E-38205 La Laguna, Tenerife, Spain
        \and
        Departamento de Astrofísica, Universidad de La Laguna, Av. del Astrofísico Francisco Sánchez s/n, E-38206 La Laguna, Tenerife, Spain
        \and
        Department of Astrophysics, University of Vienna, Türkenschanzstraße 17, 1180 Wien, Austria
        \and
        Departamento de Astronom\'{\i}a, Universidad de Concepci\'on, Concepci\'on, Chile        
}

   \date{}

  \abstract
   {As some of the oldest stellar systems in the Universe, globular clusters (GCs) are key fossil tracers of galaxy formation and interaction histories. This paper is part of the LEWIS project, which provides the first homogeneous MUSE integral-field spectroscopic survey of a complete sample of ultra-diffuse galaxies (UDGs) in the Hydra I cluster.}
   {We use MUSE spectroscopy and new VIRCAM $H$-band imaging data to study the GC populations and dark matter content in four dwarf galaxies from the LEWIS sample, which were found to host several GC candidates based on previous photometric studies.}
   {We retrieved line-of-sight velocities (LOSVs) for all the sources in the observed MUSE fields and classified them based on their spectral features and LOSVs.
   Because the spectroscopic measurements are limited to relatively bright sources ($m_H \lesssim 23.5$ AB mag), we developed a multi-band photometric procedure to identify additional GC candidates that are too faint for spectroscopic confirmation. 
   GC candidates were selected based on a combination of photometric properties (colors, magnitudes) and morphometric criteria (shape and size). The same selection criteria were applied to empty fields to estimate a statistical background correction for the number of identified GC candidates. Additionally, the $H$-band observations were used to constrain the stellar masses of the studied galaxies. }
   {Based on the spectroscopic classification, we confirm one GC in UDG\,3, two in UDG\,7 and four in UDG\,11, while UDG\,9 has no spectroscopically confirmed bright GCs. We identify four intra-cluster GCs in the vicinity of UDG\,3 and UDG\,11, and one ultra-compact dwarf (UCD) with the radial velocity only $-85\pm10\ \mathrm{km/s}$ different from that of UDG 7 and thus possibly bound to it.   Considering the completeness correction and accounting for possible contamination by unresolved background galaxies, from the photometry we estimate that the number of GCs ranges between 0 and $\sim$40 for the investigated UDGs. Their specific frequencies suggest that three out of four UDGs are either GC-rich, similar to those in the Coma cluster, or belong to an intermediate population as seen in the Perseus cluster. Dark matter content estimates, inferred from GC counts and stellar mass, indicate that these galaxies are dark matter dominated, with dynamical-to-stellar mass ratios $M_{dyn}/M_*$ ranging from $\sim10-1000$.}
{}

\maketitle

\section{Introduction}

\begin{figure*}[h!]
    \centering
    \includegraphics[width=\textwidth]{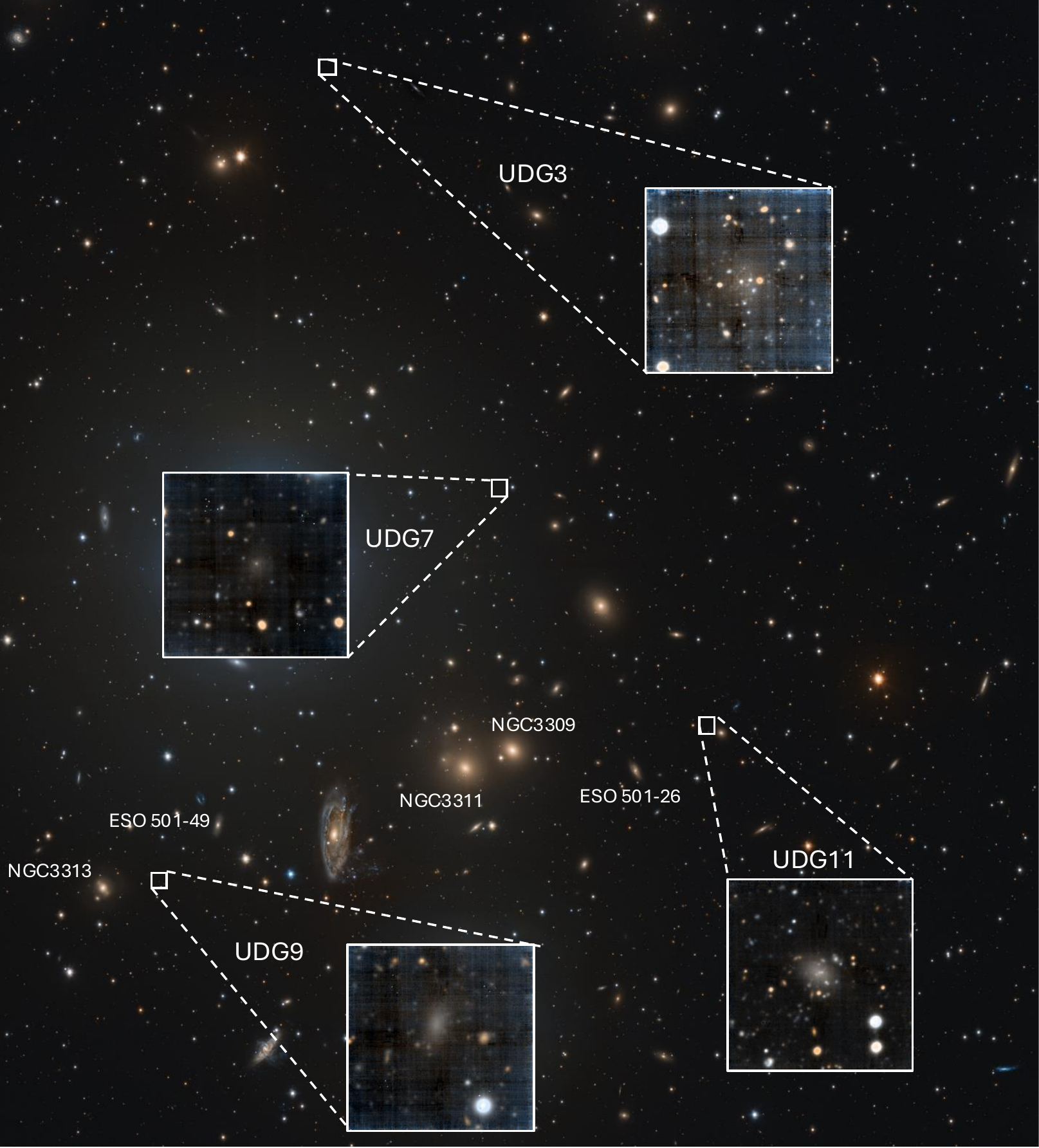}
    \caption{Zoom-in of a $34 \times 40\ \mathrm{arcmin}^2$ color composite OmegaCAM@VST image of the Hydra I cluster \citep[credit: \href{https://www.eso.org/public/italy/images/potw2437b/}{ESO}/][]{spavone24}. White boxes mark the locations of UDGs studied in this work. For each UDG, a color composite image of $1\times1\ \mathrm{arcmin}^2$ combining the $g^*$- and $r$-band MUSE data is shown. The cluster galaxies discussed in this work are labeled.}
    \label{fig:color_comp}
\end{figure*}

\begin{table*}[ht]
\centering
\caption{Properties of the LEWIS sample galaxies.}
\label{tab:galaxy_properties}
\resizebox{\textwidth}{!}{
\begin{tabular}{ccccccccccccc}
\hline\hline
Galaxies & RA [J2000] & Dec [J2000] & $M_r$ [mag]& $\mu_e$ [mag arcsec$^{-2}$] & $\mu_0$ [mag arcsec$^{-2}$] & $R_e$ [kpc] &  $N_{GC}$ &  \\
(1) & (2) & (3) & (4) & (5) & (6) & (7) & (8) \\
\hline
UDG 3  & 10:36:58.63 & $-$27:08:10.21 & $-14.70$ & $26.1 \pm 0.2$ & $25.2 \pm 0.2$ & $1.88 \pm 0.12$   & $15 \pm 6$         \\
UDG 7  & 10:36:37.16 & $-$27:22:54.93 & $-13.72$ & $26.9 \pm 0.4$ & $24.4 \pm 0.4$ & $1.66 \pm 0.12$   & $3 \pm 1$              \\
UDG 9  & 10:37:22.85 & $-$27:36:02.80 & $-15.16$ & $26.8 \pm 0.2$ & $24.2 \pm 0.2$ & $3.46 \pm 0.12$   & $7 \pm 1$              \\
UDG 11 & 10:34:59.55 & $-$27:25:37.95 & $-14.75$ & $25.7 \pm 0.1$ & $24.4 \pm 0.1$ & $1.66 \pm 0.12$   & $7 \pm 3$                \\
\hline

\end{tabular}
}
\begin{justify}
Notes: Col. (1): the target name in the LEWIS sample. Col. (2-3): the Right Ascension and Declination of the galaxies. Col. (4-7): the absolute magnitude at the distance of the Hydra I Cluster in the $r-$band, the effective and central surface brightness, and the effective radius, in the $g-$band from \citet{iodice20,lamarca22b}, respectively. Col. (8): total number of GCs candidates hosted by the galaxies estimated in VEGAS imaging data \citet{iodice20,lamarca22b}.
\end{justify}

\label{tab:lewis_sample}
\end{table*}

Dwarf galaxies are by far the most numerous galaxies in the Universe \citep{Ferguson94} and exist in all kinds of
environment \citep{Binggeli90, rekola05, kim14,lamarca22a}.
Although low surface brightness (LSB) and extended galaxies have been known since the 1980s \citep{Sandage84}, \citet{vandokku15} introduced a new subsample of so called ultra-diffuse galaxies (UDGs), defined as galaxies with central surface brightness $\mu_{g,0} \gtrsim 24.0\ \mathrm{mag/arcsec^2}$ and effective radii $R_e \gtrsim 1.5$ kpc. Since then, large number of UDGs have been identified in a wide range of environments \citep{Koda15,Martinez16,Venhola17,habas20,mueller21}.

The origin of UDGs remains a topic of active investigation, with several formation scenarios proposed. These include internal processes such as strong stellar feedback  \citep{DiCintio17} or high-spin dark matter halos fostering extended star formation \citep{Amorisco16}. Environmental interactions also play a role: early mergers may disrupt galaxies into diffuse structures \citep{Wright21}, while tidal forces in clusters or groups can transform gas-rich dwarfs into "puffed-up" UDGs through processes like ram pressure and tidal stripping 
\citep{Carleton19,Tremmel20}. Furthermore, UDGs may form from collisional debris in gas-rich interactions \citep{Roman21}. Alternatively, some UDGs may originate as failed galaxies -- systems inhabiting unexpectedly massive dark matter halos that quenched prematurely \citep{vandoku16}. Each scenario implies different dark matter (DM) content as well as different field and star-cluster stellar population properties. 

Most GCs are ancient ($t \geq 8$ Gyr), compact (half-light radius $r_e \sim 2.5$ pc) stellar systems that formed during the early stages of galaxy assembly, when conditions favored intense bursts of star formation capable of producing massive star clusters \citep{larsen00,brodie06,forbes2021}. As relics of this epoch, old GCs serve as fossil tracers of the host galaxy's evolutionary history. Their high luminosity also makes them detectable at cosmological distances \citep{Harris24}.

The number of GCs serves as a diagnostic tool to investigate the different nature of the UDGs. Indeed, some UDGs host unexpectedly rich GC systems relative to their stellar mass, implying the presence of massive dark matter halos, while others exhibit GC populations consistent with typical dwarf galaxy halos \citep{Beasley16,Amorisco18,lim20, Saifollahi21b}.
The total number of GCs in a galaxy scales approximately linearly with its total halo mass \citep{blake97,spitler09,Harris17,Burkert20}, providing an indirect method to estimate DM content. Deep imaging campaigns targeting UDGs in the Coma and Fornax clusters \citep{Prole19, Lim18}, as well as in isolated environments \citep{Jones23}, have quantified their GC populations, revealing significant scatter in specific frequencies ($S_N$). This scatter may indicate diverse formation pathways, although it is important to note that at low GC numbers, the process of GC formation naturally becomes stochastic \citep{georgiev10,harris13}.
Spectroscopic studies of GCs provide additional constraints on UDG evolution by revealing kinematic signatures indicative of tidal interactions \citep{Toloba18}. They can constrain the velocity dispersion of the GC system, which is then linked to the DM content \citep{Buzzo25, Haacke25}. 
To date, there are very few UDGs for which spectroscopic data are available \customcitep{see}{for an overview}{Gannon24}. 

Here, we present the spectroscopic and photometric analysis of the GC systems in four UDGs observed as part of the LEWIS (Looking into the faintEst WIth MUSE) project \citep{Iodice23, buttitta25, Hartke25}. The LEWIS project assembled the first homogeneous integral-field (IF) spectroscopic survey of 30 extreme LSB galaxies targeting the Hydra I galaxy cluster \citep{Iodice23}.  An overview of the central area of Hydra I, along with the locations and zoomed-in images of the four analyzed UDGs, is shown in Fig. \ref{fig:color_comp}. We selected for this study UDG\,3, UDG\,7, UDG\,9, and UDG\,11, because they were found to be among the UDGs with the highest number of GC candidates initially detected in the VST imaging of the Hydra I cluster \citep{iodice20}. 

In the present study we complement the MUSE spectroscopic dataset with new deep  VIRCAM $H-$band observations of the Hydra I cluster. The paper focus is on the study of the GC population hosted by these galaxies and their DM content.

The paper is organized as follows. In Sect. \ref{sec:dataset} we present our datasets. Section \ref{sec:method} describes the procedure used to extract and fit spectra of point sources in MUSE datacubes with the aim to identify GCs hosted by UDGs. Section \ref{sec:sel_gc} focuses on the selection criteria for GC candidates based on their photometric properties. In Sect. \ref{sec:analysis_gc}, we present the analysis of the GC candidates in the four galaxies. Our conclusions are presented in Sect. \ref{sec:conclusiocn}.



\section{Dataset}
\label{sec:dataset}

\begin{figure*}[h!]
    
    \centering
    \sidecaption
    \includegraphics[width=\textwidth]{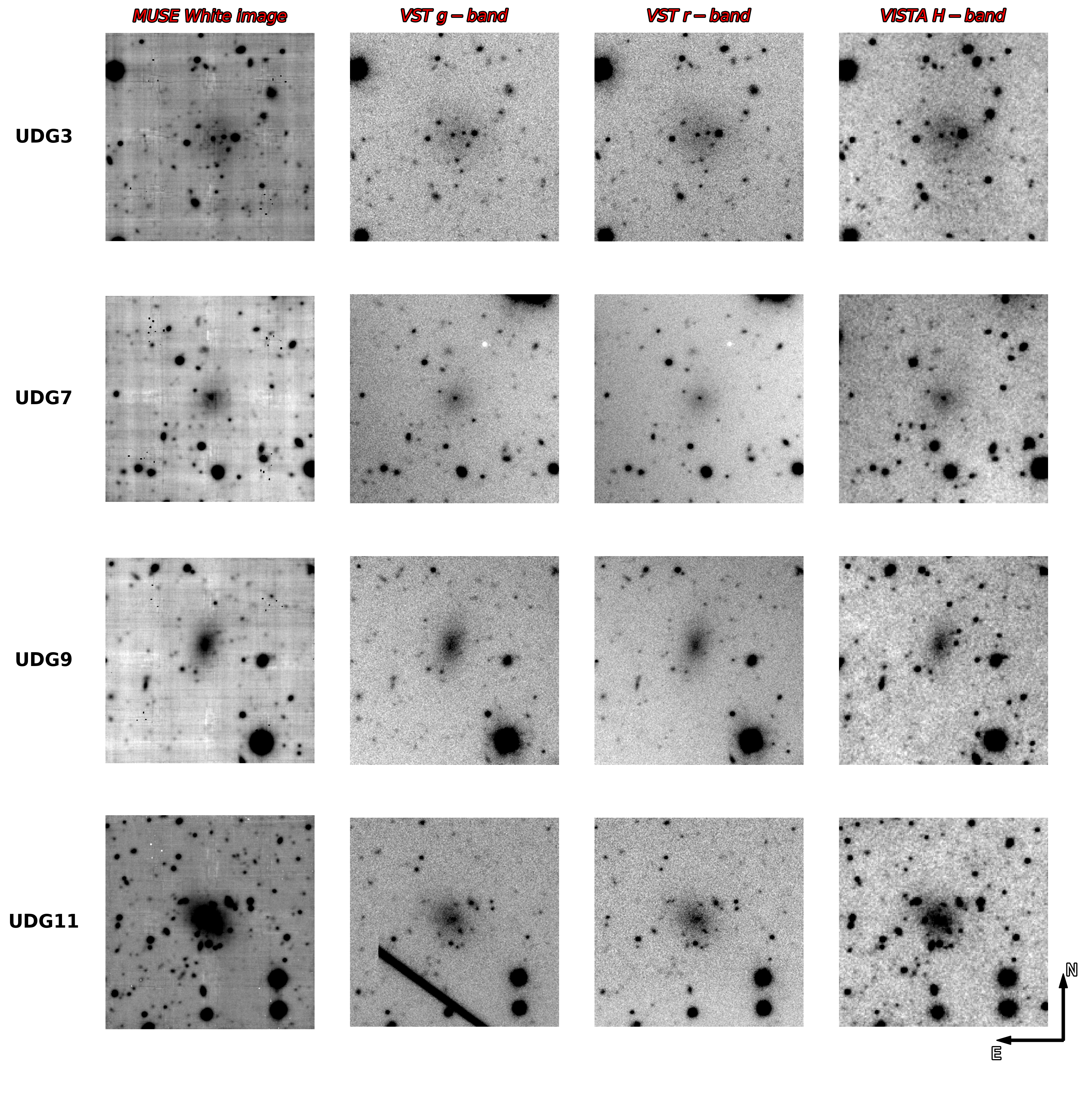}
    \caption{Observations of the four UDGs. From left to right: white image obtained from the MUSE cube, $g-$ and $r-$band from VST data, $H-$band from VISTA. The size of the cutouts is $1\times1\ \mathrm{arcmin}^2$ ($\sim15\times15\ \mathrm{kpc}^2$). North is up, east is to the left.}
    \label{fig:fields}
\end{figure*}

The main spectroscopic dataset consists of the LEWIS sample observed with MUSE@VLT \citep{Iodice23}, combined with ancillary MUSE observations of NGC\,3311 from \citet{barbosa18}. The imaging dataset is made of wide-field optical to near-infrared (IR) observations of the Hydra I cluster obtained with the VST and VISTA telescopes. Figure \ref{fig:fields} presents observations of the four UDGs analyzed in this work across different passbands. From left to right, each panel shows the white-light image extracted from the MUSE data cube, the $g$- and $r$-band images obtained from VST observations, and the $H$-band image acquired with VISTA. 



\subsection{MUSE@VLT Observations}
\label{sec:spec_data}

The observations of the galaxies from the LEWIS programme (ESO Prog. ID. 108.222P, P.I. E. Iodice)  were carried out with the ESO integral-field spectrograph \textit{Multi Unit Spectroscopic Explorer}  \citep[MUSE,][]{Bacon10} mounted on UT4 of the ESO Very Large Telescope (VLT). MUSE was configured with the Wide Field Mode providing a field-of-view (FOV) of $1\times1\ \mathrm{arcmin}^2$ and a spatial resolution of 0.2 arcsec/pixel. The nominal wavelength range of MUSE is 4800-9300 $\AA$ with a spectral sampling of 1.25 $\AA$/pixel ($\sim$50 $\mathrm{km\ s^{-1}}$/pixel) and an average nominal spectral resolution of FWHM = 2.51 $\AA$ \citep{Bacon17}. A detailed description of the LEWIS project, galaxy sample, observing strategy and data reduction procedures were presented in \citet{Iodice23} and \citet{buttitta25}.  
We used reduced data cubes from  \citet{Iodice23}
and \citet{buttitta25}, which include dedicated sky subtraction techniques required for analyzing extreme LSB galaxies. These data cubes include all LEWIS observations collected between 2021--2024, when the program reached $\sim 92$\% completion.\footnote{Our Large Programme was recently completed, and we plan to use the newly reduced and slightly deeper data cubes in a future study of GC systems of all LEWIS galaxies.} The exposure times are 3.0 hours, 4.2 hours, 3.6 hours, and 6.1 hours for UDG\,3, UDG\,7, UDG\,9, and UDG\,11, respectively. The average image quality\footnote{Image quality (IQ) is computed as the median FWHM of bright and compact sources.} (IQ) of the four data cubes is $0.75 \pm 0.1$ arcsec.
Table \ref{tab:lewis_sample} summarizes the properties of the four galaxies in the LEWIS sample analyzed here. Additionally, we also report the total number of GCs ($N_{GC}$\footnote{The $N_{GC}$ estimated in \citet{iodice20} and \citet{lamarca22b} is obtained accounting for the entire luminosity function.}) for each galaxy estimated in \citet{iodice20} and \citet{lamarca22b} using $g-$ and $r-$band photometry.

In addition to the LEWIS observations, we also analyze MUSE observations of four fields around NGC\,3311. The sources in these fields are used to build a master catalog that constrains our criteria for the selection of GC candidates (see Sect. \ref{sec:mast_cat}).
The MUSE observations of the four fields around NGC\,3311 were obtained from the ESO archive (ESO program 094.B-0711A, PI: M. Arnaboldi). Three fields were positioned along the major axis of the galaxy at increasing galactocentric distance, and the fourth was located over the extended diffuse tail of the lenticular galaxy HCC\,007 (see Fig. \ref{fig:grasser_field}). The observations of the four fields were carried out under varying seeing conditions, with average seeing values of $0\farcs92$, $1\farcs4$, $1\farcs3$, and $1\farcs8$ for fields A, B, C, and D, respectively.
For the purposes of this work, we decided to use only fields A, B, and C and exclude field D due to its poorer seeing conditions (see Sect. \ref{sec:mast_cat}). For more details about the dataset and the data reduction procedures see \citet{barbosa18} and \citet{Grasser24}. 

We exploited the wide wavelength coverage of MUSE, spanning from optical ($\sim4800\ \AA$)  to near-IR ($\sim9300\ \AA$), to extract three passbands from the data cube. This step was essential to fully exploit the MUSE sample, including the faintest objects detected within the 1 sq. arcminute field surrounding the four UDGs (see Sect. \ref{sec:sel_gc}). The FWHM of the point sources was more compact in the MUSE@VLT cubes than in the OmegaCAM@VST images, allowing for a more reliable cleaning of the sample from interlopers (see Sect. \ref{sec:VST_imaging}). Using the MPDAF Python package \citep{Piqueras17}, we derived images equivalent to the SDSS $g$, $r$, and $i$ filter passbands, along with integrated error images for each filter. While the MUSE cube fully covers the $r$- and $i$-bands, it captures only $\sim$51\% of the $g$-band. Consequently, we refer to the $g$ filter as a pseudo-band, denoted as $g^*$ hereafter.
Although MUSE cubes also cover part of the SDSS $z$ filter range, we opted not to use this passband due to the increased noise from the sky subtraction residuals in that spectral region. 






\subsection{Imaging dataset: VST \& VISTA observations}
\label{sec:sec_phot_data}

Here, we introduce the two imaging datasets: optical and near-IR observations. Throughout this work, we adopt the AB magnitude system for all the photometric measurements.

\subsubsection{OmegaCAM@VST optical imaging}
\label{sec:VST_imaging}

The $g-$ and $r-$band observations of the Hydra I Cluster used in this work are part of the \textit{VST Elliptical GAlaxy Survey} (VEGAS, P.I. E.Iodice; \citealt{capaccioli15,iodice21}). The survey was performed with the 2.6m INAF VLT Survey Telescope (VST) at Cerro Paranal, Chile \citep{Schipani2010} using OmegaCAM wide field imager that provides a 1 sq. deg FOV with a pixel scale of $0\farcs21$.
A detailed description of the survey and procedures adopted for data acquisition and reduction can be found in \citet{grado12}, \citet{capaccioli15}, and \citet{iodice16fds}. Details of the Hydra I cluster dataset have been described in \citet{iodice20,iodice21b,lamarca22a,lamarca22b,spavone24}. The average IQ is $0.84 \pm 0.03$ arcsec for the $g$-band and $0.79 \pm 0.02$ arcsec for the $r$-band.

\subsubsection{VIRCAM@VISTA near-IR imaging}

\begin{figure}[ht]
    \centering
    \includegraphics[width=\columnwidth]{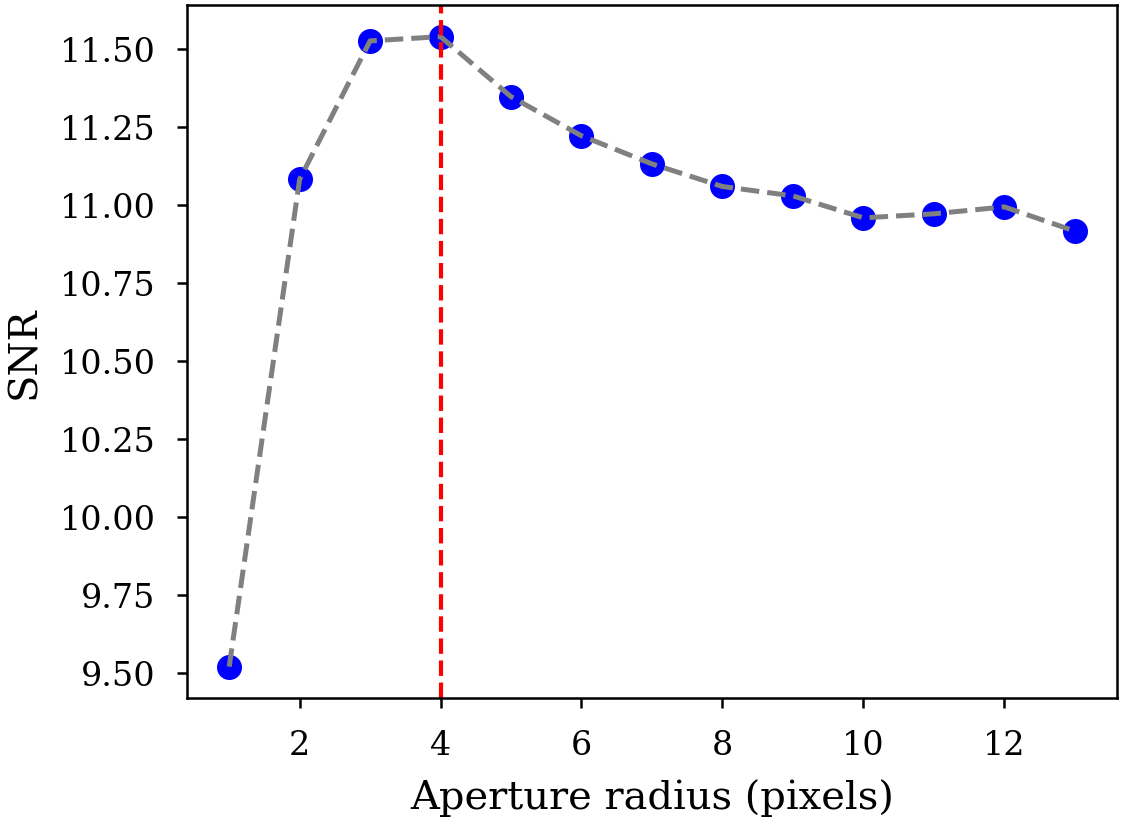}
    \caption{Signal-to-noise ratio estimates as a function of aperture radius for a compact source of $m_H\sim20$ mag. The red dashed line indicates the adopted aperture radius and the point where the SNR profile peaks before decreasing due to the inclusion of additional background noise. The pixel scale is $0\farcs 2$/pixel.} 
    \label{fig:rad_sn_trend}
\end{figure}

The identification of GCs based on optical $g$ and $r$ images suffers from large contamination from foreground stars and background galaxies \citep[see][]{lamarca22b}. We therefore obtained near-IR images to reduce the effect of possible interlopers \citep{Daddi04,Bielby12,Merson13,munoz14}.
 Deep $H-$band observations (ESO Prog. ID. 109.231E.00, P.I. M. Cantiello) of the Hydra I galaxy cluster have been taken using the ESO Paranal 4 m-VISTA (Visible and Infrared Survey Telescope for Astronomy) telescope \citep{Dalton06,Emerson06,Sutherland15}. The telescope was equipped with the VIRCAM detector (VISTA InfraRed Camera), offering a 1.65 sq. degrees FOV with a pixel scale of 0.34 arcsec/pixel. 
A minimum of six exposures, known as pawprints, are required to survey a contiguous area of 1.65 sq. degrees, called a tile\footnote{see \href{https://www.eso.org/sci/facilities/paranal/decommissioned/vircam/doc/VIRCAM_VISTA_User_Manual_20100701.pdf}{VIRCAM/VISTA User Manual}}. 

The observations were reduced by the Cambridge Astronomy Survey Unit (CASU), which takes care of the pre-reduction, calibration, sky subtraction, and stacking of the observation, on a pawprints base \citep{Irwin01}. The final products retrieved from CASU are the 6 reduced and calibrated pawprints, as well as their combination into a tile based on each $\sim 1$h-long Observation Block (OB). 
ESO observers classified each OB according to the fulfillment of the requested observing conditions (seeing, sky transparency, airmass, etc.)\footnote{The observation requirements were as follows: image quality $\leq 1$", sky transparency clear (i.e., $<10$\% cloud coverage), airmass$<1.4$.}. The executed OBs of our observing program consist of 44 tiles graded A, fulfilling all requested constraints, 16 B-graded tiles taken mostly within constraints and a parameter varying up to 10\% of the requested value, and 13 C-graded observations, during which some conditions were violated by more than 10\%.

In this work, we used pawprints from the A-graded OBs. We identified the pawprints containing the UDGs and extracted a $1\times1\ \mathrm{arcmin}^2$ cutout around them, matching the MUSE FOV. The cutouts were then stacked using SWarp \citep{bertin10} to get a deep $H-$band image for each UDG. We also combined all the grade A pawprints to have a preliminary wide-field image of the Hydra I cluster with an average IQ of $0.90\pm0.03$ arcsec. The resulting deep tile served to verify our photometry (see Sect. \ref{sec:phot_approach}), and to build the GCs master catalog (see Sect. \ref{sec:mast_cat}). 
More details about these observations and the data reduction strategies will be presented in an upcoming paper (Mirabile et al. in prep).




\section{Selection of GCs using MUSE Spectroscopy}
\label{sec:method}

Here, we describe the routine and methods used to extract and analyze spectra from the LEWIS MUSE cube, to identify and characterize the GC populations in the four UDGs. We examine the radial velocities and assess the membership of GC candidates identified in \citet{iodice20} and \citet{lamarca22b}. Furthermore, we search for other sources that could be classified as GCs based on their radial velocities and spectra.

\subsection{Source detection across the MUSE cube}

We first run SExtractor \citep{bertin96} on the MUSE white image (i.e., the mean collapsed cube across the wavelength axis) to detect sources for which spectra can be extracted. We also model and subtract the galaxy light profile using the Elliptical Isophote Analysis package\footnote{\url{https://photutils.readthedocs.io/en/stable/api/photutils.isophote.Isophote.html}} to improve the analysis of central sources. A detailed description of the model generation process can be found in \citet{Hazra2022}; here, we provide a brief summary. Initially, we create a preliminary model that helps us better constrain the ellipticity and position angle and to mask all bright sources that could interfere with the modeling of the light profile. Subsequently, we perform the final elliptical isophote fit on the masked image, generating a refined galaxy profile model that is subtracted from the original image.

Subsequently, we exclude all detected sources at the edges of the cubes, as these are mostly false positives due to image artefacts, or, if real, have low  signal-to-noise ratio (SNR), unusable for further inspection.



\subsubsection{Spectrum extraction}
\label{sec:extraction}

\begin{figure}[ht!]
    \centering
    \includegraphics[width=\columnwidth]{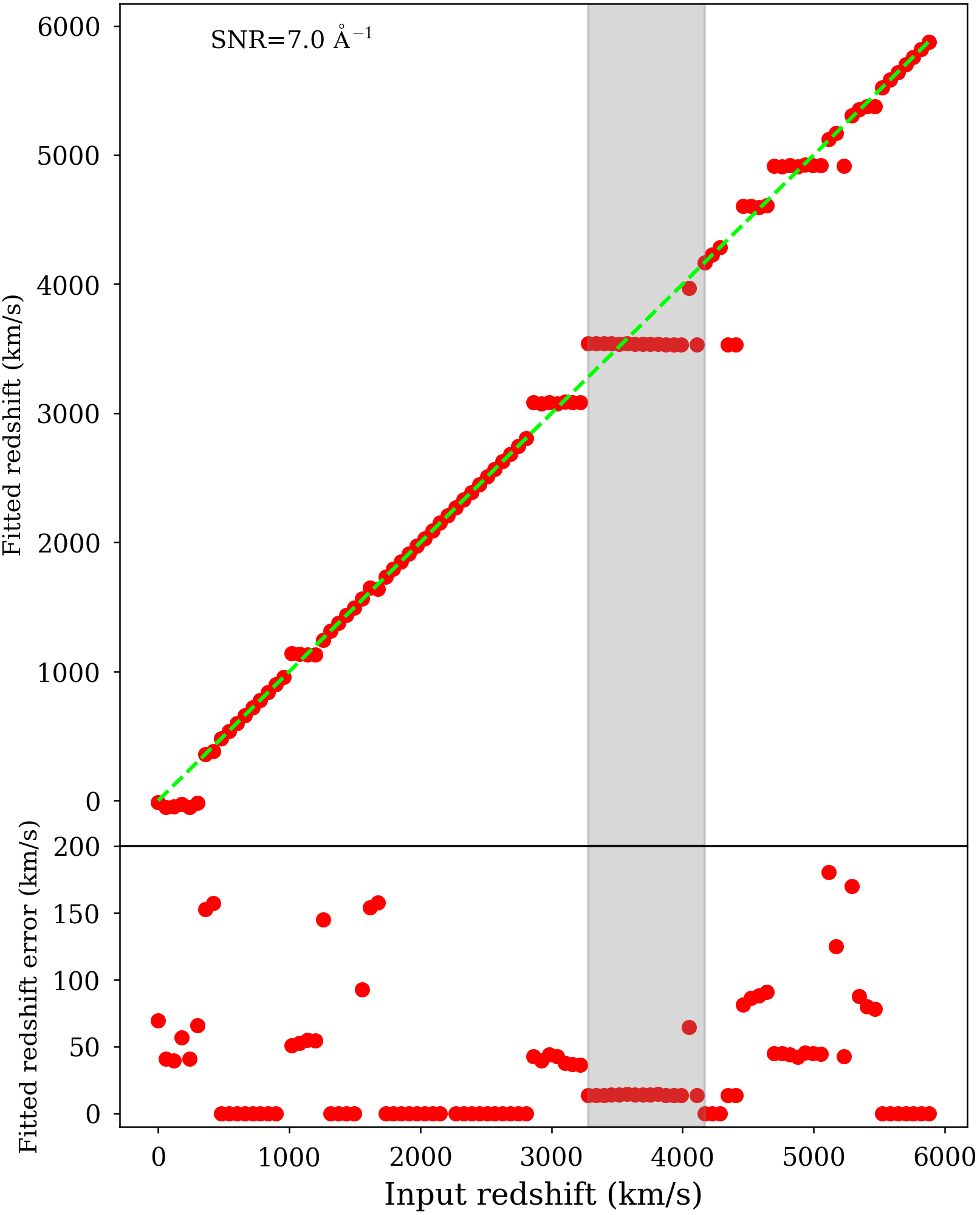}
    \caption{Relation between the the initial input redshift ($z_{\rm input}$) and fitted redshift ($z_{\rm fit}$) obtained from the pPXF. The green dashed line shows the one-to-one relation. The gray shaded region indicates the adopted redshift.}
\label{fig:z_guess_z_fit}
\end{figure}

At the adopted distance of $\sim$50 Mpc for the Hydra I cluster \citep{Christlein03}, GCs are not spatially resolved at the IQ of our entire dataset. For reference, the median effective radius of GCs in the Milky Way (MW) is  $\sim$2.5pc \citep{brodie06}. At the adopted distance, the MUSE spatial resolution of $0\farcs2$ per spectral pixel (spaxel) corresponds to $\sim$50pc. Therefore, even under the optimal image quality conditions in our data, GCs are point-like sources. 
We used the PyMUSE package \citep{Pessa18} to extract the spectra from the sources within a circular aperture. The spectral SNR was determined in a continuum region around 6500 $\AA$ using the $estimateSNR$ function of PyAstronomy \citep{Czesla19}.

We tested in PyMUSE all the available methods for extracting spectra while varying the radius of the apertures and compared the SNR as a function of these different extraction methods and aperture radii. Our analysis identified that the optimal aperture radius is 4 pixels (0$\farcs$8), and the most effective method for combining the spectra involves weighting the flux using both the information from the white image and the variance map\footnote{The adopted extraction method is 'wwm\_ivar'. Check the PyMUSE webpage \url{https://pymuse.readthedocs.io/en/latest/tutorial.html} for more details.}. This aperture is applied to all sources, regardless of their magnitude. We confirmed that, across a wide magnitude range, the optimal aperture radius remains basically unchanged.
Figure \ref{fig:rad_sn_trend} illustrates the behavior of the SNR as a function of the aperture radii for the selected extraction method for one source in the field of UDG\,11. The SNR reaches a peak, followed by a decline at larger radii, primarily due to adding background noise to the spectra. 

The local background for each source was derived from an annular aperture with an inner radius of 8 pixels and an outer radius of 11 pixels, centered on the source. To prevent contamination from neighboring sources during the background spectrum extraction, we masked the positions of all surrounding sources.
For sources located at small galactocentric distances, extracting the background spectrum can be particularly challenging due to the varying surface brightness profile of the galaxy. Consequently, for sources between $3\farcs0$ and $10\farcs0$ from the UDG photocenter, we adopted a smaller annular aperture with inner and outer radii of 6 and 9 pixels, respectively. 
In the case of compact and bright sources close to the galaxy core ($<3\farcs0$), we chose not to subtract the local background. This decision is based on the observation that nuclear sources in UDGs and dwarf-like galaxies typically stand out significantly against the galaxy light profile \citep[see Appendix \ref{sec:app_central_cource} and][]{Fahrion20,Fahrion25}. To justify this choice, we examined the spectra before and after background subtraction. The latter revealed insufficient SNR to estimate the radial velocity. For these sources, we extracted spectra within a circular aperture of 4 pixels instead of 8. Additional details regarding the sources for which this approach was applied, along with the tests conducted to verify the reliability of the radial velocity estimates, are provided in Appendix \ref{sec:app_central_cource}.

\subsubsection{Spectrum fitting}
\label{sec:spec_fitting}

We fitted the source spectra using the penalized Pixel-fitting code \citep[pPXF,][]{Cappellari04,Cappellari17}, which applies full spectral fitting via penalized maximum likelihood, combining input template spectra to analyze stellar systems. We adopted the single stellar population (SSP) model spectra from the Extended Medium resolution INT Library of Empirical Spectra\footnote{\url{http://miles.iac.es/}} \citep[E-MILES,][]{Vazdekis10,Vazdekis16}. These models provide extensive wavelength coverage from 1680 to 50,000 $\AA$ and include BaSTI isochrones \citep{pietrinferni04,pietrinferni06}, offering a grid of ages ranging from 30 Myr to 14 Gyr, and total metallicities [M/H] values between -2.27 dex and +0.04 dex. The E-MILES library includes only the so-called baseFe models, which are based on empirical spectra and thus reflect the abundance patterns of the stars in the library. These models maintain [Fe/H] = [M/H] at higher metallicities but incorporate $\alpha$-enhanced spectra at lower metallicities.
We adopted a MW-like double power-law initial mass function (IMF) with a high-mass slope of 1.30 \citep{Vazdekis96}. The model spectra have a spectral resolution of 2.51 $\AA$ in the wavelength range covered by MUSE \citep{FalcBarroso11}, which is nearly the same as the mean instrumental resolution \citep[$\sim2.5\ \AA$,][]{Bacon17}.


We used only the spectral templates with age $t\geq8\ \mathrm{Gyr}$, which is typical for GCs in the nearby universe \citep[e.g.,][]{brodie06,Fahrion20}.

We fitted the spectra of the sources with $\mathrm{SNR}   \gtrsim2.5\ \mathrm{\AA^{-1}}$  to estimate the line of sight velocity (LOSV) assuming an additive polynomial of degree 12 and no multiplicative polynomial \citep{Fahrion20,Iodice23}. For each fit, the velocity dispersion was set at $10\ \mathrm{km\ s^{-1}}$. At the spectral resolution of MUSE, the internal dispersions of our GCs cannot be resolved. To evaluate the impact of this assumption, we re-analysed the spectra with different initial dispersion values ranging from 1 to $30\ \mathrm{km\ s^{-1}}$ (Beasley et al. 2024). The resulting radial velocities have a standard deviation of 0.1 $\mathrm{km\ s^{-1}}$, demonstrating that our results are insensitive to the choice of initial dispersion. We limited the fitting to $\lambda\leq7000\ \mathrm{\AA}$ to avoid contamination from residual skylines at redder wavelengths and masked background subtraction residuals during the fitting procedure.  
To obtain an initial guess of the LOSV, we provided pPXF with a series of input redshift ranging from  z=0 to z = 0.02 ($\sim 6000\ \mathrm{km\ s^{-1}}$, larger than $ V_{\text{Hydra}} + 3\sigma_{\text{Hydra}}$\footnote{The systemic velocity and velocity dispersion of the cluster are 3683 $\mathrm{km\ s^{-1}}$ and $\sim700\ \mathrm{km\ s^{-1}}$, respectively \citep{Christlein03,Lima-Dias21}.}). We then analyzed the relationship between $z_{\text{input}} $ and $z_{\text{fit}} $. 

\begin{figure}[ht]
    \centering
    \includegraphics[width=\columnwidth]{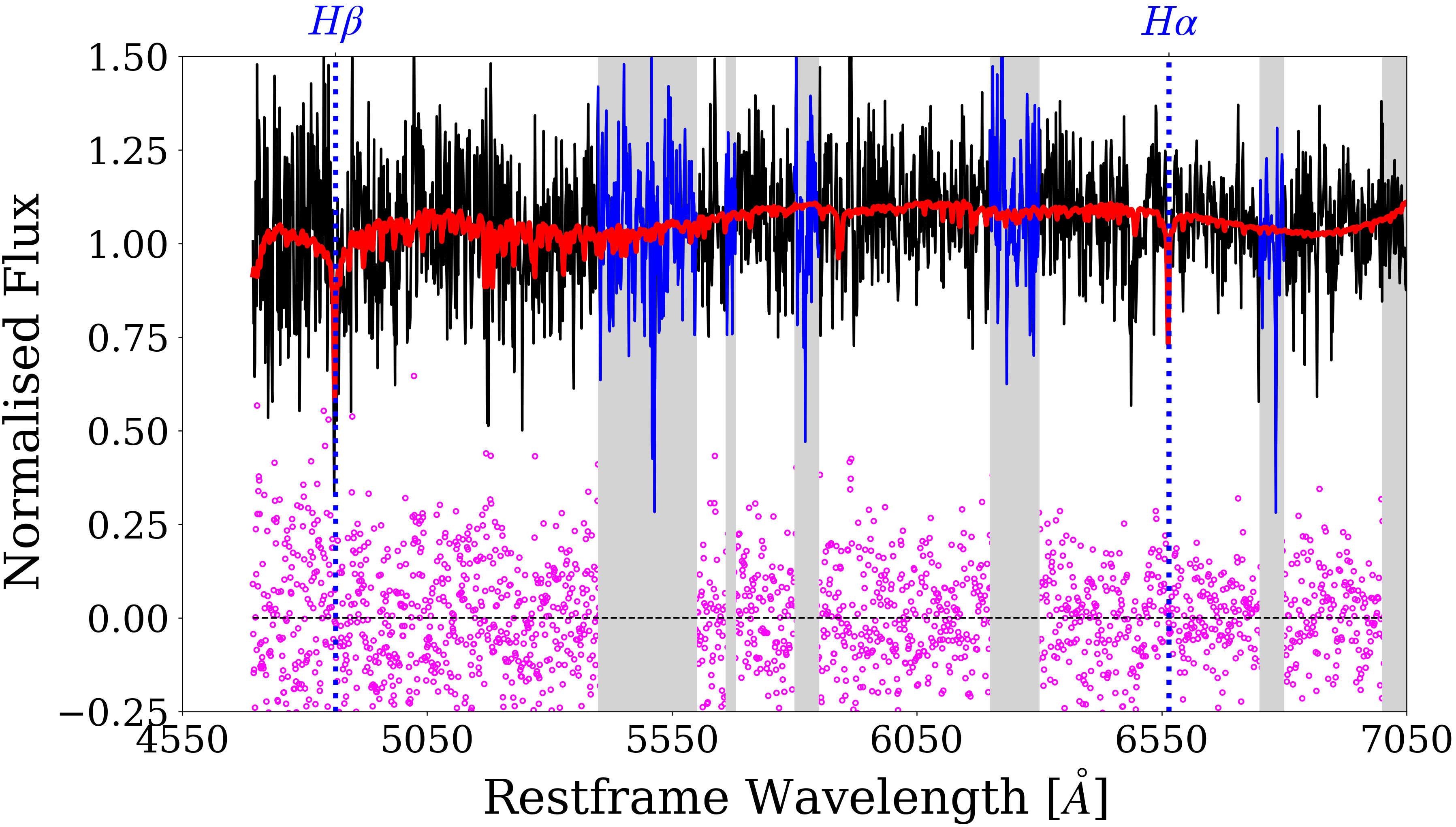}
    \caption{Spectrum of a GC (black line) with SNR of $7\ \AA^{-1}$, along with the pPXF best-fit model (red solid line). Regions with strong sky residual lines are masked from the fit (grey shaded areas). Purple points show the residual spectrum (observed-model).}
    \label{fig:GC_spectra}
\end{figure}

\begin{figure}[ht]
    \centering
    \includegraphics[width=\columnwidth]{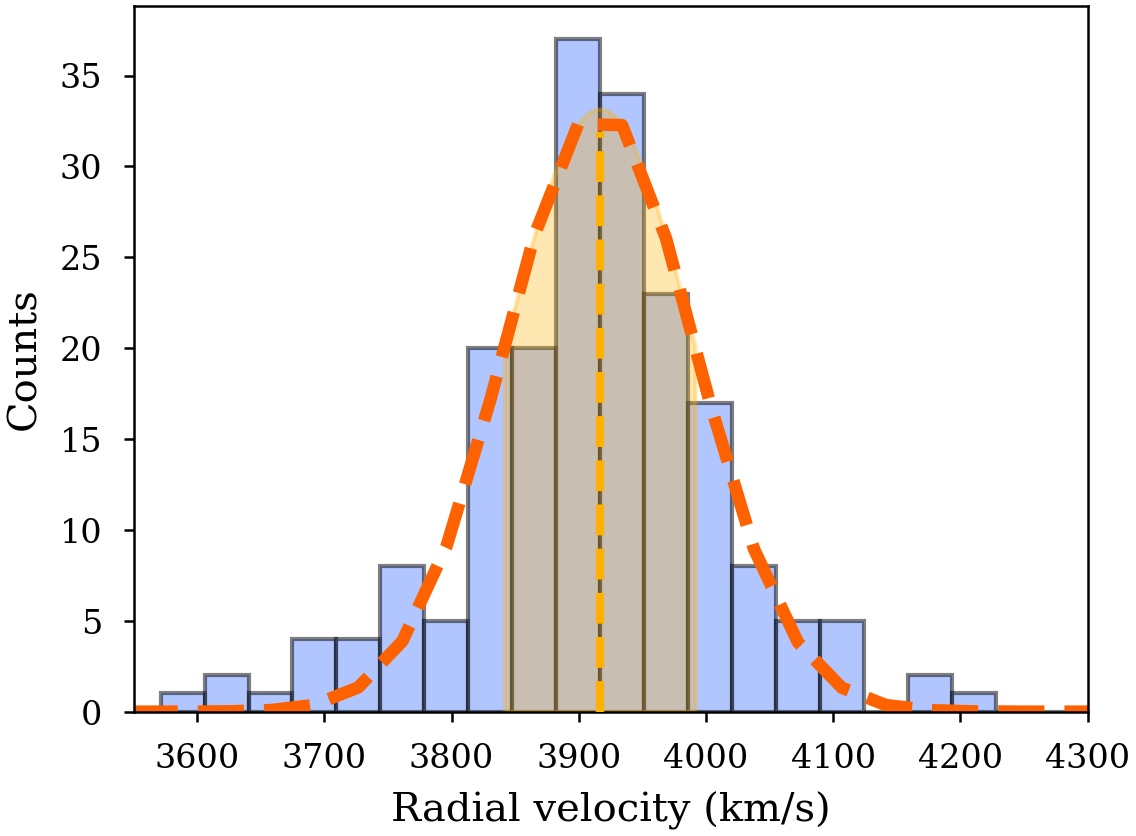}
    \caption{The distribution of radial velocity values derived from the MC simulations is shown as blue histogram. The dark orange dashed curve represents a Gaussian fit to the distribution. The mean and one standard deviation around the mean are shown as an orange vertical line and shaded region, respectively. These values correspond to the adopted radial velocity and its uncertainty.}
    \label{fig:ex_fit_mc}
\end{figure}

\begin{table*}[h!]
\centering
\renewcommand{\arraystretch}{1.2} 
\begin{tabular}{c c c c c c c c}
\hline
\hline
Galaxy & RA [J2000] & Dec [J2000] & $v_{LOS}$ & $\Delta v$ & SNR & $m_H$ & Classification \\
 & (deg) & (deg) & (km/s) & (km/s) & (Å$^{-1}$) & (mag) & \\
\hline
\hline

\hline
\multirow{3}{*}{UDG3} & 159.244433 &-27.1359946 & $3581\pm19$ & $-30\pm20$ &$4.3$ & $22.600\pm0.10$ & GC$^{\square}$ \\
                       & 159.243494 & -27.637910 & $5065\pm29$& $-1454\pm30$ & $3.2$ & $23.14\pm0.20$ & ICGC$^{\square}$ \\
                       & 159.2440753 & -27.1358455 & $2706\pm24$& $+905\pm25$ & $2.90$ & $23.46\pm0.17$ & ICGC$^{\square}$ \\
\hline
\multirow{3}{*}{UDG7} & 159.1548137 & -27.3819035 & $4137\pm24$ & $-11\pm24$ & $6.3^*$ & $22.58\pm0.10$ & GC \\
                    & 159.1562253& -27.3813961 & $4100\pm25$& $+26\pm25$ & $2.8$ & $23.23\pm0.17$ & GC$^{\square}$ \\
                       
                       & 159.1611879 & -27.3874636 & $4211\pm9$ & $-85\pm10$ & $20.0$ & $21.36\pm0.07$ & UCD \\

\hline

\multirow{6}{*}{UDG11} & 158.7486937 &  -27.4272033 & $3470\pm30$  & $+37\pm30$  & $6.14^*$& $22.96\pm0.16$ & GC\\
                       & 158.7483019 & -27.4269613& $3537\pm14$  & $-30\pm15$ & $7.04^*$ & $22.47\pm0.10$ & GC \\
                       & 158.748077 & -27.4269608 & $3519\pm16$  & $-12\pm17$  & $6.50^*$ & $22.54\pm0.10$  & GC \\
                       
                        & 158.7478277 & -27.4292176  & $3443\pm41$  & $+64\pm41$  & $5.00$ & $23.21\pm0.5$ & GC \\
                       
                       & 158.7522136 &  -27.4338378 & $3624\pm22$  & $-114\pm23$  & $3.3$ & $23.31\pm0.13$ & ICGC \\
                       & 158.7475517 & -27.4279218  & $3927\pm87$  & $-414\pm87$  & $4.20$ & $22.8\pm0.3$ & ICGC \\
\hline
\hline

\end{tabular}

\caption{ Properties of spectroscopically identified GCs associated with UDG3, UDG7, and UDG11. Columns list: (1) host galaxy; (2) and (3) celestial coordinates (RA, DEC); (4) line-of-sight velocity ($v_{LOS}$); (5) velocity offset with respect to the systemic galaxy velocity ($\Delta v$); (6) spectra signal-to-noise ratio (SNR); (7) total $H$-band magnitude; and (8) classification as GC, intracluster GC (ICGC), or ultra-compact dwarf (UCD). Asterisks next to SNR values indicate the sources for which the local background was not subtracted. Squares next to the classification of the sources indicate whether the sources were selected as GC candidates in \citet{iodice20} and \citet{lamarca22b}, based on optical $g$ and $r$ photometry.}
\label{tab:spectra_gc}
\end{table*}

Figure \ref{fig:z_guess_z_fit} shows the adopted procedure highlighting relationship between $ z_{\text{fit}}$ ($\Delta z_{\text{fit}}$) and $z_{\text{input}}$ for a source with $\mathrm{SNR}=7\ \mathrm{\AA^{-1}}$. When the $z_{\text{input}}$ produces no useful solution, the fitted output LOSV aligns along the $z_{\text{fit}} = z_{\text{input}}$ line (green dashed line in Fig. \ref{fig:z_guess_z_fit}). Conversely, as we approach the correct solution (gray shaded region), we observe a flattening of the $z_{\text{fit}}$-$z_{\text{input}}$ relation. 
In cases where more than one broad flattening is observed, we use the uncertainties of the $z_{\text{fit}}$ parameter, which correlate with the goodness of fit \citep{Cappellari17}, selecting solutions with the smallest uncertainties. This approach serves as the main criterion for selecting the appropriate flattening region. Additionally, we visually inspect the fitted spectra to evaluate the possible solutions.
In the case of sources with $\mathrm{SNR} < 2.5\ \mathrm{\AA^{-1}}$, regardless of the input value provided to pPXF, the output
remains the same as the input. Hence, we cannot derive velocities for spectra below such SNR.

Figure \ref{fig:GC_spectra} shows the spectrum of a GC and the
corresponding pPXF fit as example.
We adopted a Monte Carlo approach to obtain a realistic estimate of the LOSV and its uncertainty \citep{Cappellari04,Fahrion20,Fahrion22}. 
Once the redshift is identified, we generated 200 realizations that preserved the same SNR as the original spectra. By perturbing the noise-free best-fit spectrum (red line in Fig. \ref{fig:GC_spectra}) using random draws from the residuals (the difference between the original spectrum and the best fit, magenta dots in Fig. \ref{fig:GC_spectra}) in each wavelength bin. We then repeated the fitting process using the redshift identified with the above procedure as input. The mean LOSV and its random uncertainty are obtained by fitting a Gaussian to the resulting distribution (Figure \ref{fig:ex_fit_mc}).

Several studies of UDGs use the criterion $|\Delta v = v_{\text{galaxy}} - v_{\text{source}}| \leq 200\ \mathrm{km\ s^{-1}}$ to identify sources as being bound to their host galaxies \citep[e.g.,][]{Toloba18, Doppel23, Gannon24}, while other have a more conservative threshold of $100\ \mathrm{km\ s^{-1}}$ \citep{Muller20, Haacke25}. We adopt the conservative criterion $|\Delta v| \leq 100\ \mathrm{km\ s^{-1}}$ to decide whether a source is gravitationally bound to a galaxy. Compact sources with velocities within the range $V_{\text{Hydra}} \pm 3\sigma_{\text{Hydra}}$ and $|\Delta v| > 100\ \mathrm{km\ s^{-1}}$  are classified as intra-cluster GCs (ICGCs). Sources with a $|\Delta v|$ consistent with 100 km/s within the error bars were classified as either GCs or ICGCs based on their distance from the galaxy. Specifically, sources located within 1.5 $R_e$ of the galaxy were considered GCs, while those beyond this radius were classified as ICGCs. All those sources that showed convergence around zero are classified as stars, while those that did not converge within the inspected intervals are considered to be background galaxies.


\subsection{Spectroscopic sample of GCs}
\label{sec:spe_cat}

Table~\ref{tab:spectra_gc} presents the properties of spectroscopically confirmed GCs and ICGCs identified in the MUSE datacubes of the four UDGs. Sources marked with a square in the "classification" column are those selected as GC candidates in \citet{iodice20} and \citet{lamarca22b}. About  $\sim$36\% of spectroscopically confirmed GCs were previously tagged as GC candidates. In some cases, the sources selected by \citet{iodice20} and \citet{lamarca22b} are too faint to be spectroscopically analyzed. With our procedure, we cannot measure radial velocities for sources fainter than $m_H \simeq 23.5$ mag.
For three of the four studied galaxies we confirm the presence of bound GCs. UDG\,9 is the only galaxy in our sample for which we did not detect any GCs or ICGCs brighter than $m_H\sim23.5\ mag$ in the field (see also Sect. \ref{sec:tot_n_gc}). This galaxy resides in a dynamically active region of the cluster, at a projected distance of approximately $35\ \mathrm{kpc}$ from both NGC\,3313 and ESO\,510-49, aligned along its major axis. NGC\,3313 also exhibits evidences of past interaction(s) such as stellar streams \citep{spavone24}. Both neighboring galaxies are members of the cluster and have a velocity difference of $\Delta V \sim 300\ \mathrm{km\ s^{-1}}$ relative to UDG\,9 \citep{Smith04}. Past interactions in this environment may have stripped away the GC population from UDG\,9.
Additionally, in the field of the UDG\,7 we identified an ultra-compact dwarf (UCD) with a velocity consistent to be gravitationally bound to the UDG. The UCD differs from GCs in its intrinsically brighter magnitude ($M_H\simeq-12\ mag$, see Sect.~\ref{sec:mag_cut} and Tab.~\ref{tab:spectra_gc}) and its higher SNR of $20\ \AA^{-1}$ (see Tab.~\ref{tab:spectra_gc}).


\section{Selection of GC candidates from multi-band photometry}
\label{sec:sel_gc}

\begin{figure*}[h!]
\sidecaption
\includegraphics[width=6cm]{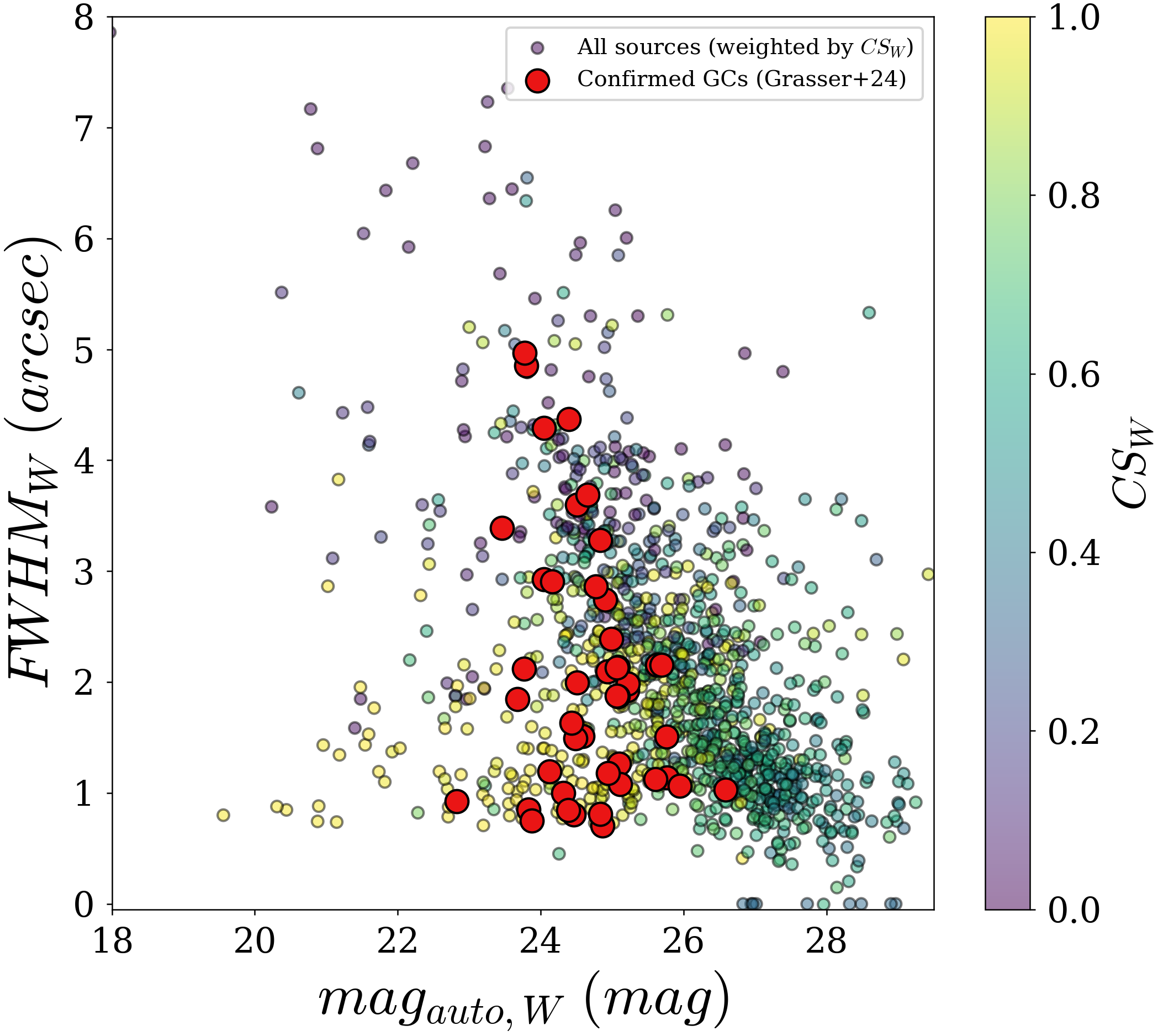}
\includegraphics[width=6cm]{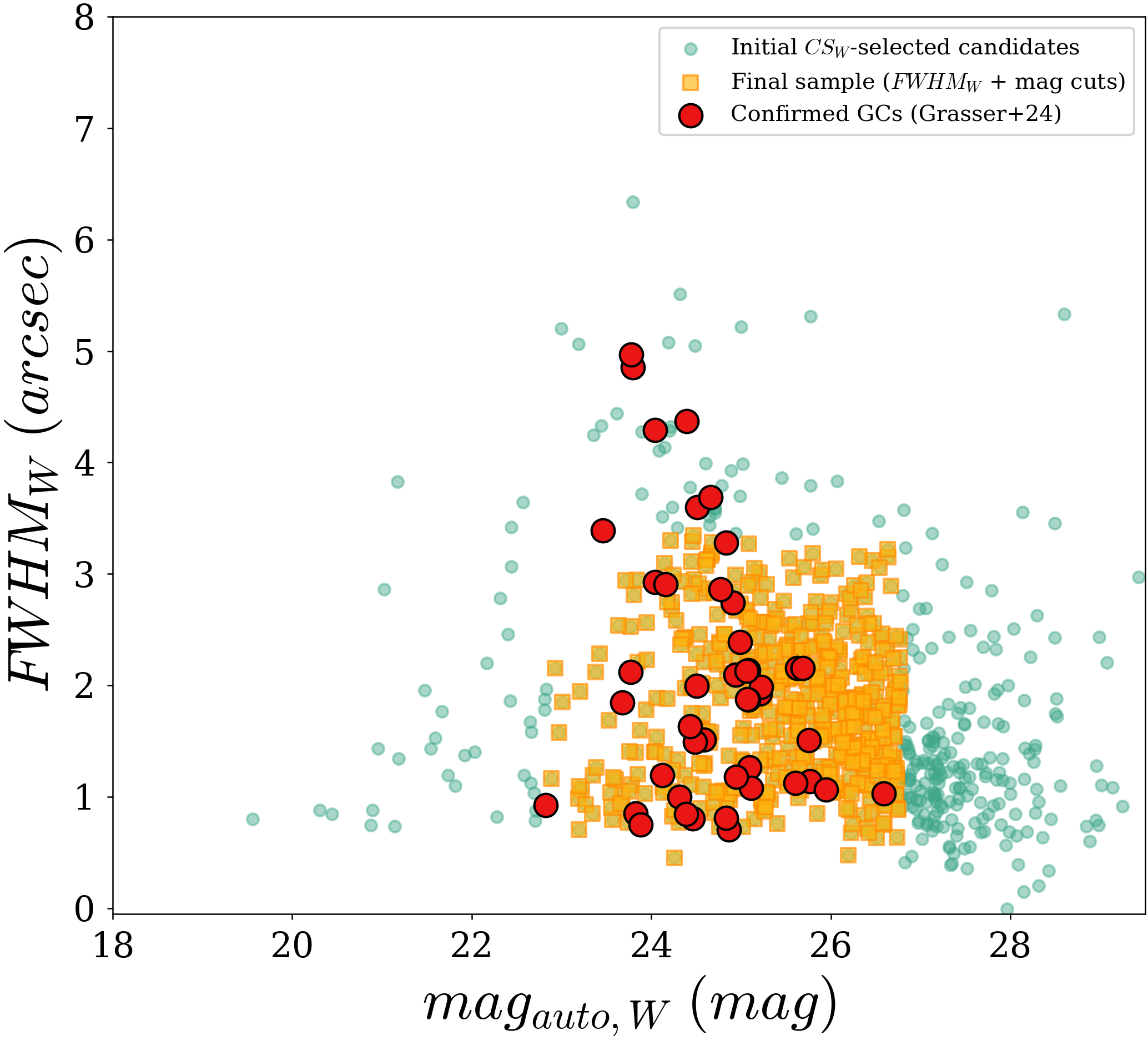}
     \caption{Selection of sources for constructing the master catalog of GCs. Left panel: $\mathrm{FWHM}_W$ versus $mag_W$ for all sources in the three ancillary MUSE fields, weighted by their $\mathrm{CS}_W$ values. Spectroscopically confirmed GCs from \citet{Grasser24} are shown as red dots. Right panel: Source selection process, showing initial candidates selected based on CS parameter (cyan dots), final sample after applying $\mathrm{FWHM}_W$ and magnitude cuts (orange squares), and spectroscopically confirmed GCs (red dots).}
     \label{fig:master_cat}
\end{figure*}

We favour the spectroscopic data for the identification of GCs candidates. However, reliable LOSV measurements require $\mathrm{SNR}\geq2.5\ \mathrm{\AA^{-1}}$ \citep{Fahrion20} which limits this method to the relatively bright sources ($m_H\lesssim23.5$ mag), while MUSE images provide good photometry for significantly fainter sources (see Fig. \ref{fig:fields}). 

We introduce a procedure that integrates the wide wavelength coverage of the MUSE observations and combines it with H-band data to identify a sample of GC candidates too faint for spectroscopic confirmation.

\subsection{Photometric analysis}
\label{sec:phot_approach}

We outline the photometric procedure implemented to obtain the properties of the sources in the $g^*$-, $r$-, $i$- and white (W) images from the MUSE cube, along with the $H$-band dataset. The MUSE spectroscopy is used to further clean the sample even if low SNR prevents reliable LOVS measurements and membership confirmation. Our objective is to use near-IR and MUSE-based optical photometry to create a catalog of faint GC candidates. 

The $g$ and $r$ observations from the VEGAS survey are used to check if the photometry from the MUSE data cubes is consistent and stable from one observation to another. 

\subsubsection{Photometry and photometric calibration}
\label{sec:phot_cal}

We used SExtractor in \textit{dual-image} mode for photometry. In this mode, photometric measurements are performed for all the sources detected on a reference image. Our reference image is the white-band image, the deepest and sharpest one in our dataset (see Fig. \ref{fig:fields}). We refined the detection parameters for this reference image and then measured the $g^*$, $r$, and $i$ band properties of all the sources detected in the white image. We also use as weight image for the detections and photometric measurements the integrated error images extracted along with the $g^*$, $r$, and $i$ band.
As a reference, we adopted the magnitudes obtained within an aperture of 8 pixels (1\farcs6 at the MUSE pixel scale) to measure the colors of the sources. We did not use the automatic magnitudes computed by SExtractor to avoid biasing the flux measurement, especially for faint sources. The magnitudes derived from MUSE images lack photometric calibration. We applied an arbitrary ZP of 30 to the instrumental magnitudes.
The $g^*riW$ matched catalog is then used to select the GC candidates (see Sect. \ref{sec:sel_gc}).


Due to the different pixel scale between MUSE and VIRCAM, it is not possible to run SExtractor in \textit{dual-image} mode across the two dataset. We aligned the MUSE images with the $H-$band image and then used the $Photutils$ package to build a routine that performs aperture photometry on VIRCAM images at the same location of the sources in the MUSE images. First, we generated a 2D background map using the $Background2D$ class with a mesh size of $32\times32 $ pixels, applying a median-sigma-clipping to estimate and subtract the background. We subtracted the local background around each source estimated within an annulus before calculating the final flux. This step ensures that small-scale variations in the background are accounted for.
 The aperture photometry was performed using the $aperture\_photometry$ function of $Photutils$, which integrates the flux within each aperture while dividing pixel contributions proportionally for those lying on the aperture boundary. 
The magnitude of all sources in each aperture was computed as: 

\begin{equation}
    m_H=-2.5\cdot log_{10}(F_{aperture}) +ZP + AB_{factor}
\end{equation}

where the photometric ZP is 26.13 mag. The term $AB_\mathrm{factor} = 1.384$ is used to convert from the Vega magnitude system to the AB magnitude system in the $H$-band \citep{Gonzalez18}, while $F_\mathrm{aperture}$ (in counts) represents the background-subtracted flux measured within the specified aperture.
We adopted as our reference magnitude the one estimated within an aperture of 1\farcs6, ensuring that colors of sources are computed from fluxes derived from within the same physical area for all bands\footnote{The image quality of our datasets are comparable within the errors (see Secs. \ref{sec:spec_data} and \ref{sec:sec_phot_data}).}. 

To account for flux outside this aperture, we applied a curve of growth approach and determined the aperture correction necessary to obtain the total magnitude of each source. Given the small FOV around the UDGs, the aperture correction was estimated for a sample of bright, compact, and isolated sources from the full tile of the Hydra-I cluster (see Sect. \ref{sec:sec_phot_data}).
By comparing the magnitude within the reference aperture to those at larger radii ($\Delta mag = m_{1\farcs6} - m_i$, where $i$ represents larger apertures), we obtained the asymptotic aperture correction value of $0.49\pm0.02$ mag. This correction was then applied to derive the total apparent magnitude for each source. For the foreground extinction correction, we obtained the mean extinction value across 2 sq. degrees around NGC\,3311 from the IRSA (NASA/IPAC Infra-Red Science Archive) archive, which provides $E_{B-V}=0.0684\pm0.0007$ mag. We derived the $H$-band extinction as $A_H$=0.449$\cdot E_{(B-V)}$ adopting UKIRT $H$-band reddening value from Table 6 in \citet{sf11} assuming a $R_V$=3.1.
The complete $g^*riWH$ matched catalog for each galaxy is available on CDS, containing a total of 288, 357, 321, and 435 detected sources for UDG\,3, UDG\,7, UDG\,9, and UDG\,11, respectively.

%



As a sanity check, we compared our $H$-band photometry with the 2MASS one \citep{Skrutskie06}. After matching our catalog with 2MASS and selecting compact bright sources, we converted the 2MASS $H$ magnitude into the VISTA system using Eq. 8 in \citet{Gonzalez18}: 

\begin{equation}
    H_V = H_{2M} + (0.015 \pm 0.005) \cdot (J-K_s)_{2M}
\end{equation}
where $H_{2M}$ and $H_V$ are the 2MASS and VISTA magnitude, respectively, and $(J-K_s)_{2M}$ is the 2MASS color. Table \ref{tab:phot_check} reports the median magnitude offset of the matched sources together with the rms median absolute deviation (MAD)  of the magnitude offset ($\mathrm{rms}_{\mathrm{MAD}}\simeq1.48\cdot \mathrm{MAD}$). Our data are in excellent agreement with 2MASS.


The OmegaCAM@VST observations were used to verify whether the MUSE photometry was consistent across the dataset. We first checked the photometric calibration of the OmegaCAM data by comparing our measurements with the photometry of bright and isolated stars in PanSTARRS \citep{Chambers16}.
We compared the aperture corrected  magnitudes\footnote{We estimated the aperture correction as the difference between the magnitude measured within a $1\farcs6$ aperture and that measured within a $4\farcs0$ aperture (i.e. the pipeline calibration diameter). The resulting aperture corrections are $0.37 \pm 0.01$ mag and $0.33 \pm 0.01$ mag for the $g$ and $r$ bands, respectively.}from our dataset with the Pan-STARRS catalog. For the $g$-band, sources were also selected based on color due to differences between the two photometric systems \citep{Tonry12}. The adopted color interval used to compare the two photometric systems for the $g-$band was $|g-r|<$ 0.5 mag \citep{Tonry12}. The results of this comparison are shown in Tab. \ref{tab:phot_check}. We corrected our VST $r$-band measurements by applying the offset relative to the PanSTARRS photometry reported in Tab. \ref{tab:phot_check}. The small discrepancy between the two photometric systems will be further investigated in an upcoming paper (Mirabile et al. in prep).

Finally, we compared the photometry from the $g^*$- and $r$-bands of the LEWIS MUSE observations with the $g$- and $r$-band measurements from the VST. We thus estimated the median difference in magnitude ($\Delta \mathrm{mag} = m_{\mathrm{VST}} - m_{\mathrm{MUSE}}$) and the $\mathrm{rms}_{\mathrm{MAD}}$ for the four fields.
A consistent offset was observed among the fields, with a small scatter ($rms_{MAD}$=0.1 mag), indicating appropriate photometric quality of the dataset. 
We also attempted to verify if the same offset and variation were observed in the three MUSE fields (A-C, see Sect. \ref{sec:dataset}) around NGC\,3311. However, this analysis did not produce useful results, possibly because of the high galaxy background in the A-C fields. Nonetheless, the photometry derived from the ancillary MUSE dataset remains comparable to that from the four LEWIS fields inspected here, as demonstrated when using source colors instead of individual magnitude values (see Sec.\ref{sec:mast_cat} and Sect. \ref{sec:gc_sel_morph_col}).

\begin{table}
    \small
    \centering
    \caption{Photometry quality check.}
    \begin{tabular}{cc}
    \hline \hline
    
      $\Delta mag$ & Offset \\
\hline
    VISTA-2MASS & $0.0\pm0.08$ \\ 
    $(VST-PanSTARRS)_g$& $0.01\pm 0.01$ \\ 
    $(VST-PanSTARRS)_r$&\ $-0.03\pm0.02$ \\

    \hline
    \end{tabular}

    
    \label{tab:phot_check}
\end{table}

\subsubsection{GC master catalog}
\label{sec:mast_cat}

To improve the selection of the GC candidates, we constructed a master catalog by exploiting the properties of spectroscopically confirmed GCs identified by \citet{Grasser24}. We collapsed the MUSE observations taken on NGC\,3311 (fields A, B and C) into three passbands ($g^*-$, $r-$ and $i-$band), modeled and subtracted the galaxy light profile in all the filters, and applied the same photometric procedure described in Sect. \ref{sec:phot_cal}. For all the MUSE-based photometry, we did not apply any foreground extinction correction, as the negligible spatial variability of the extinction across the field (see Sect.  \ref{sec:phot_cal}) implies that such a correction would be uniform and would therefore not affect or improve the relative approach we adopted in Sect. \ref{sec:sel_gc}.
We derived the morphometric (i.e. shape) and photometric (i.e. magnitude) properties of spectroscopically confirmed GCs identified in \citet{Grasser24}. Specifically, we selected sources based on the CLASS\_STAR\footnote{CLASS\_STAR is SExtractor output parameter that provides a stellarity index for classifying astronomical objects. It ranges from 0 (likely extended object) to 1 (compact).} (CS) and FWHM parameters measured by SExtractor in the white image. We applied broad cuts, considering all the sources with $\mathrm{CS}_W\geq0.5$, combined with $0.39\leq \mathrm{FWHM}_{W}\leq 3.35$ and $22.88\leq \mathrm{mag}_{W}\leq 26.78$ mag. The latter intervals were defined using the median $\pm\ 2\ (3)\mathrm{rms}_{MAD}$ of the $\mathrm{FWHM}_{W}\ (\mathrm{mag}_{W})$ of spectroscopically confirmed GCs. We also applied broad color cuts $|g^*-i|<5$ and $|r-i|<5$ to further clean the catalogs from spurious detections. The left panel in Fig. \ref{fig:master_cat} displays the $\mathrm{FWHM}_{W}$ versus $\mathrm{mag}_{W}$ for all sources in the three fields on NGC\,3311, color coded based on their $\mathrm{CS}_W$ values, along with the spectroscopically confirmed GCs (red dots) by \citet{Grasser24}. The right panel in the same figure illustrates the source selection process. It shows the sources initially selected based on the $\mathrm{CS_{W}}$ parameter (light blue dots), the final sample of selected sources after applying cuts in $\mathrm{FWHM}_{W}$ and magnitude (orange squares), and the spectroscopically confirmed GCs (red dots). It is interesting to note that the spectroscopically confirmed GCs from \citet{Grasser24} show a wide range of measured $\mathrm{FWHM_W}$. In particular, some sources are measured to be very extended (i.e., $\mathrm{FWHM_W} > 3.5\ \mathrm{arcsec}$). These sources are either located close to the core of the galaxy, near the edges of the images, or are very faint objects affected by image artifacts in the MUSE data. The observed wide range further supports our choice to adopt conservative selection criteria.

\begin{figure}
    \centering
    \includegraphics[width=\columnwidth]{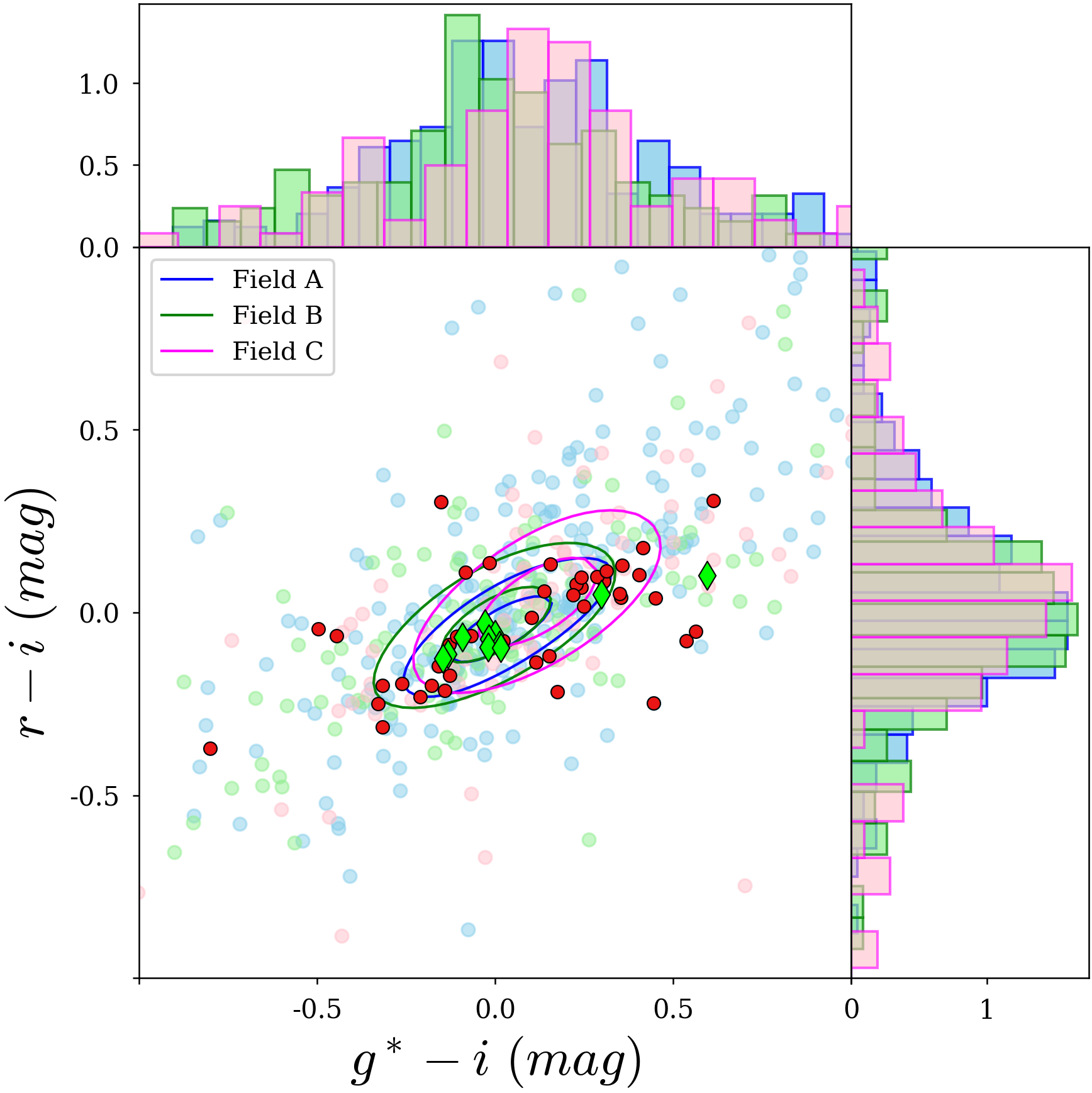}
    \caption{Optical $(g^*-i)$ vs.\ $(r-i)$ color-color diagram. The 20\% (outer) and 5\% (inner) isodensity contours from the master catalog are shown separately for each of the three ancillary fields. Individual sources are plotted in light blue, light green, and light magenta for fields A, B, and C, respectively. Spectroscopically confirmed GCs from this work are marked as green diamonds.
    Spectroscopically confirmed GCs from \citet{Grasser24} are shown as red dots.}
    \label{fig:magda_field_contour}
\end{figure}

Figure \ref{fig:magda_field_contour} shows the isodensity contours\footnote{The isodensity levels of the master catalog are obtained using a kernel density estimator\footnote{\url{https://docs.scipy.org/doc/scipy/reference/generated/scipy.stats.gaussian_kde.html}} (KDE). The adopted parameters were:  bw\_method='scott' and bW\_adjust=1.} of the sources selected in the three NGC\,3311 fields in the color-color plane, along with projected histograms along each axis. Red dots are the spectroscopically confirmed GCs from \citet{Grasser24}. The contour levels represent the region containing 20\% (outer) and 5\% (inner) of the selected sources in each field. The density peaks in these regions align with the locations of spectroscopically confirmed GCs and ICGC identified around UDGs in this work (green diamonds). All but one of these sources are located around the peaks of the distribution. The single exception still falls within the 68\% density contour used to select sources (see Sect. \ref{sec:sel_gc}). This demonstrates that the sources in the master catalog can be used to identify broad regions for selecting GC candidates based on their color information (see also Sect. \ref{sec:gc_sel_morph_col} and Fig. \ref{fig:color_sel}). Finally, we matched this catalog with the $H$-band catalog, thus defining a master catalog containing also the near-IR information. We further refined the catalog removing sources with poor SNR having $m_H \geq 26$ mag. It contains 32/49 confirmed spectroscopic\footnote{The master catalog contains only 32 out of 49 spectroscopically confirmed GCs from \citet{Grasser24} as a consequence of the detection and selection procedure. We miss 3 sources because they are close to the edge of the images, and 14 are discarded mainly during the morphological cleaning process of the catalog.} GCs from \citet{Grasser24}. Considering that the observations were taken on NGC\,3311, which has a very rich estimated GC population of $\sim 16.000$ \citep[][, see also Sect. \ref{sec:tot_n_gc}]{Wehner08}, and assuming a uniform distribution of contaminants\footnote{Assuming the average contamination density estimated in Sect. \ref{sec:tot_n_gc} over the area covered by the three MUSE pointings, we expect a contaminant population of  $\sim$100 sources. In comparison, integrating the radial profile of GC candidates studied by \citet{Wehner08} for NGC\,3311 out to $3\ \mathrm{arcmin}$ yields an expected population of  $\sim$ 2600 GC candidates, which overwhelmingly dominates the contaminant population.} in the observed area, we can safely assume that the fraction of interlopers (unresolved background galaxies and faint MW stars) among the GC population is negligible.

The position and properties of these sources are given in Tab. \ref{tab:mast_cat}. The complete catalog is available on CDS. The final $g^*riWH$ master catalog contains 426 GCs around NGC\,3311.

\subsection{Photometric selection of GC candidates }
\label{sec:gc_sel_morph_col}
Here, we present the procedure adopted to select GC candidates based on a combination of photometric tracers. The approach adopted to identify these compact stellar systems relies on a procedure similar to our previous works \citep{cantiello15,cantiello18fds,Mirabile24,Lonare24,Saifollahi25}.

\subsubsection{Morphometric selection}

\begin{figure*}[ht]
    \sidecaption
    \centering
    \includegraphics[width=12cm]{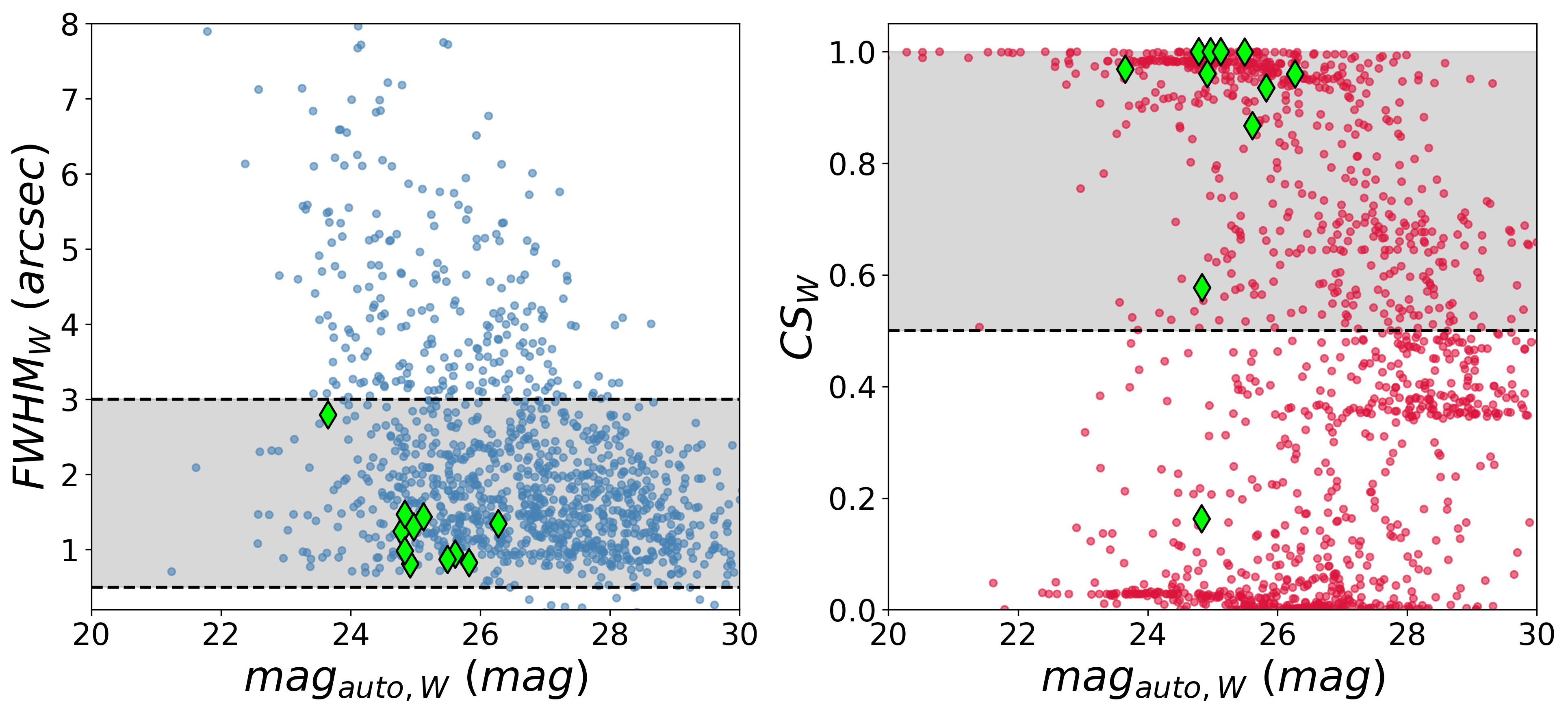}
    \caption{ $\mathrm{FWHM}_W$ and $\mathrm{CS}_W$ versus magnitude for all sources in the four analyzed UDG fields. Spectroscopically confirmed GCs are marked with green diamonds. The adopted selection cuts are indicated by the dashed black lines and the gray shaded region.}
    \label{fig:morpho_sel}
\end{figure*}

 As already noted, GCs at the distance of the Hydra I cluster are basically point-like sources. We relied on two compactness indicators derived from the MUSE white-image, because it provides the deepest and best image quality observations (see Fig. \ref{fig:fields} and Sect. \ref{sec:dataset}). We combined morphometric selection based on the measured $\mathrm{CS}_W$ and $\mathrm{FWHM}_W$. In all the observed fields, we decided to adopt conservatively wide selection ranges for GC candidates: $\mathrm{CS}_W\geq 0.5$  and $0.5\leq \mathrm{FWHM}_W\leq3$ arcsec. These limits were chosen to account for artifacts in the MUSE images that may affect the morphometric properties of the sources (see Sect. \ref{sec:mast_cat}). Figure \ref{fig:morpho_sel} shows the $\mathrm{CS}_W$ and $\mathrm{FWHM}_W$ vs magnitude plot for all the sources in the four analyzed UDG fields, along with the spectroscopically confirmed (IC)GCs (green diamonds) and the adopted selection cuts (dashed black line and gray shaded region). One of the spectroscopically confirmed GCs was excluded by the $\mathrm{CS}_W$ selection criterion. This source, located in UDG\,3, lies very close to the core of the galaxy and to a bright star, which may have affected its $\mathrm{CS}_W$ measurement (see Fig. \ref{fig:sel_pros}).

\subsubsection{Color selection}

The selection of GC candidates can be improved by adding near-IR to optical data. The near-IR (NIR) information is essential to further clean the catalog from MW stars and background galaxies \citep{cantiello18fds,Saifollahi21b}. Indeed, the wider the wavelength coverage is, the better we can distinguish between different populations (e.g., stars, galaxies, GCs, etc.) in a color-color plane \citep{munoz14}. 

Following the approach from \citet{cantiello20} and relying on our master catalog (see Sect. \ref{sec:mast_cat}), we used a color-color plot to refine the selection of GC candidates.
First, we selected sources using the optical ($r-i$) vs ($g^*-i$) and then further cleaned the sample inspecting sources in the ($r-H$) vs ($g^*-i$) color-color plane.

\begin{figure*}[ht]
    \centering
    \sidecaption
    \includegraphics[width=6cm]{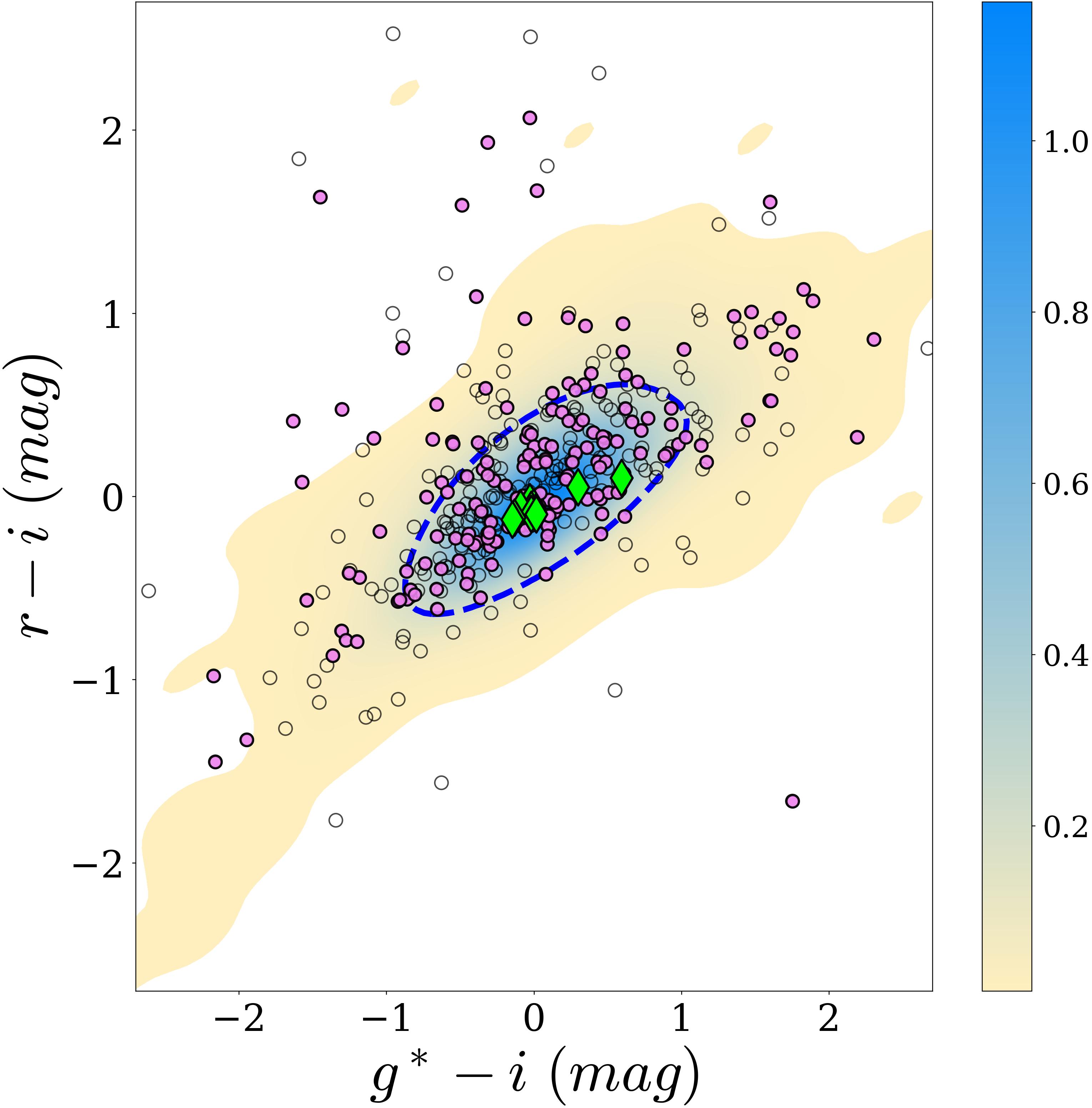}
    \includegraphics[width=6cm]{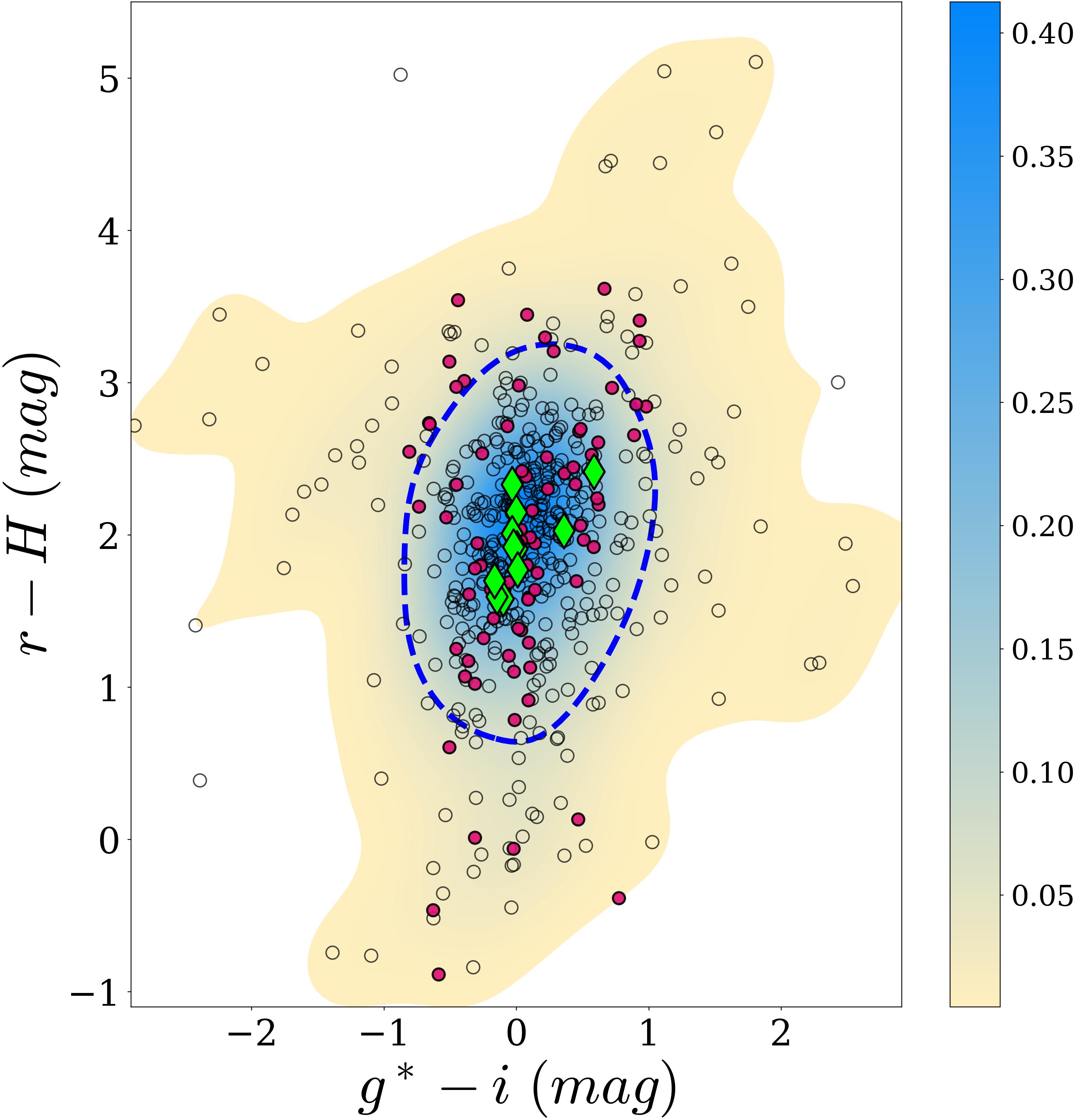}
    \caption{Color-color diagrams highlighting selection cuts for GC candidates in UDG 11. In the left panel, open circles represent all cataloged sources, while violet dots indicate those selected based on morphometric criteria ($\mathrm{CS}_W$ \& $\mathrm{FWHM}_W$). In the right panel, red dots correspond to sources meeting both morphometric and optical-color selection criteria. Spectroscopically confirmed GCs are shown as green diamonds in both panels. Colored regions show the isodensity contours derived from the master catalog. The blue contour encloses 68\% of the master catalog sources.}
    \label{fig:color_sel}
\end{figure*}

Figure \ref{fig:color_sel} illustrates the optical color-color plane (left panel) and the optical-NIR plane (right panel) for UDG\,11. In the left panel, open circles represent all cataloged sources (see Sect. \ref{sec:phot_approach}), while violet dots highlight those passing the morphometric criteria. The right panel shows, as red dots, the sources selected by morphometric criteria and by optical-color criteria. Spectroscopically confirmed (IC)GCs identified in this study are marked with green diamonds.
In both panels, the colored regions show the isodensity contours of the catalog. The blue line represents the contour level that contains 68\% of the sources in the master catalog (see Sect. \ref{sec:mast_cat}). All sources within this blue contour were selected for further analysis.
The consistency between our field and the master catalog in source colors validates our selection method and supports the reliability of color-based GC identification.

\subsubsection{Magnitude selection}
\label{sec:mag_cut}

After the selection of sources based on their morphometric and color properties, we can further clean the sample based on the magnitudes. 
GCs in bright galaxies usually show a universal luminosity function \citep[GCLF, ][]{villegas10,rejkuba12}, which can be used to select GC candidates based on their magnitudes. For UDGs, there is still no consensus if the GCLF is the same as for bright galaxies \citep{shen21,Janssens22,Romanowsky24,Tang25,Saifollahi25}. Nevertheless, to identify a magnitude range, we adopted a Gaussian luminosity function for these galaxies. Adopting as the turn-over magnitude (TOM, peak of the distribution) a value of $H_{TOM}=-8.3$ AB mag \citep{Nantais06} and a distance modulus of $m-M=33.5\ mag$ \citep{Christlein03}, the expected GCLF peak would be  $m_H^{TOM}\sim 25.2$ mag. For the width of the GCLF ($\sigma_{GCLF}$), we adopted a value of $\sigma_{GCLF}=1$ mag. Although $\sigma_{GCLF} \simeq 0.8$ or smaller is more in line with expectations for UDGs/dwarf galaxies \citep{villegas10,lamarca22b, Marleau24a}, we opted for a larger value due to the depth of the cluster \citep{Marleau24a}. We selected all sources within $\pm3\sigma_{GCLF}$ around $m_H^{TOM}$, namely $22.2\leq m_H\leq 28.2$ mag as GC candidates. The faint magnitude limit is beyond the depth/completeness of our dataset (see Sect. \ref{sec:tot_n_gc} and Appendix \ref{sec:comple}), hence we applied only a bright limit. As reported in Tab. \ref{tab:sel_int}, we used a more generous lower limit ($m_H \geq 20$ mag) to include potential ultra-compact dwarfs (UCDs) and nuclear star clusters. Based on the literature, we adopted an absolute magnitude threshold in the $g$-band of $M_g = -10.5$ mag \citep{mieske04, hilker07, cantiello20} as separation criteria between GCs and UCDs. Assuming a color of $g-H\sim1$ AB mag, the separation criteria correspond to an apparent magnitude of $m_H \sim 22$ mag. Therefore, we selected sources with $m_H \geq 20$ mag to ensure the inclusion of possible UCDs\footnote{We adopted a two-magnitude range for UCD selection following \citet{mieske12}.}.

\begin{table}[h!]
    \centering
    \caption{Photometric and morphometric parameters limits adopted for GCs selections.}
    \begin{tabular}{ccc}
    \hline
    \hline
         Parameter& min &max \\
         \hline
         $m_H$ (mag)& 20.0 &....\\
         $\mathrm{CS}_W$&0.5& 1.0\\
         $\mathrm{FWHM}_W$ (arcsec)&   0.5  & 3  \\ 
         \hline
         \hline
    \end{tabular}
    
    \label{tab:sel_int}
\end{table}

\subsection{Spectro-photometric final cleaning}


After identifying a preliminary list of GC candidates using the methods described above, we cross-matched this catalog with our spectroscopic dataset. This step enabled additional refinement of the catalog by removing foreground stars and background galaxy sources meeting the criteria of $\mathrm{SNR}\geq2.5\ \mathrm{\AA^{-1}}$ and $m_H \lesssim 23.5$ mag (see Sect. \ref{sec:spe_cat}). Emission-line sources were removed even if they had lower SNR and fainter magnitudes. At this stage, we also removed the residual fake detections due to image artifacts.

\begin{figure*}
    
    \centering
    \includegraphics[width=\textwidth]{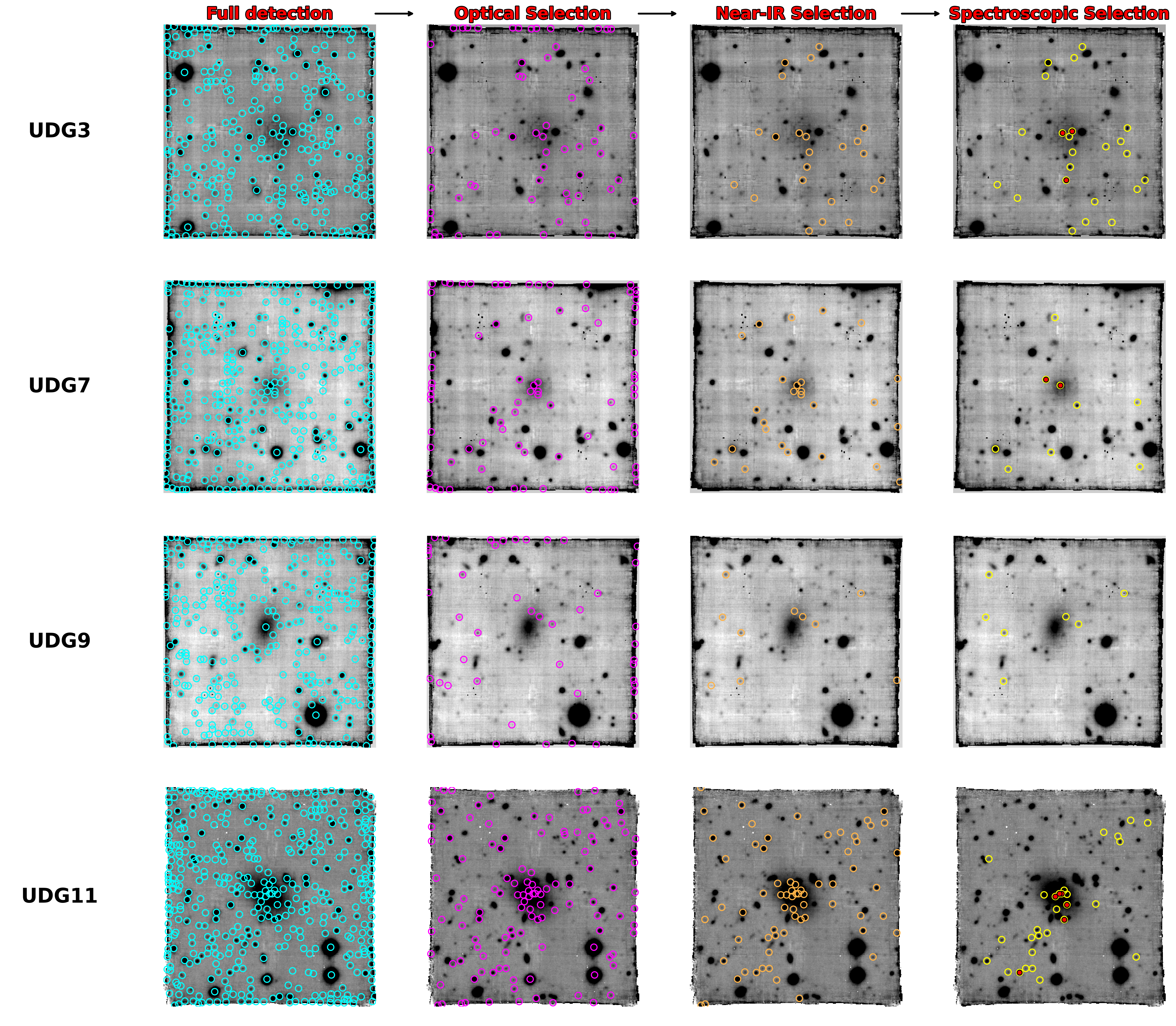}
    \caption{Step-by-step selection process for GC candidates.  Left to right: (1) MUSE white image with all sources detected (light blue circle); (2) sources filtered based on morphometric and photometric properties from optical/MUSE data (violet); (3) further selection incorporating near-IR data (orange); (4) final catalog of GC candidates (yellow). Red filled circles in the last panels indicate spectroscopically confirmed GCs.}
    \label{fig:sel_pros}

\end{figure*}

Figure \ref{fig:sel_pros} illustrates the comprehensive selection process for GC candidates for the four galaxies. The figure is divided into four columns, each representing a stage in the refinement of our catalog. The leftmost panel displays all sources initially detected. Moving to the right, the second panel shows the sources selected based on morphometric and photometric properties derived from the optical/MUSE images. The third panel presents the further refined selection after incorporating $H$-band phootmetry. Finally, the rightmost panel exhibits the final catalog of sources after cross-matching and cleaning with spectroscopic data. In this last panel, red-filled dots denote spectroscopically confirmed GCs. The average rejection rate of sources across the four galaxy fields, when cleaning using spectroscopic information, is  $\sim$50\%. In other words, roughly half of the sources initially selected through our photometric and morphometric procedures were rejected once spectroscopic data were included. 
The large overlap (92\%\footnote{This percentage indicates the fraction of spectroscopically confirmed GCs that were successfully retrieved through the photometric selection procedure.}) between photometrically selected candidates and spectroscopically confirmed GCs further validates the precision of our selection criteria. 
We also note that for UDG\,11, the distribution of the final GC candidates appears to be spatially aligned along the direction of ESO\,501-26, an edge-on disk galaxy, apparently morphologically disturbed at projected distance of $\sim95kpc$ from UDG\,11, with a velocity difference of $\Delta v\sim 500km/s$ relative UDG\,11 \citep{Smith04}. Further analysis of such a possible substructure will be presented in Mirabile et al. (in prep.).

\section{GC population: sample analysis}
\label{sec:analysis_gc}


In the following, we discuss the properties of GC systems in our target galaxies, after applying completeness correction and statistical background decontamination \citep[see][]{cantiello15,dabrusco16} to our final list of GC candidates.

\subsection{Total number of GCs}
\label{sec:tot_n_gc}




Estimating the total population of GCs ($N_{\mathrm{GC}}$) in any galaxy requires correcting for incompleteness and accounting for contaminants, including foreground stars and background galaxies. For galaxies in groups or clusters, one must also consider the population of ICGCs.
The first step requires detailed knowledge of the GCLF, including its shape, peak, and width. The second step entails characterizing the contaminant population.

We begin by counting GC candidates within a specified radius from each galaxy. Following \citet{Marleau24a}, we adopted an outermost radius of 1.5$R_e$, where $R_e$ is the effective radius of the galaxy (see Tab. \ref{tab:lewis_sample}). Any GC identified as a spectroscopic ICGC within the specified radius \citep[see][]{iodice20} was excluded from the count.

To estimate background contamination, we combined all sources from the four observed fields that lie at distances between 15$\farcs$0 and 30$\farcs$0 from the galaxies, well beyond the 1.5$R_e$ limit for all the UDGs (see Fig. \ref{fig:anulus_background}). This approach allowed us to better constrain the contaminant population by increasing our statistical sample.

The number of GCs for each galaxy was calculated as $N_{\mathrm{GC}}^{raw} = A_{\mathrm{gal}} \cdot (\rho_{\mathrm{gal}} - \rho_{\mathrm{background}})$. Here, $\rho_{\mathrm{gal}}$ is the density of GC candidates on-galaxy, and $\rho_{\mathrm{background}}$ is the background density, both expressed in number of sources per square arcsec and $A_{\mathrm{gal}} = \pi \cdot (1.5R_e)^2$. The error associated with each value was determined by considering Poisson errors in both the on- and off-galaxy counts. Sources identified spectroscopically as GCs were excluded from the error propagation. Table \ref{tab:usg_prop} reports the estimate of $N_{GC}^{raw}$ for each galaxy.

The estimated $N_{\mathrm{GC}}^{raw}$ must be corrected for the completeness fraction ($N_{GC}=N_{GC}^{raw}*(1/f_{completeness})$). The most common approach to account for completeness involves obtaining a magnitude-dependent completeness function and using it to correct the GCLF.

However, determining the overall completeness of our GC candidate catalog is not straightforward due to the combination of magnitude-dependent detection completeness, the use of different bands and instruments, and our adopted selection criteria. The detection completeness is mainly dominated by the $H-$band observations, which have the brightest limiting magnitude (See Appendix \ref{sec:comple}). 
Combining GCLFs from multiple UDGs, which are sparsely populated, could improve the statistical significance of the sample, therefore allowing us to determine the combined limiting magnitude.





\begin{figure*}[ht]
    \sidecaption
    \centering
    \includegraphics[width=12cm]{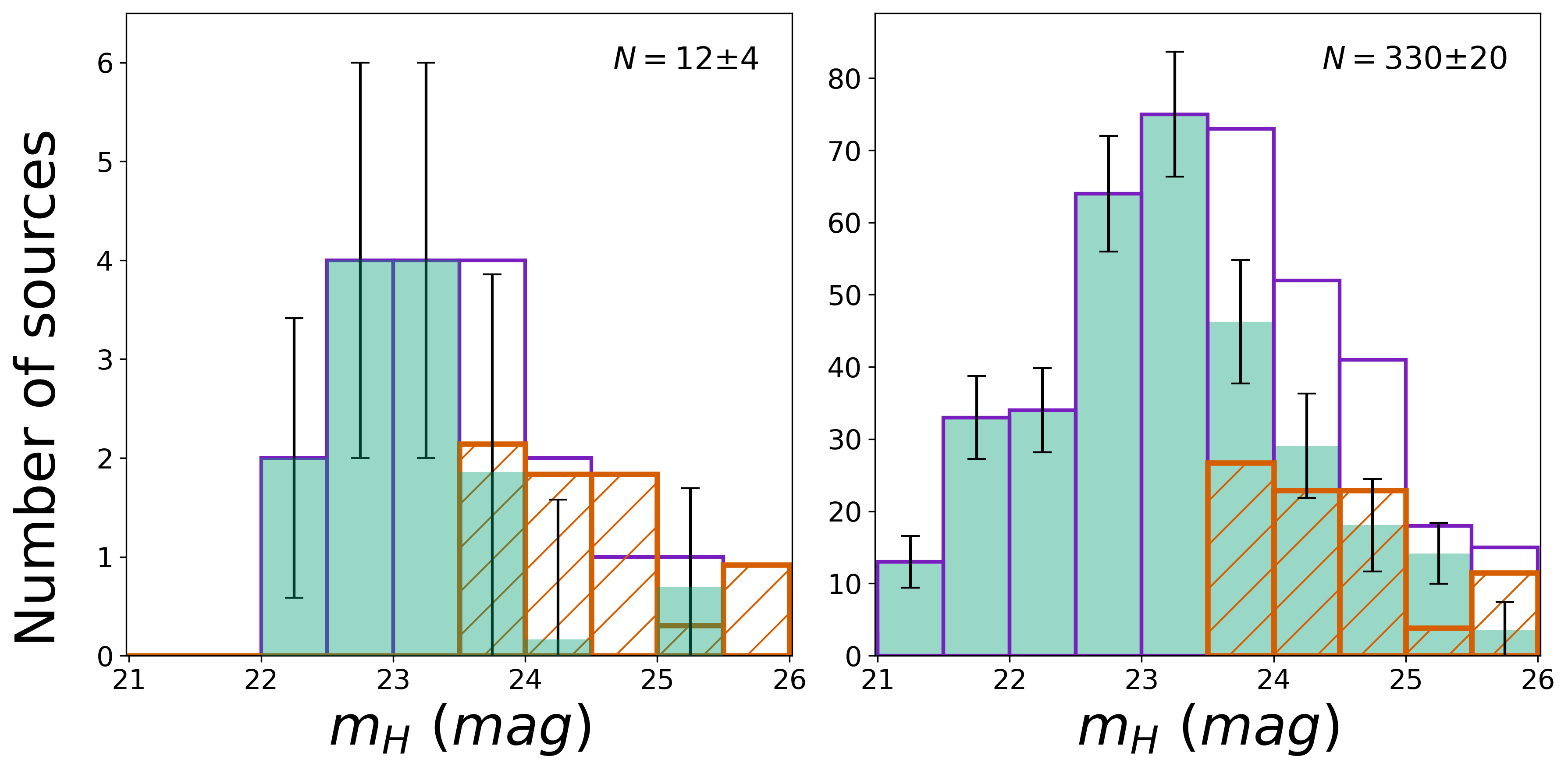}
    \caption{Left panel: Combined LF of GC candidates around all UDGs (violet histogram), shown alongside the assumed background distribution (orange hatched histogram) and the background-subtracted LF (teal histogram). Right panel: LF distributions derived from the full master catalog. The total number of GCs after background subtraction is indicated in each panel.}
    \label{fig:LF}
\end{figure*}

The left panel in Fig. \ref{fig:LF} displays the luminosity function (LF) of GC candidates around all UDGs in this study (violet histogram), along with the background distribution (hatched orange histogram) and the background-subtracted LF (teal histogram). Poissonian uncertainties are reported for the background-subtracted distribution. The background distribution is not populated at $m_H\lesssim23.5$ mag, showing that our spectroscopic cleaning procedure of the GC candidate catalogs is efficient up to $m_H\simeq23.5$ mag. The right panel shows the corresponding distributions for the GCs master catalog, which refers to a total area of 9 $\mathrm{arcmin}^2$ at a median galactocentric distance of 0.6 $\mathrm{arcmin}$ from NGC\,3311.
Although the master catalog was not constructed using the same spectroscopic cleaning procedure for bright sources as the source catalogs around the UDGs, we chose to subtract the same background distribution as adopted for the UDGs. This decision is motivated by the fact that the two dominant limiting factors are the $H$-band VIRCAM data and the spectroscopy from MUSE, with the former being identical for the UDGs and the master catalog, and the latter negligible for the master catalog. Indeed, the master catalog is based on pointings close to NGC\,3311, a galaxy known to host a very rich GC system \citep{Wehner08}.
For this reason, we expect the master catalog to be dominated by GCs associated with NGC\,3311, rather than by MW stars or background galaxies. 

In each panel of Fig. \ref{fig:LF}, the total number of sources after background subtraction is reported along with the uncertainty. The sample derived from the master catalog is clearly richer and relatively less contaminated due to its location relative to NGC\,3311, thus providing robust statistics for constraining our detection limit. In contrast, the UDG GC candidates sample is dominated by low-count uncertainties ($30\%$ of the measure).




The analysis of the two distributions in Fig. \ref{fig:LF} can be used to obtain the detection level factor which we define as the ratio of observed to expected GC numbers ($N_{\mathrm{GC,obs}}/N_{\mathrm{GC,exp}}$).

The background-decontaminated luminosity function from the master catalog (right panel in Fig.~\ref{fig:LF}) is the convolution of the intrinsic GCLF of NGC\,3311 and a completeness function that arises from a convolution of the observational properties of our datasets and the several selection stages adopted. The dominant effect stems from the most incomplete dataset, namely the $H-$band data (see Appendix \ref{sec:comple}). To identify both the intrinsic GCLF and completeness function, we implemented a Monte Carlo (MC) simulation designed to reproduce the observed distribution.

We adopted a Pritchet function \citep{McLaughlin94} to describe the completeness of our dataset: 

\begin{equation}
    f(m)=\frac{1}{2}\cdot\Bigg[1-\frac{\alpha\cdot(m-m_{lim})}{\sqrt{1+\alpha^2\cdot(m-m_{lim}})^2}\Bigg]
\end{equation}

where $m_{lim}$ is the magnitude at which the completeness is 50\%, and $\alpha$ determines the steepness of the cut.
The intrinsic GCLF was assumed to be well represented by a Gaussian function with three parameters: the expected turnover magnitude ($\mu^{\text{TOM}}$), dispersion ($\sigma_{\text{GCLF}}$), and the GCLF amplitude. 
In our MC simulations, we allowed the five parameters to vary ($\alpha$, $m_{lim}$, $\mu^{\text{TOM}}$, $\sigma_{\text{GCLF}}$, $N_{GC}$), but, when possible, we adopted scientifically motivated priors. For instance, $\mu^{\text{TOM}}$ and $\sigma_{\text{GCLF}}$ were allowed to vary around estimates defined by the well-established galaxy distance/luminosity relations \citep{villegas10}.
 We simulated 30000 realizations by sampling parameters within the following ranges: the turnover magnitude $\mu^{\text{TOM}}$ was varied by $0.2$ mag around the expected peak value, $m_H=25.2$ mag (see Sec.~\ref{sec:sel_gc}), the range for $\sigma_{\text{GCLF}}$ was set between 1.2 and 1.6 mag typical for GC systems in bright galaxies\footnote{We estimated the $\sigma_{\text{GCLF}}$ for NGC\,3311 using Eq. 5 from \citet{villegas10}, adopting a $z$-band magnitude $m_z = 8.6$ mag, obtained by converting the total magnitude in the $H$-band from \citet{Jarrett03} and assuming a color term $z - H \sim 0.2$ \citep[see][]{Hazra2022}. Finally, we obtained $\sigma_{GCLF}\simeq1.4\ mag$.}, $N_{\text{GC}}$ was allowed to range from 300 to 2000\footnote{We adopted as a lower limit approximately the size of the master catalog after background subtraction (see Fig. \ref{fig:LF}), while the upper limit was determined through several runs of the Monte Carlo simulation, during which we observed that the $N_{GC}$ parameter began to converge at values around a few thousand objects.}, and completeness parameters $\alpha \in [0.1:1.5]$ and $m_{\text{lim}} \in [23:24.2]$.\footnote{The range for $m_{\text{lim}}$ was defined based on the observational requirements of the $H$-band dataset, with the upper limit set to one magnitude brighter than the magnitude corresponding to a point-like source with a SNR$\sim$5.}

During each realization of the MC simulation, we obtained a simulated luminosity to be compared with the observed one. The resulting simulated luminosity function was then compared with the observed one. The goodness of fit was evaluated using the $R^2$ statistic\footnote{Implemented using the $R^2$ scikit-learn package: \href{https://scikit-learn.org/stable/modules/generated/sklearn.metrics.r2_score.html}{scikit-learn.org}.}, which quantifies how well a model reproduces the observed distribution. The closer this metric is to 1, the better the model explains the dataset; the closer to zero, the poorer the fitting. During each realization, we also obtained an estimate of the average detection level factor.


After running all the realizations, we determined the best-fit parameters by calculating the median and rms$_{\text{MAD}}$ values across all simulations with $R^2$ within $\pm1\sigma$ of the maximum $R^2$ (i.e., 0.97). Here, $\sigma$ is the standard deviation of the $R^2$ distribution, obtained from the rms of the $R^2$ scores across all simulations. The best-fit values obtained are presented in Tab.~\ref{tab:sim_gclf}. The resulting $\mu^{\text{TOM}}$ and  $\sigma_{\text{GCLF}}$ of the GCLF are consistent within the error with the expected values for a galaxy at the distance and mass of NGC\,3311\footnote{We also repeated the MC simulation using a wider range for the adopted values of $\mu^{\text{TOM}}$ and $\sigma_{\text{GCLF}}$, and observed no difference in the resulting values; they remain consistent with the expected ones.}.

We estimated that the population required to match the observed GCLF of the master catalog contains $\sim1650$ sources within the three fields (i.e., $\sim3 \times 3\mathrm{arcmin}^2$). We derived the expected total number of GCs over the larger spatial extent of NGC\,3311 \citep{Wehner08,spavone24} by integrating a flat GC density radial profile until $\sim9$ kpc \citep{Capuzzo09} and then a $r^{1/4}$ de Vaucouleurs profile beyond that \citep{faifer11,Cantiello18} out to a radius of $5R_{\mathrm{e}}$ \citep{alamo13}. We adopted the effective radius for NGC\,3311 $R_{\mathrm{e}} = 99\ \mathrm{arcsec}$ from \citet{spavone24}. This yielded a total expected population of $12700\pm 1800$ GCs, which is consistent within 1.5 sigma  with \citet{Wehner08}, who reported $N_{GC}=16500 \pm 2000$. Given the difference in the approaches used to derive the total GC number, this can be considered a satisfactory agreement. Our result is also in agreement with the estimated total population of GCs around NGC\,3311 obtained using FORS data (Lohmann et al., in prep). They found an estimated population of  $\sim$12000 by integrating the density profile presented in \citet{spavone24}.

\begin{figure}
    \centering
    \includegraphics[width=\columnwidth]{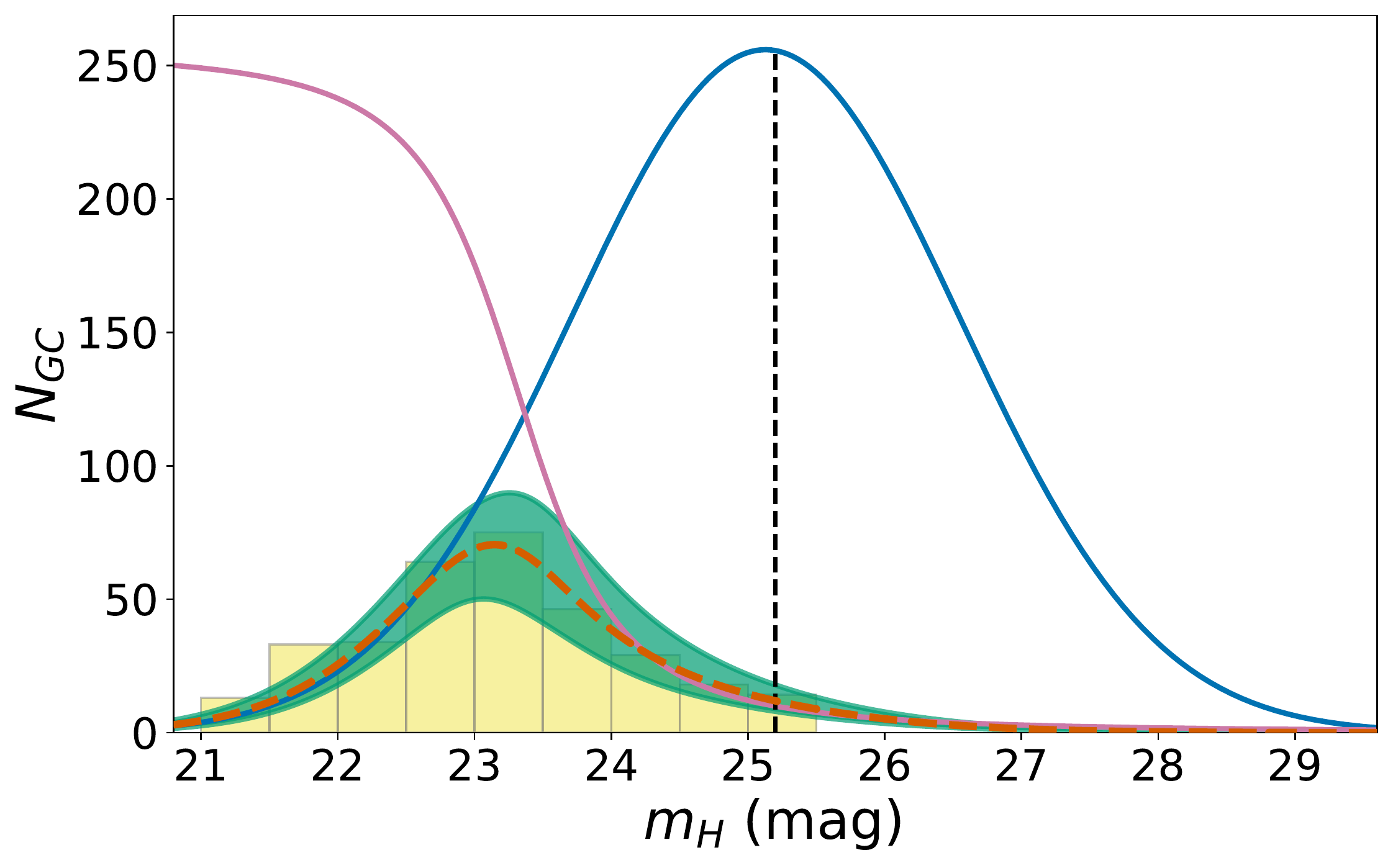}
    \caption{Results of the modeling MC-based procedure. The blue solid line represents the best fit GCLF, while the pink solid line shows the fit to the completeness function. Yellow histogram indicate the background-corrected observed GCs distribution, and the orange dashed line corresponds to the best-fit model. The green-shaded region denotes the 1$\sigma$ uncertainty envelope.}
    \label{fig:gclf}
\end{figure}

Figure \ref{fig:gclf} shows the results of the procedure described above. The blue solid line is the best fit expected GCLF, the completeness function is shown as a pink solid line, the background-corrected observed distribution is represented by yellow histogram, and the best-fit model reproducing the observed data is shown by the orange dashed line. The green-shaded area shows the $1\sigma$ uncertainty range, calculated as the mean plus or minus the rms of all acceptable curves based on the best-fit parameters and their $\pm1\sigma$ errors.

These simulations show that the sample of GC candidates is complete at the 19\% level, meaning that, given our dataset and selection procedure, we recover only 19\% of the expected total GC population.  Under the assumption that such completeness factor derived for the master catalog is roughly consistent with the same for the UDGs, we can now estimate the $N_{GC}$ for our four targets. Table \ref{tab:usg_prop} reports the $N_{GC}$ for each galaxy. 
UDG\,9 is the only galaxy of the sample with GC counts statistically consistent with zero and it is also the only galaxy with no spectroscopically identified GCs (see Fig. \ref{fig:sel_pros} and Tabs. \ref{tab:spectra_gc}, \ref{tab:usg_prop}).

\begin{table}[]
    \centering
    \begin{tabular}{ccc}
    \hline
    \hline
    
         Parameter& MC interval &Best fit  \\
         \hline
           $\mu^{TOM} (mag)$ & [25.0,25.4] & $25.1\pm0.1$\\
           $\sigma_{GCLF} (mag)$& [1.0,1.6]    &$1.4\pm0.1$\\
           $N_{GC}$&   [500,2000]        &$1656\pm267$\\
           $\alpha$&    [0.1,1.5]          &$1.3\pm0.2$\\
           $m_{lim} (mag)$ & [23.0,24.2]    &      $23.3\pm0.2$\\
           $N_{\mathrm{GC,obs}}/N_{\mathrm{GC,exp}}$ (\%)&  ...  & $19\pm3$\\

        \hline
        \hline
    \end{tabular}
    \caption{Adopted intervals for the MC simulations and the best fit values for each parameter.}
    \label{tab:sim_gclf}
\end{table}

\subsection{Specific frequency}

The specific frequency ($S_N$) is the total number of GCs normalized to the galaxy luminosity \citep{harris81}. It is calculated as: 
\begin{equation}
    S_N=N_{GC}\cdot10^{0.4(M_V+15)}
\end{equation}

\begin{figure}
    \centering
    \includegraphics[width=\columnwidth]{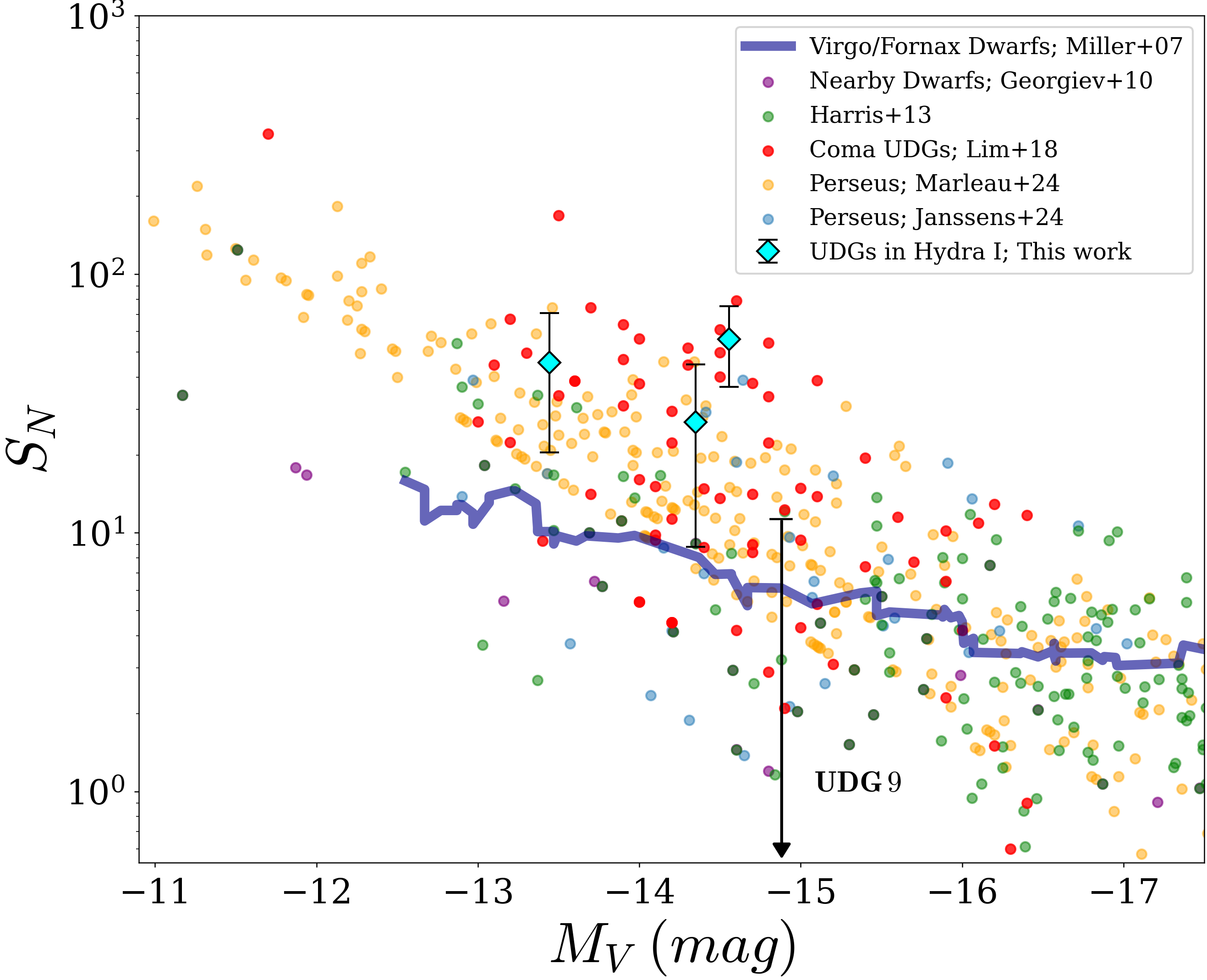}
    \caption{Specific frequency versus magnitude plot comparing Hydra I UDGs $S_N$  with literature values of dwarfs and UDGs from  different environments \citet{miller07,georgiev09,harris13,lim24,Marleau24a,Janssens24}. All literature measurements are shown as colored dots, except for the \citet{miller07} data, which is displayed as a running mean (solid blue line). }
    \label{fig:SN}
\end{figure}

We estimate the $V-$band ($M_V$) absolute magnitude of the UDGs using the relation by \citet{Kostov18}: $M_V=M_r-0.017+0.492\cdot(g-r)$.
The absolute magnitude in the $r-$band ($M_r$) is taken from \citet{iodice20} (see also Tab. \ref{tab:galaxy_properties}). Table \ref{tab:usg_prop} reports the estimated $S_N$ for each galaxy. Figure \ref{fig:SN} shows our values compared to other measurements from the literature \citep{miller07,georgiev10,harris13,Lim18,Marleau24a,Janssens24}.

Multiple studies of dwarf galaxies and UDGs \citep{Lim18, Prole19} have revealed significant scatter in the $S_N$ at the faint end of the luminosity distribution. This scatter reflects both sample size effects due to the small number of GCs and the diverse evolutionary pathways of these systems. \citet{Lim18} and \citet{Forbes20} noted that Coma cluster UDGs of comparable luminosity exhibit lower surface brightness and higher specific frequencies than typical dwarf galaxies. \citet{Lim18} also observed that galaxies with an effective surface brightness of $\mu_e \sim 25$ mag arcsec$^{-2}$ tend to have systematically higher $S_N$ values than those with brighter surface brightnesses. However, this trend is accompanied by substantial scatter: reported $S_N$ values range from zero to hundreds, with error bars comparable to the values themselves.

The sample of galaxies analyzed in this work reflects these broad trends. All galaxies in our sample have effective surface brightness fainter than $25$ mag arcsec$^{-2}$, and we observe both GC-poor systems (one galaxy has no detected GCs) and GC-rich systems. This difference highlights the challenge of quantifying $S_N$ in UDGs. Furthermore, uncertainties in GC counts -- driven by methodological differences in selection criteria, background subtraction techniques, and observational limitations -- complicate direct comparisons across studies.

The uncertainties associated with $S_N$ measurements are substantial. For UDGs with $S_N \gtrsim 30$ studied in \citet{Lim18}, the average uncertainty is $\sim 35\%$. 
In our sample, uncertainties average $\sim 50\%$, primarily driven by our low completeness for the GC candidates sample. 
Based on our analysis three out of four Hydra I UDGs are consistent with a rich GC population (as seen in Coma), or an intermediate population as in the Perseus cluster (see Fig. \ref{fig:SN}). This observation aligns with the established trend that low-mass galaxies in denser environments tend to exhibit high $S_N$ values \citep{peng08,Lim18}.


\subsection{Dark matter content}
\label{sec:dm}

The number of GCs is a good indirect tracer of the total mass ($M_h$) of galaxies \citep{blake97, Beasley16, Harris17}. The total mass scales with $N_{GC}$ over 6 orders of magnitude with a large scatter at low masses \citep{harris13,Burkert20} and flattens out for galaxies with $M_h\lesssim10^{10}M_{\odot}$. Following \citet{Harris17}, the total mass of the galaxies is connected to the mass of the host GC system through the relation: $M_{GC}/M_h\sim2.9\times10^{-5}$. 
Assuming a mean GC mass of $\sim10^5\ \mathrm{M_{\odot}}$ \citep{Harris17}, we obtained the total mass of the GC system by multiplying the mean GC mass with the $N_{GC}$, and then derived the virial mass of the UDGs in our sample. Table \ref{tab:usg_prop} reports this estimate for the four UDGs.

We obtained an independent estimate of the total mass of the galaxies using their total stellar mass ($M_\star$) based on the $H-$band observations. The $H-$band light is primarily emitted by old ($t\geq$5 Gyr), low-mass stars, which dominate the total stellar mass of galaxies \citep{Dib22}, providing a more robust estimate of the overall stellar mass compared to the optical bands. Using the $(g-r)$ color for UDGs from \citet{iodice20} and the $H$-band M/L relation $log(M/L)_H=1.06\times(g-r)-0.884$ \citep{Into13} we retrieved the stellar mass of the galaxies.  In Appendix \ref{sec:hband_tot_mag} we provide more details about the procedure adopted to estimate the total $H$-band magnitude for each galaxy. 

The order of magnitude of the stellar masses we obtained is consistent with the estimates reported in \citet{iodice20} for all the galaxies, apart from the UDG\,9 for which we find a significantly lower $M_*$.
Given the stellar masses, we estimated the total mass of the galaxies using two complementary methods. First, we applied Eq. 2 from \citet{Moster13}, which provides a stellar mass–halo mass relation derived from a multi-epoch abundance matching model. Second, we used the empirical relation from \citet{Zaritsky23}:

\begin{equation}
\label{eq:dm}
M_h = 10^{10.35 \pm 0.02} \left(\frac{M_*}{10^8 M_{\odot}}\right)^{0.63 \pm 0.02}
\end{equation}

This power-law relation was obtained by fitting a heterogeneous dataset of measured stellar and dark matter masses, and is valid for galaxies with $M_* < 10^{9} M_\odot$ and $M_h < 10^{12} M_\odot$. All the galaxies in our sample fall within this range.
The halo mass estimates derived from Eq.~\ref{eq:dm} are consistent, within the uncertainties, with those obtained using the relation from \citet{Moster13}. Table \ref{tab:usg_prop} lists the derived total masses, $H$-band photometry, and mass-to-light ratios for each galaxy. 

In Tab. \ref{tab:usg_prop}, we also report the $V-H$ colors\footnote{The $V$ magnitude was derived from the $g$ magnitude estimate in \citet{iodice20}, assuming a color term of $g - V \sim 0.4$ mag \citep{cantiello20}.}. As a sanity check of our $H$-band magnitudes, these colors were compared to predictions from the Yonsei Evolutionary Population Synthesis (YEPS) model \citep{chung13}, selecting stellar populations with ages $\geq 5$ Gyr and metallicities $-2 \leq \mathrm{[Fe/H]} \leq 0$ \citep{Pandya18}. Three galaxies exhibit $V-H$ values consistent within the error with the median and $rms_{MAD}$ of the models ($V-H$=2.2$\pm$0.3 mag).
UDG\,9 is a notable exception: its $V-H$ color is $\sim1$mag bluer than the model predictions. We checked with the stellar population model adopted if the presence of a young population could explain such blue color and we found out that a young (< 5 Gyr) population can have color $V-H\sim$1.2 mag, consistent within the error with the observed one. The color offset for this galaxy might alternatively arise from its location in a region of the Hydra I cluster where contamination from bright stars and nearby galaxies (see Sect. \ref{sec:spe_cat}) complicates background subtraction in the stacked $H$-band observations. The $g-r$ color values of the four UDGs are on the other hand consistent within the uncertainties. 
To verify whether this difference impacts the results presented in this work, we derived an alternative stellar mass estimate for this galaxy. Specifically, we adopted the average $V-H$ color predicted by models for galaxies with the same $g-r$ color as UDG\,9, and then derived the corresponding $H$-band magnitude. This approach yielded $m_H = 17.8 \pm 0.3$ mag and $V-H = 2.1 \pm 0.4$, which implies $M_* \sim 8 \times 10^7M_\odot$ and $M_h = 2.6 \pm 1.1 \times 10^{10}M_\odot$. The resulting stellar/halo mass is consistent, within the uncertainties, with the value estimated using the $H$-band magnitude measurement (see Tab. \ref{tab:usg_prop}). 



We observe agreement between virial mass estimates derived from the $N_{GC}$ and those based on stellar mass with two exceptions: UDG\,9 and UDG\,11. For UDG\,9, no virial mass estimate was possible due to the lack of detected GC candidates, while the discrepancy for UDG\,11 may stem from the large scatter at low GC counts in the adopted relation \citep{Harris17,Burkert20}

\begin{figure}
    \centering
    \includegraphics[width=\columnwidth]{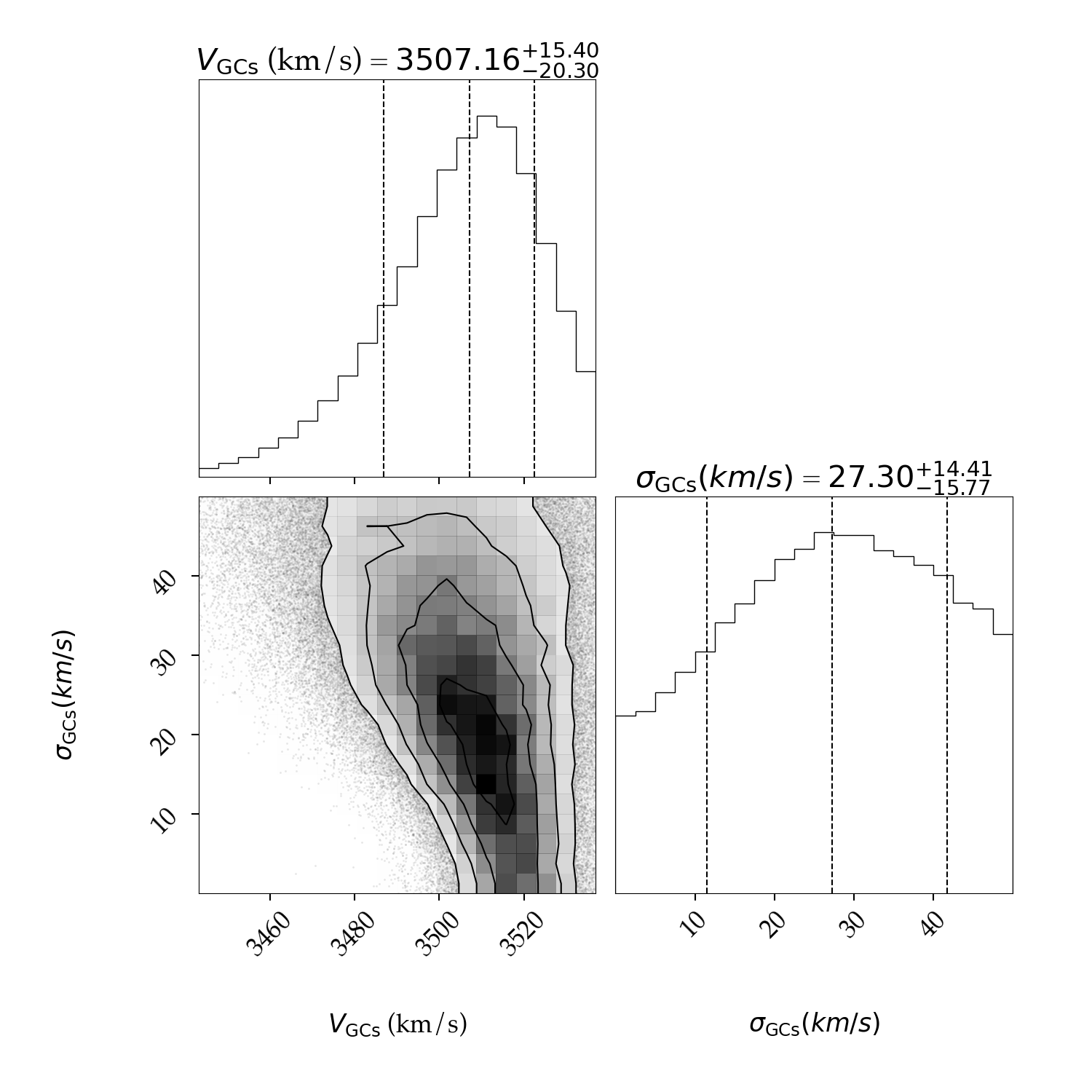}
    \caption{Inference of the systemic velocity and velocity dispersion of UDG\,11 using its four GCs as dynamical tracers. The corner plot presents the results of an MCMC fit to the GC velocities, assuming uniform priors on both systemic velocity and velocity dispersion ($\sigma_{GC}$). The top panel shows the posterior distribution of the systemic velocity, the bottom right panel shows the velocity dispersion distribution, and the bottom left panel displays the correlation between the two parameters.}
    \label{fig:cornerplot}
\end{figure}

For UDG\,11, for which four GCs were spectroscopically confirmed, we also obtain an estimate of the DM content using the velocity dispersion of the spectroscopically confirmed GCs, following the approach described in \citet{Buzzo25}. We adopted an MCMC method \citep{Foreman-Mackey13} to estimate the velocity dispersion ($\sigma_{GC}$) of the system assuming a uniform prior, and obtained a $\sigma_{GC}=27^{+14.}_{-16}\ \mathrm{km\ s^{-1}}$ (Fig. \ref{fig:cornerplot}). The velocity dispersion of the galaxy obtained through the GCs and the velocity dispersion obtained from the stellar kinematics \citep[$\sigma_{star}=20\pm 8$ $\mathrm{km\ s^{-1}}$, see][]{buttitta25} agree within the errors. The DM mass was then estimated using Eq. 2 from \citet{Wolf10}: $M_{dyn}=4R_{e,c}
\sigma^2/G$, where $R_{e,c}=R_e\sqrt{q}$ is the circularized
half-light radius calculated through the galaxy axial ratio $\mathrm{q}$, and
$\mathrm{G}$ is the gravitational constant.
Thus, we obtained a dynamical mass within half light radius of 
$M_\mathrm{dyn} =6.2^{+8.4}_{-5.1} \times 10^8\,M_\odot$.
This value agrees within uncertainties with estimate from stellar kinematics in \citet{buttitta25}, $M_{dyn} = 5.9 \pm 4.8 \times 10^8\,M_\odot$.
To test for systematics, we repeated the procedure by iteratively removing one GC candidate at a time (i.e., recalculating $\sigma_\mathrm{GC}$ using the remaining three sources per iteration). The resulting dynamical masses has a weighted mean of $8.5\pm3 \times 10^8\,M_\odot$. 
Even in this case, the estimates remain consistent with those derived from stellar kinematics.
This further strengthens our conclusion that the four GCs belong to the galaxy. 


From \citet{buttitta25}, other estimates of the dynamical masses are available using the stellar kinematics for the other galaxies. Such estimates were obtained  using Eq.2 in \citet{Wolf10},  
$M_{dyn}=1.04 \pm 0.70 \times 10^9 M_\odot,$ and  
$M_{dyn}=4.91 \pm 1.51 \times 10^9 M_\odot$  
for UDG\,9 and UDG\,7, respectively. Such values do not directly compare with the total masses estimated through the $N_{GC}$ and $M_*$ because they are based on different assumptions. Indeed, the total mass inferred from $N_{GC}$ and the stellar mass both trace the total mass (baryon+dark matter) content of the galaxy. The dynamical mass derived from stellar kinematics \citep[e.g.,][]{buttitta25} probes the mass within $1R_e$. To make a fair comparison, we computed the total magnitude within the circularized half-light radius, and obtained an estimate of $M_h$. We found agreement between the two independent total mass quantities. 
The values obtained here, despite being derived under different assumptions, independently indicate that these galaxies are DM dominated, with $M_{h}/M_*\sim10-1000$ similar to typical dwarfs of the same magnitude \citep{Battaglia22, Buzzo25}.




\begin{table*}[h!]
\scriptsize
    \centering
    \caption{UDG parameters.}
    \begin{tabular}{cccccccccccc}
    
    \hline
    \hline
       Galaxy  & $m_H$ &  $g-r$& $V-H$ &  $(M_*/L)_H$&$M_*$& $N_{GC}^{raw}$&$N_{GC}$ & $S_N$ & $M_{h}$& $M_{h}$&$M_{h}$\\
       & (AB mag) &(AB mag)&(Vega mag) & &   ($10^7M_\odot$)   &  &  &  & ($10^{10}M_\odot$) &($10^{10}M_\odot)$&($10^{10}M_\odot$) \\
        (1)& (2) &(3) &(4) & (5)&(6)&(7)&(8) &(9) &(10)&(11)&(12)\\
       \hline
        
        UDG3  &$17.96\pm0.04$& $0.75\pm0.3$   & $2.61\pm0.39 $  &  $0.81\pm0.59$& $11\pm8.0$&     $3\pm2$     &$15\pm9$  &  $27\pm18$ &  $5.1\pm3.3$  &$2.97\pm1.47$   & $2.36\pm1.10$  \\
        UDG7  & $19.26\pm0.03$ &  $0.6\pm0.3$   & $2.14\pm0.35$ & $0.56\pm0.41$ & $2.28\pm1.67$&    $2\pm1$   &$11\pm6$   &  $46\pm25$ &   $3.8\pm1.9$ &$1.42\pm0.8$  & $0.88\pm0.41$   \\
        UDG9  & $19.01\pm0.08$ &  $0.6\pm0.2$  & $0.97\pm0.27$ & $0.56\pm0.28$ & $2.87\pm1.42$&       $0\pm2$  & $0\pm11$  &  $0\pm12$  &   ....           &  $1.58\pm0.77$  & $1.02\pm0.32$    \\
        UDG11 & $18.19\pm0.03$&  $0.43\pm0.11$  & $2.00\pm0.18$& $0.37\pm0.1$& $4.03\pm1.09$&    $7\pm2$      &$37\pm11$  &  $56\pm19$ &  $12.8\pm4.1$  &$1.86\pm0.80$  & $1.26\pm0.22$   \\

         \hline
         \hline
    \end{tabular}
     \begin{justify}
    Notes:  Col. 1 galaxy ID; Col. 2 total $H$-band magnitude; Col. 3 $(g - r)$ color from \citet{iodice20}; Col. 4 derived $(V - H)$ color used to validate our photometry (see Sect. \ref{sec:dm}); Col. 5 $M_*/L$ ratio derived from \citet{Into13}; Col. 6 stellar mass derived assuming the absolute solar magnitude $M_H$ from \citet{Into13};  Cols. 7-9 reports the raw number of GC ($N_{GC}^{raw}$), the corrected for completeness number of GCs ($N_{GC}$) and the specific frequency ($S_{N}$); Cols 10-12  reports the dark halo mass estimated using \citet{Harris17}, \citet{Moster13}, \citet{Zaritsky23} relations, respectively.
    \end{justify}
    
    \label{tab:usg_prop}
\end{table*}

\section{Conclusions}
\label{sec:conclusiocn}

In this work, we analyzed the GC population hosted by four UDGs observed within the LEWIS program. The methodology adopted combines integral-field spectroscopy from MUSE with deep optical-to-NIR imaging, to refine the GC selection. Our analysis focuses on a preliminary sample of four out of the 25 UDGs in the LEWIS sample, selected from among the UDGs with he largest number of GC candidates from optical photometry alone  \citep{iodice20}. The analysis of the remaining targets will be presented in a forthcoming paper.

We first identified GC candidates through spectroscopy. We identified and found 2, 3, 0, and 4 GCs for UDG\,3/UDG\,7/UDG\,9 and UDG\,11, respectively. We also identified a total of three ICGC and one UCD. To extend the analysis to fainter GC candidates, we constructed a photometric catalog using optical ($g^*riW$) and near-IR ($H$-band) imaging. Contaminants --including foreground stars, background galaxies, and emission-line sources-- were removed using spectroscopic priors and color-magnitude-shape cuts. The final catalog of GCs around each UDG contains residual contamination from faint MW stars and unresolved background galaxies. We mitigated this by applying a statistical background correction when deriving GC system properties (e.g., GC counts, specific frequencies).

UDG\,9 is the only galaxy in the sample that shows no GC candidate from spectroscopy nor photometry. This UDG resides in a relatively dense environment, surrounded by bright galaxies such as NGC\,3316 and ESO\,501-49. The latter lies at a projected distance of $\sim 35\ \mathrm{kpc}$ along the major optical axis of the galaxies and exhibits a LOSV difference of $\sim 300\ \mathrm{km/s}$. The former presents stellar streams and exhibits the same LOSV difference. The UDG\,9 is the only galaxy in the sample exhibiting elongation, whereas the remaining galaxies display predominantly round morphologies. These factors suggest that UDG\,9 may have lost its GC population through tidal interactions \citep{Jones21, Marleau24b}. In the LEWIS sample, UDG\,32 is a galaxy previously studied spectroscopically that does not exhibit any GCs \citep[see][]{Hartke25}. The authors speculate that the absence of GCs may be related to the galaxy having formed from stripped material, possibly originating from interactions within the Hydra I cluster environment.

For the remaining three UDGs in the sample, the GC population is estimated to number between $\sim10$ and $\sim40$ GCs with a relative uncertainty of $\sim50\%$. The total number of GC candidates in this work is not directly comparable to that obtained using only the $g$- and $r$-band imaging from VST  \citep{iodice20}. This is because, although the wide-field imager enables a more accurate background characterization, by using only two optical passbands -- especially ones close in wavelength -- does not allow for effective removal of interlopers. While our dataset covers a small area of the sky, the combination of a broad wavelength coverage (optical to NIR) and spectroscopic information allows us to reliably constrain the  GC population associated with these galaxies.

The specific frequency of our sample shows a significant scatter,  with GC-poor UDG\,9 and the other three galaxies comparable to the GC-rich UDGs found in clusters such as Coma or having an intermediate population, as in Perseus. 
UDG\,11 shows a distribution of GC candidates aligned along the line connecting it with the ESO\,501-26 galaxy, which lies at a projected distance of $\sim\ 95\pm \mathrm{kpc}$, hinting at a possible past interaction between the two galaxies. 

By using the $H$-band magnitude of the UDG and the total number of GCs, we obtained three independent estimates of the DM halo mass for these galaxies. These estimates agree with each other within the error bars for two out of three galaxies. We note that the discrepancy in mass estimates for UDG\,11 is likely caused by the large scatter at low GC counts of the adopted relation. All the galaxies in this work are consistent with being dark matter dominated, with dynamical-to-stellar mass ratios ($M_{\mathrm{dyn}}/M_*$) in the range of $\sim10$ to $1000$.

In a future work, we will use the procedure presented here to analyze the GC systems for the full LEWIS sample, linking the properties of GCs in UDGs to their stellar kinematics \citep{buttitta25} and stellar populations (Doll et al., in prep).

\section*{Acknowledgements}

We thank the anonymous referee for the comments and
constructive suggestions. Part of this work was supported through the INAF “Astrofisica Fondamentale” GO-grant 2024 n. 12 (PI: M. Cantiello). MM acknowledges support from the ESO Short-Term Visiting Program (SSDF 22/18, PI: S. Mieske) and the ESO Studentship Program. MC acknowledges support from the ASI–INAF agreement “Scientific Activity for the Euclid Mission” (n. 2024-10-HH.0; WP8420) and from the ESO Scientific Visitor Programme.
KF acknowledges funding from the European Union’s Horizon 2020 research and innovation programme under the Marie Sk\l{}odowska-Curie grant agreement Number 101103830.
DAF thanks the ARC via DP220101863.
GD acknowledges support by UKRI-STFC grants: ST/T003081/1 and ST/X001857/1.
GR acknowledges support  INAF “Astrofisica Fondamentale” Mini-grant 2024 n.17/RSN1 (PI G. Riccio) and support from the INAF LSST in-kind project (ITA-INA S6, PI:Michele Cantiello).
J.F-B. acknowledges support from the PID2022-140869NB-I00 grant from the Spanish Ministry of Science and Innovation.
TR acknowledges support from ANID via Nucleo Milenio TITANs (NCN 2023-002).
MS acknowledge the support by the Italian Ministry for Education University and Research (MIUR) grant PRIN 2022 2022383WFT “SUNRISE”, CUP C53D23000850006 and by VST funds.

This work is based on observations collected at the European Southern Observatory under ESO programmes 108.222P.001, 108.222P.002, 108.222P.003, 097.B-0806(A), 094.B-0711A, 109.231E.001. 
This research has made use of the NASA/IPAC Extragalactic Database (NED), which is funded by the National Aeronautics and Space Administration and operated by the California Institute of Technology. We also acknowledge the usage of the Extragalactic Distance Database (EDD, \url{https://edd.ifa.hawaii.edu/}). We made extensive use of the softwares of SExtractor \citep{bertin96} and Topcat \citep[\url{https://www.star.bris.ac.uk/~mbt/topcat/};][]{taylor05}). This research has made use of the VizieR catalogue access tool, CDS, Strasbourg, France \citep{10.26093/cds/vizier}. The original description of the VizieR service was published in \citet{vizier00}.

This publication makes use of data products from the Pan-STARRS1 Surveys (PS1) and the PS1 public science archive have been made possible through contributions by the Institute for Astronomy, the University of Hawaii, the Pan-STARRS Project Office, the Max-Planck Society and its participating institutes, the Max Planck Institute for Astronomy, Heidelberg and the Max Planck Institute for Extraterrestrial Physics, Garching, The Johns Hopkins University, Durham University, the University of Edinburgh, the Queen's University Belfast, the Harvard-Smithsonian Center for Astrophysics, the Las Cumbres Observatory Global Telescope Network Incorporated, the National Central University of Taiwan, the Space Telescope Science Institute, the National Aeronautics and Space Administration under Grant No. NNX08AR22G issued through the Planetary Science Division of the NASA Science Mission Directorate, the National Science Foundation Grant No. AST-1238877, the University of Maryland, Eotvos Lorand University (ELTE), the Los Alamos National Laboratory, and the Gordon and Betty Moore Foundation \citep{Chambers16,Magnier_2020,Waters_2020,Magnier_2020b,Magnier_2020c,Flewelling_2020}.

This publication makes use of data products from the Two Micron All Sky Survey, which is a joint project of the University of Massachusetts and the Infrared Processing and Analysis Center/California Institute of Technology, funded by the National Aeronautics and Space Administration and the National Science Foundation \citep{Skrutskie06}.

This work made use of Astropy (\url{http://www.astropy.org}) a community-developed core Python package and an ecosystem of tools and resources for astronomy \citep{astropy:2013, astropy:2018, astropy:2022}.

\bibliographystyle{aa.bst}

\bibliography{Bibliography.bib}

\appendix

\section{MUSE field over NGC3311}

\begin{figure}
    \centering
    \includegraphics[width=8cm]{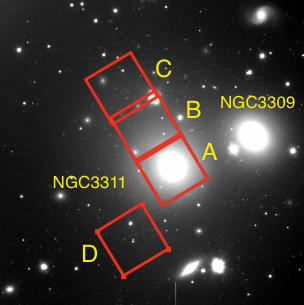}
    \caption{Spatial coverage of the MUSE observations of NGC\,3311. North is up, East is left.} 
    \label{fig:grasser_field}
\end{figure}

Here, we report the location of the MUSE pointing on NGC\,3311, as described in \citet{barbosa18}. Figure \ref{fig:grasser_field} shows the VST $g$-band image with the observed MUSE pointing indicated by red boxes.

\section{Master catalog table}

Here, we report an extract of the GC candidates master catalog (see Tab. \ref{tab:mast_cat}) used to select GC candidates based on their optical and near-IR colors.

\begin{table*}[htbp]
\centering
\caption{Master catalog of source properties.}
\begin{tabular}{ccccccccccc}
\hline
\hline
RA & DEC &  $m_{H}$ & .... &$\mathrm{FWHM}_{W}$ &$\mathrm{CS}_{W}$&Note \\
(deg) & (deg) & (mag)&....& (arcsec)&  \\
(1) & (2) & (3) & (4)& (5)& (6) \\

\hline
159.1944107 & -27.5036652 &  21.8247 & ....  &0.870 & 0.982  & 0     \\
159.1818764 & -27.5229102 &  25.1065 & ....  & 1.508 & 0.641 &  1    \\
159.1935036 & -27.4950128 &  23.8158 & ....  &1.718 & 0.808  & 0     \\
159.1921676 & -27.5053451 &  23.7597 & ....  &1.892 & 0.685  & 0     \\
159.1953659 & -27.5032052 &  24.7400 & ....  &2.478 & 0.703  & 0     \\
159.1732144 & -27.5231591 &  21.5933 & ....  & 0.852 & 0.980 &  1    \\
159.1952968 & -27.5040165 &  24.8653 & ....  &2.552 & 0.751  & 0     \\
....&....&....&....&....&....&....\\  
159.1926967 & -27.5090813 &  24.6071 & ....  &1.014 & 0.834  & 0     \\
159.1796688 & -27.5246777 &  24.3485 & ....  & 1.140 & 0.946 &  1    \\
159.1971671 & -27.5104511 &  23.9491 & ....  &2.506 & 0.870  & 0     \\
159.1758011 & -27.5229259 &  23.8625 & ....  & 1.926 & 0.928 &  1    \\
159.1743646 & -27.5234490 &  23.5881 & ....  & 2.134 & 0.923 &  1    \\
....&....&....&....&....&....&....\\  
159.1783292 & -27.5241470 &  21.9223 & ....  & 2.388 & 0.957 &  1    \\
159.1848142 & -27.5255859 &  22.8897 & ....  & 1.514 & 0.941 &  1    \\
\hline &  
\end{tabular} 
\begin{justify} 
   Note. The columns list the following information: (1) Right Ascension; (2) Declination; (3) $H$-band magnitude; (4) additional photometric properties retrieved with SExtractor, such as aperture photometry and Kron radius; and (5--6) morphological properties derived from SExtractor applied to the MUSE white-light image. Notes: 0 for photometric GC candidates and 1 for spectroscopically confirmed GCs from \citet{Grasser24}.
\end{justify}

\label{tab:mast_cat}
\end{table*}

\section{Redshift estimates for central sources}
\label{sec:app_central_cource}

To estimate the redshift of sources near the galaxy core, we extracted the spectra within a 4-pixel aperture instead of 8 pixels and omitted local background subtraction.
Sources for which the radial velocity was estimated using this procedure are marked with an asterisk in Tab. \ref{tab:spectra_gc}. We applied this procedure to four sources: three in the UDG\,11 field and one in the UDG\,7 field. We intend to apply this procedure in future papers as well. 
We performed several tests to ensure that the extracted spectra are not dominated by galaxy light. As a case study, we present here results for three sources located in the core of UDG\,11.

\begin{figure}[H]
    \centering
    \includegraphics[width=\columnwidth]{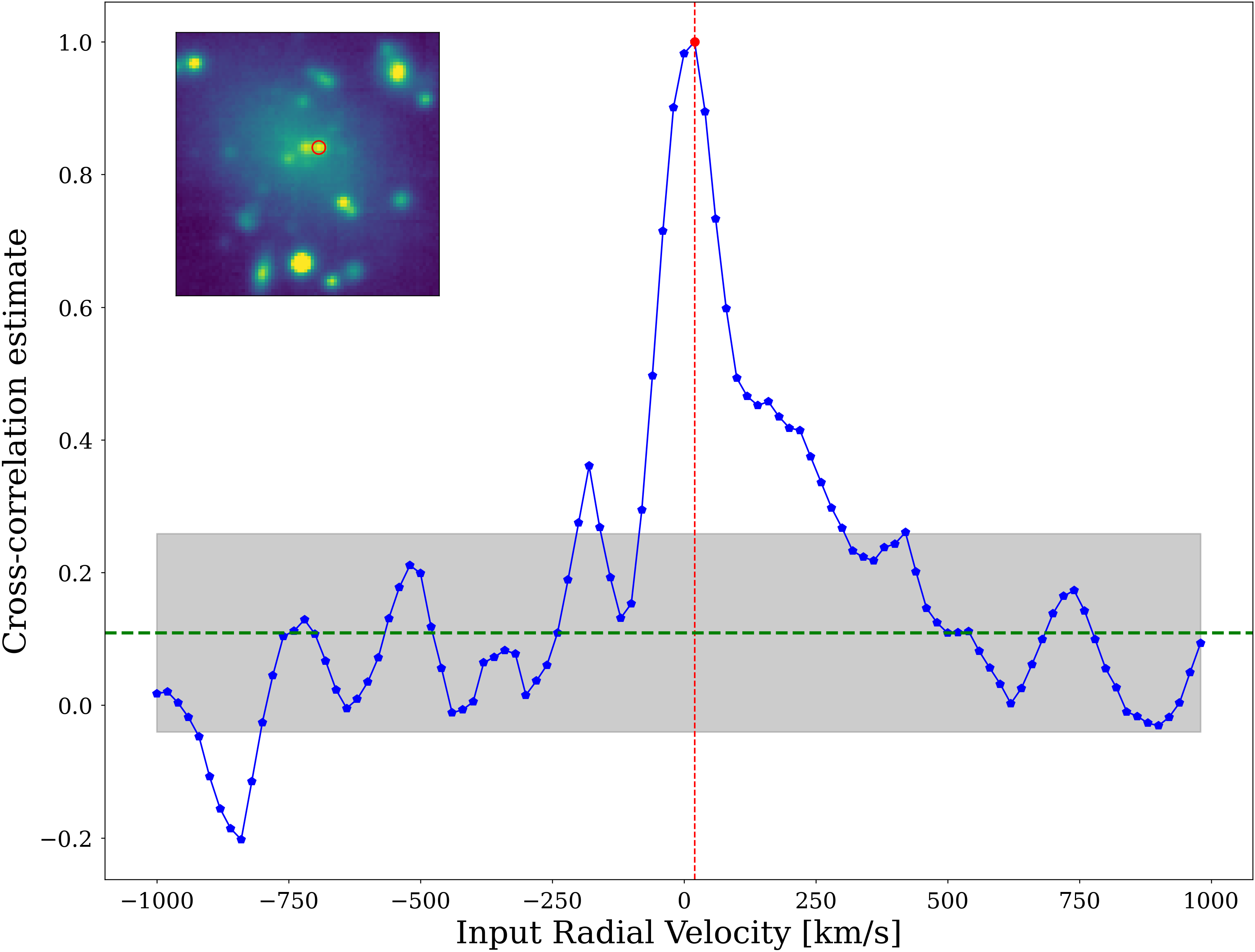}
    \caption{Cross-correlation results. The red dashed line marks the maximum of the correlation, indicating the relative shift between the galaxy spectrum and the source spectrum. The green dashed line represents the median of the cross-correlation estimates. The gray shaded region denotes the $\pm1rms_{MAD}$ around the median. In the upper left corner, an 80$\times$80 pixels zoom-in of the white-light image of the central region of UDG\,11 is shown, with a red circle marking the specific source that is analyzed.}
    \label{fig:cross_corr}
\end{figure}

\begin{figure*}[ht]
    \centering
    \includegraphics[width=6cm]{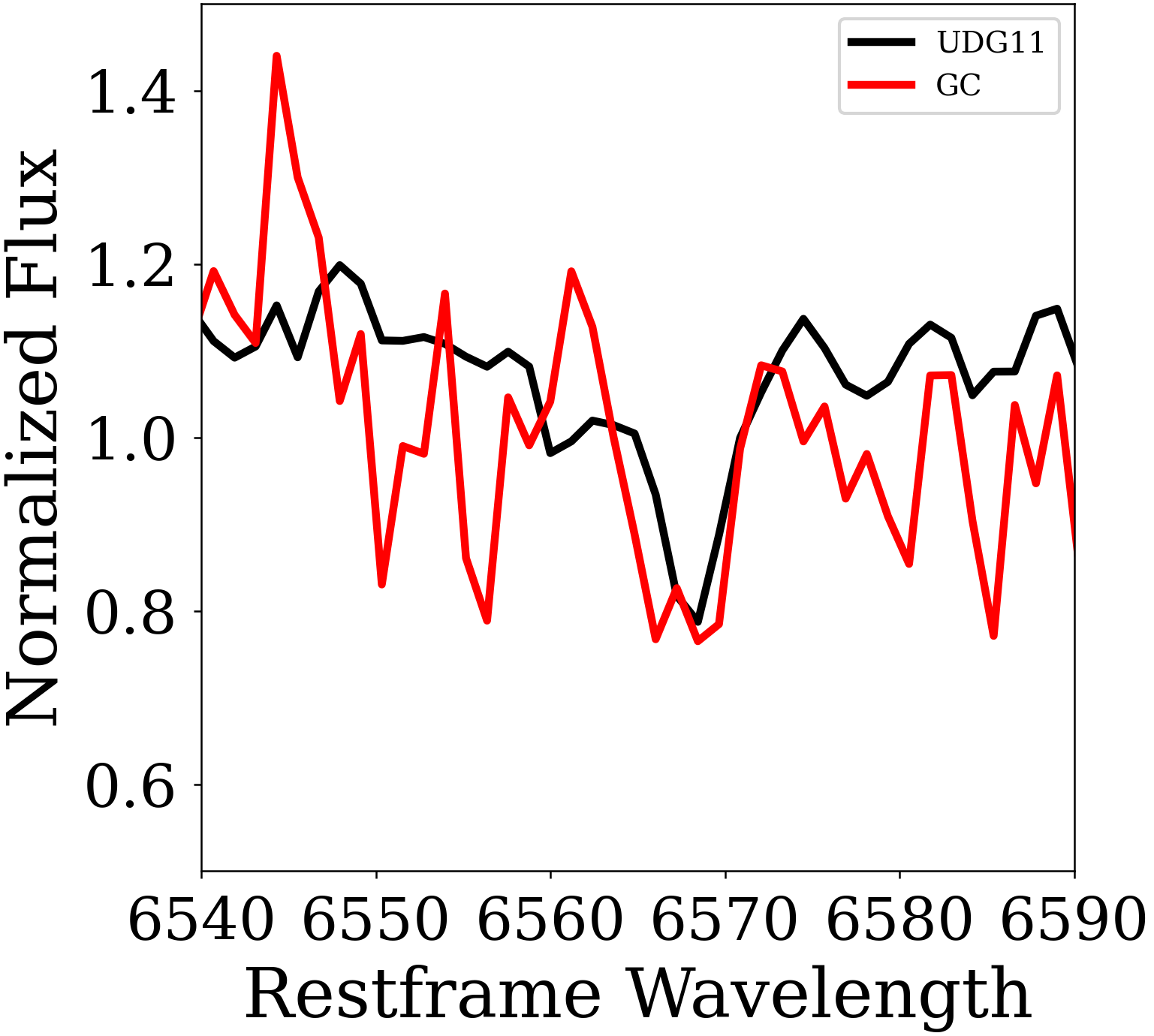}
    \includegraphics[width=6cm]{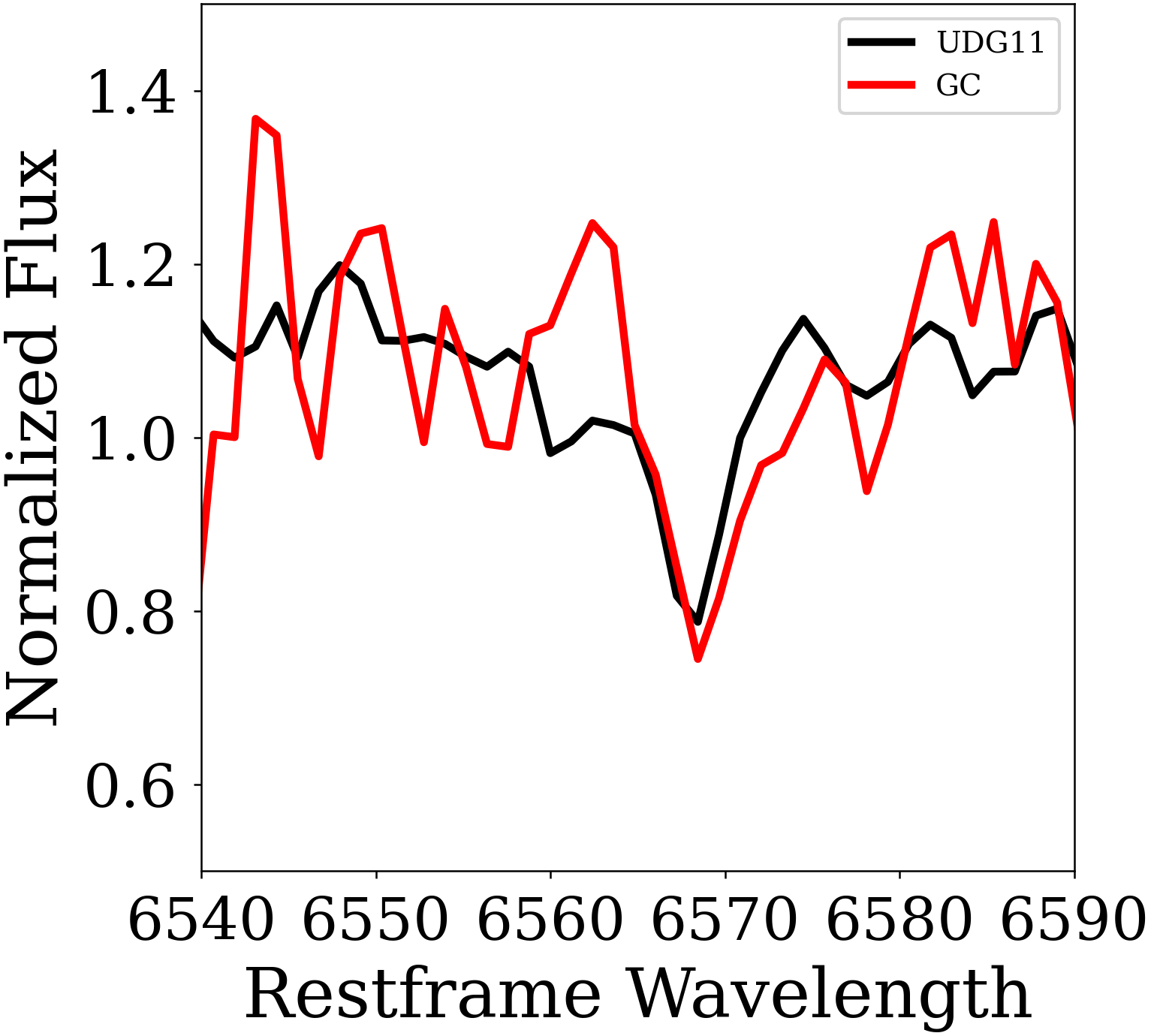}
    \includegraphics[width=6cm]{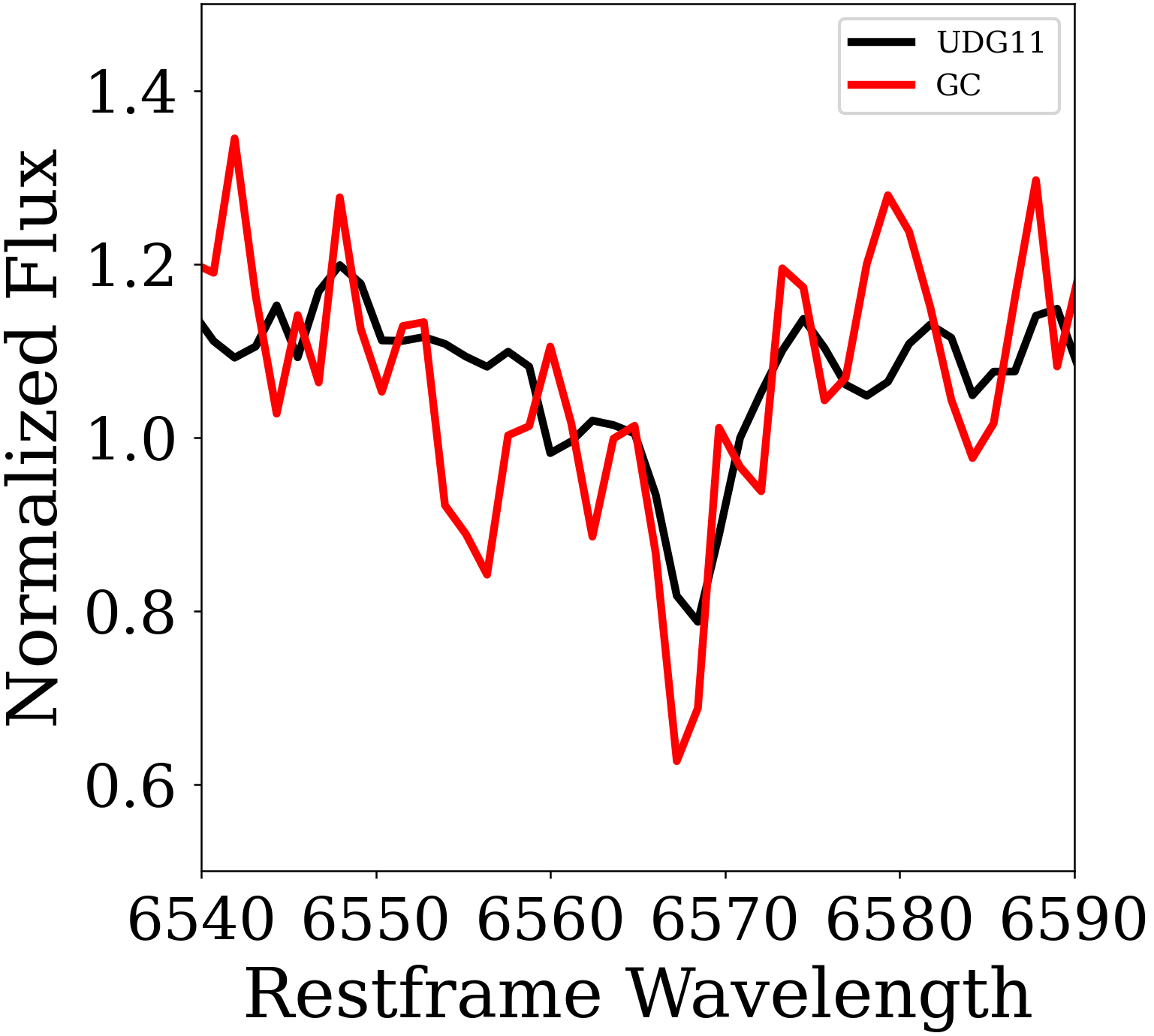}
    \caption{Comparison between the $H_{\alpha}$ line of UDG\,11 and the three GCs in the central region (see Fig. \ref{fig:gal_ligh_percentage}). The black solid line represents the UDG spectrum extracted within $1R_e$, masking all the compact and extended sources present in the aperture. The red solid line shows the spectra of the sources, extracted using a 2-pixel aperture radius.}
    \label{fig:halpha_com}
\end{figure*}

\begin{figure*}[ht]
    \centering
    \includegraphics[width=\textwidth]{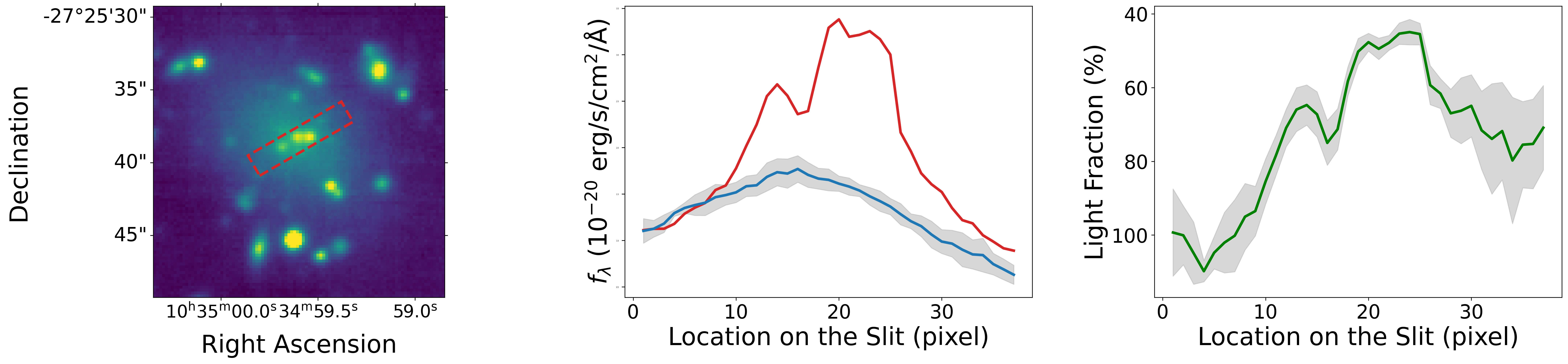}
 
    \caption{Left panel: Red dashed rectangle indicates the slit used to extract the flux profile of the sources. Additional slits were placed around the galaxy center to estimate the galaxy light contribution. Middle panel: Flux profile along the slit containing the sources (red curve) compared to the median galaxy flux profile (blue curve). The gray shaded region represents the associated uncertainties. Right panel: Estimated fraction of galaxy light for each of the three sources.}
    \label{fig:gal_ligh_percentage}
\end{figure*}

We verified that the extracted spectra exhibited a velocity shift relative to the galaxy spectrum\footnote{The galaxy spectrum was extracted by masking all sources within $1R_e$ from the galaxy core, as this aperture maximizes the SNR \citep[see][]{buttitta25}.}. This was done using the cross-correlation method \citep{Tonry79} as implemented in the $PyAstronomy$ package.
Figure \ref{fig:cross_corr} presents the results of this procedure for one of the sources. The estimated $1rms_{MAD}$ noise (gray region) is significantly lower ($\leq3\sigma_{\mathrm{rms}}$) than the peak height, leading us to conclude that the observed shift is real and not an artifact of galaxy light extraction \citep[see][]{arnaboldi12}.
To further assess the robustness of our redshift measurements, we carried out two more experiments.
First, we compared the $H_{\alpha}$ lines of the galaxy with those of the sources (see Fig.~\ref{fig:halpha_com}). This comparison revealed differences in the shape, width, and position of the $H_{\alpha}$ lines. Although some sources may be contaminated by galaxy light, the observed spectral differences suggest that these spectra have distinct characteristics, indicating they are not solely dominated by galaxy emission. Second, we quantified the fraction of flux contributed by the galaxy in the extracted source spectra. 
We constructed flux profiles along the slit for the sources (red dashed rectangle in Fig.~\ref{fig:gal_ligh_percentage}) and in three adjacent slits around the galaxy center. The middle panel of Fig.~\ref{fig:gal_ligh_percentage} displays the profile of the slit on the sources (red curve) and the median galaxy flux profile (blue curve), along with uncertainties (gray shaded region). The right panel shows the fraction of galaxy light for the three sources. The faintest source (left panel in Fig. \ref{fig:halpha_com}), which is the one most dominated by the galaxy light, still exhibits an $H_{\alpha}$ line distinct from the galaxy. For the other sources, where the galaxy contribution is less than 50\%, the $H_{\alpha}$ line is less broadened.
Taken together, these results confirm that the three objects are physically associated with the galaxy (see Tab.~\ref{tab:galaxy_properties}).

\section{Background sources}

\begin{figure*}[ht]
    \centering
    \includegraphics[width=\textwidth]{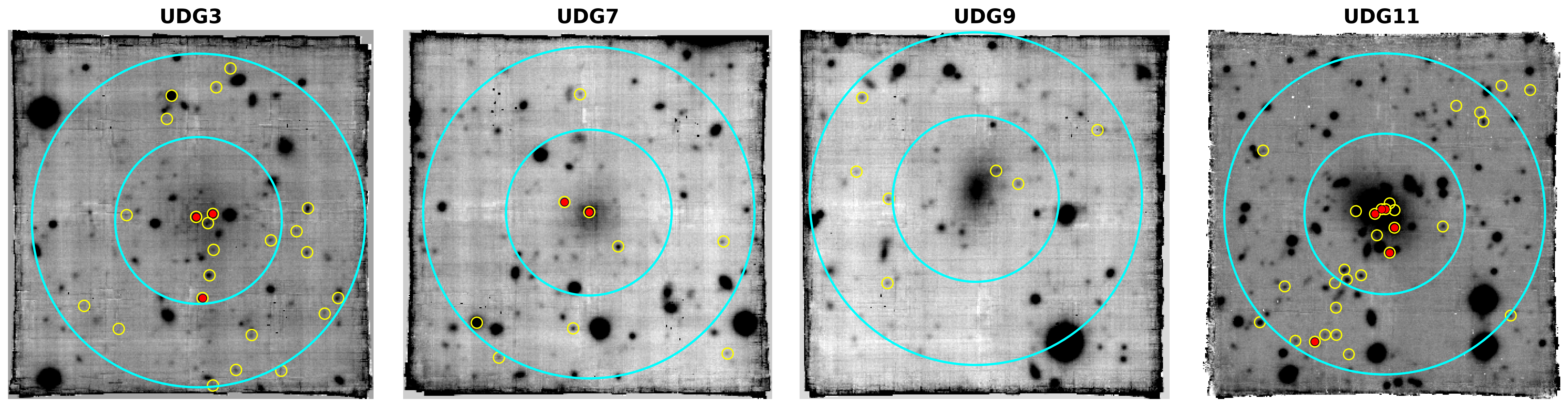}
    \caption{Final sample of GC candidates (yellow circles) and spectroscopically confirmed GCs (red dots) overplotted on the four target galaxies. The cyan annulus marks the background region used for contamination estimates, defined between 15$\farcs$0 and 30$\farcs$0 from each galaxy centre.}
    \label{fig:anulus_background}
\end{figure*}

As discussed in Sect. 
\ref{sec:analysis_gc}, to estimate the total number of GCs for each galaxy, we first identify a sample of sources used to characterize the residual contamination from foreground stars and background galaxies. In Sect. \ref{sec:tot_n_gc}, we estimated this background contamination by combining all sources located between 15$\farcs$0 and 30$\farcs$0 from each galaxy center. Figure \ref{fig:anulus_background} shows the final sample of GC candidates (yellow circles), the spectroscopic ones (red dots), and the overplotted adopted background selection annulus for each galaxy (cyan annulus).

\section{GC velocity dispersion}
\label{sec:gc_sig}

\begin{figure*}[h!]
    \centering
    \includegraphics[width=\columnwidth]{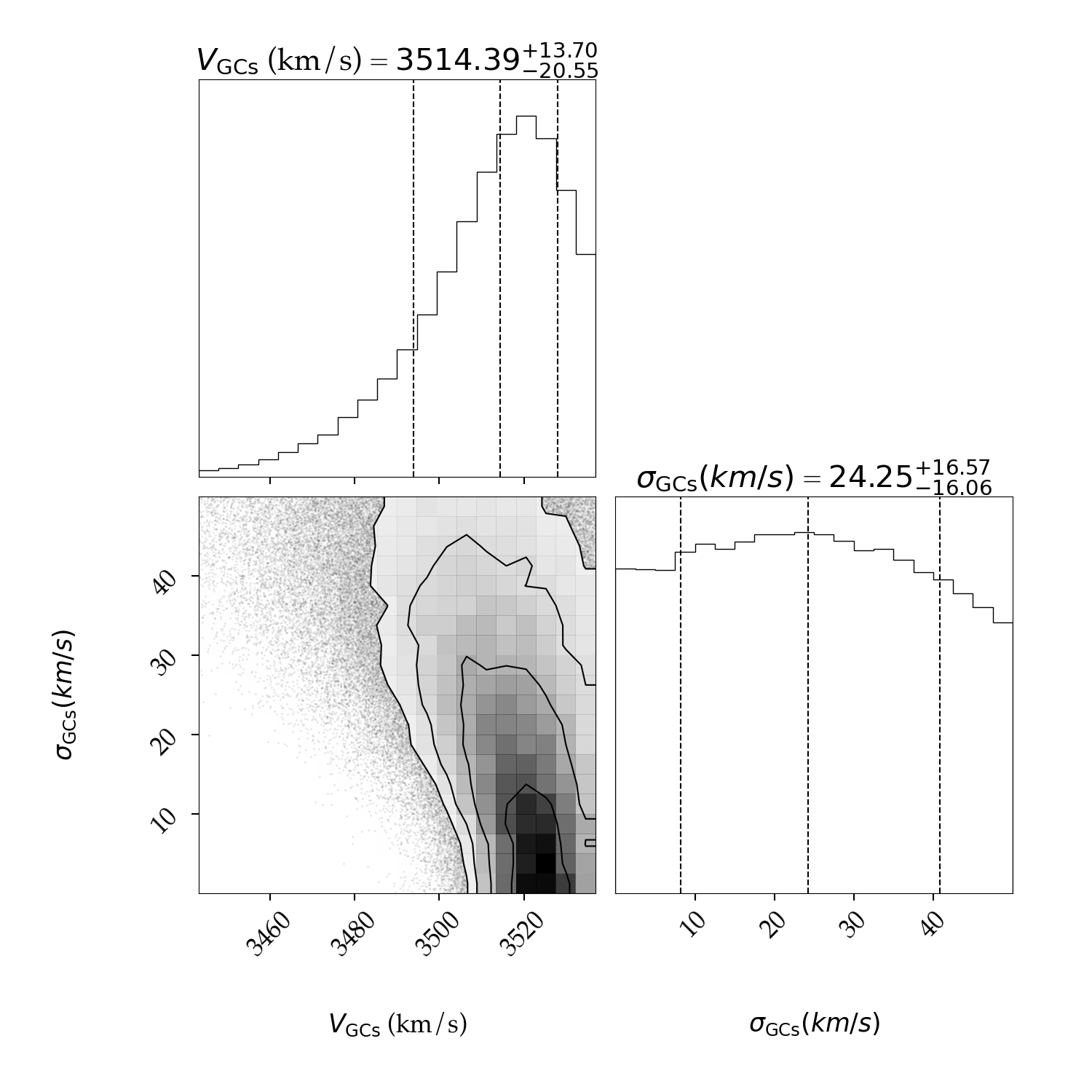}
    \includegraphics[width=\columnwidth]{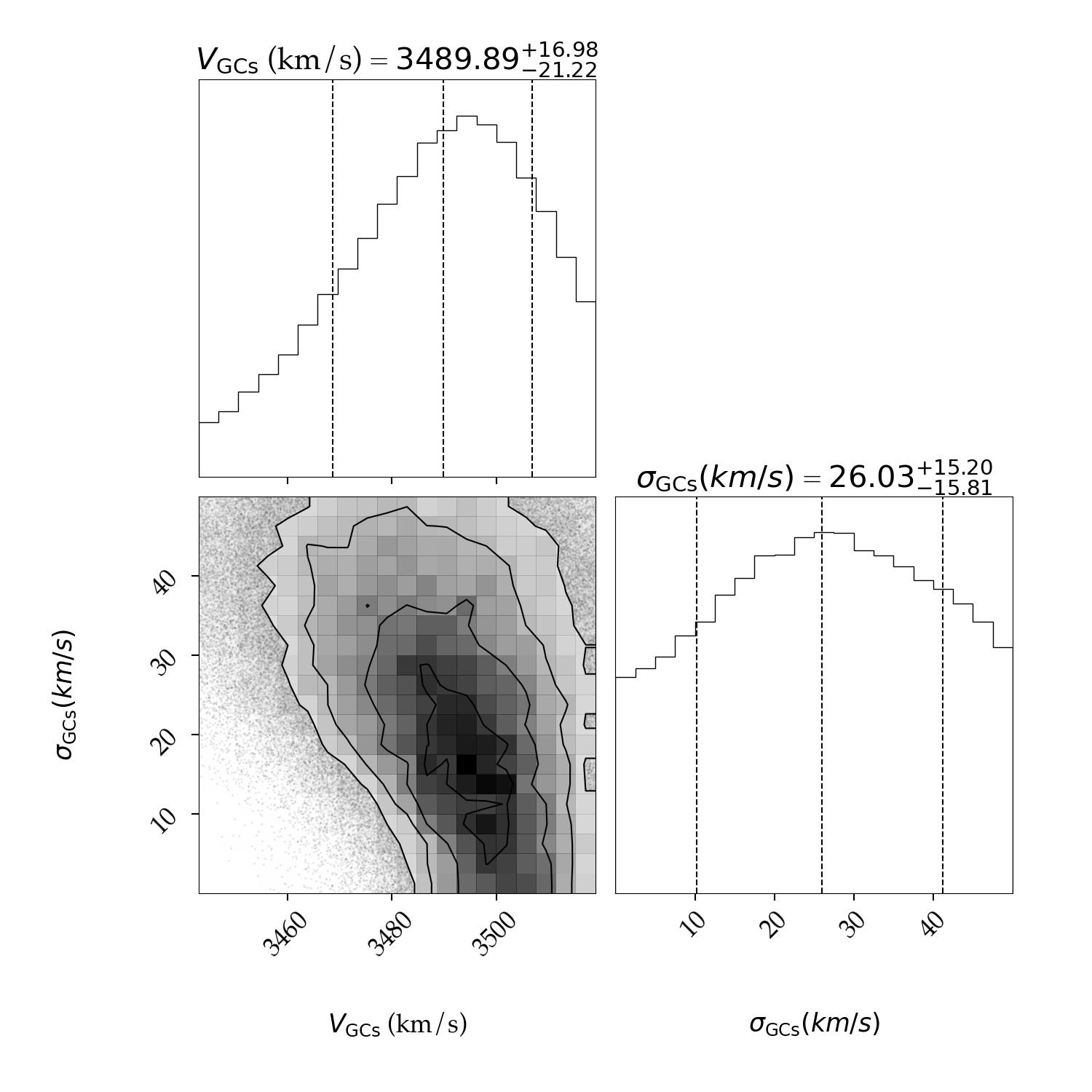}
    \includegraphics[width=\columnwidth]{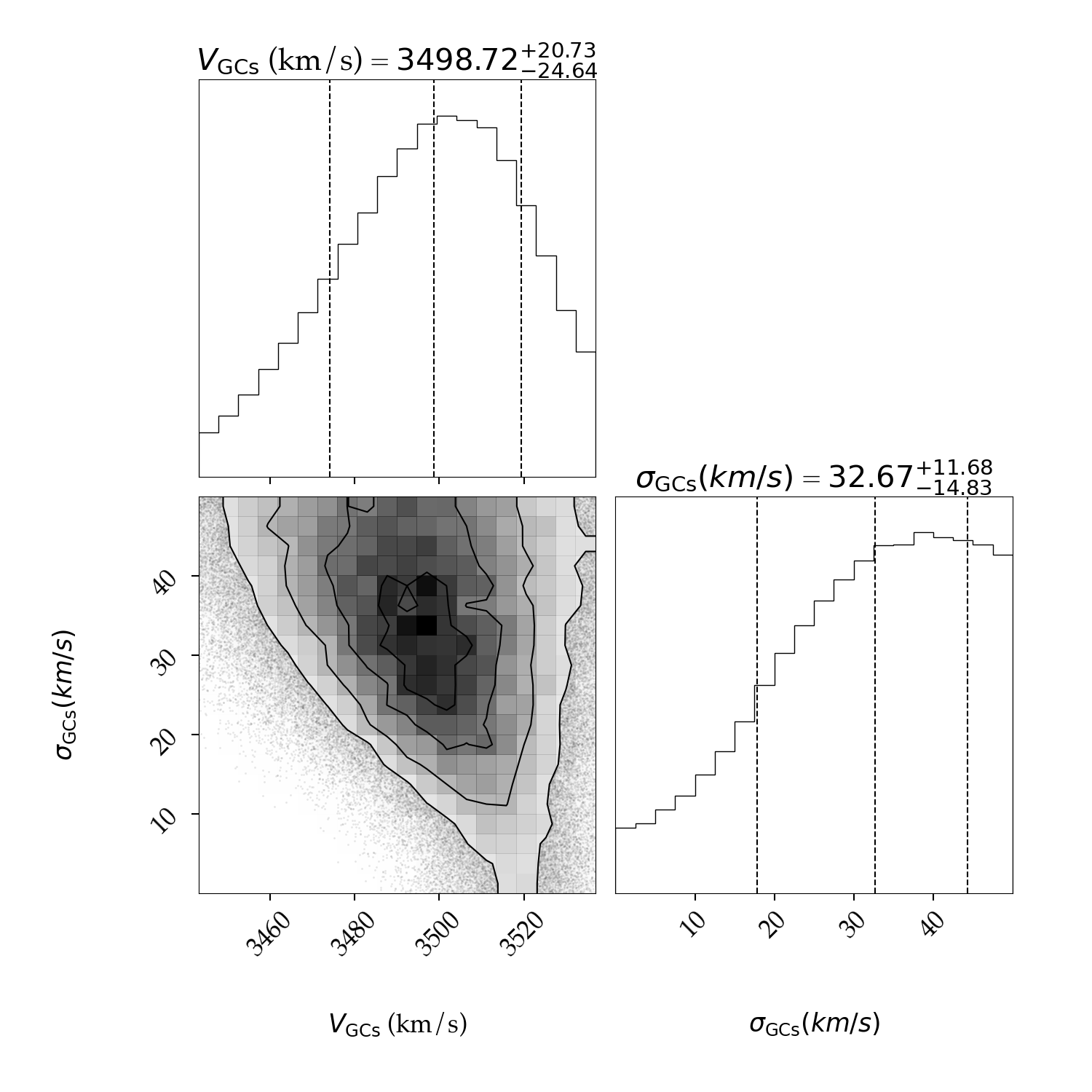}
    \includegraphics[width=\columnwidth]{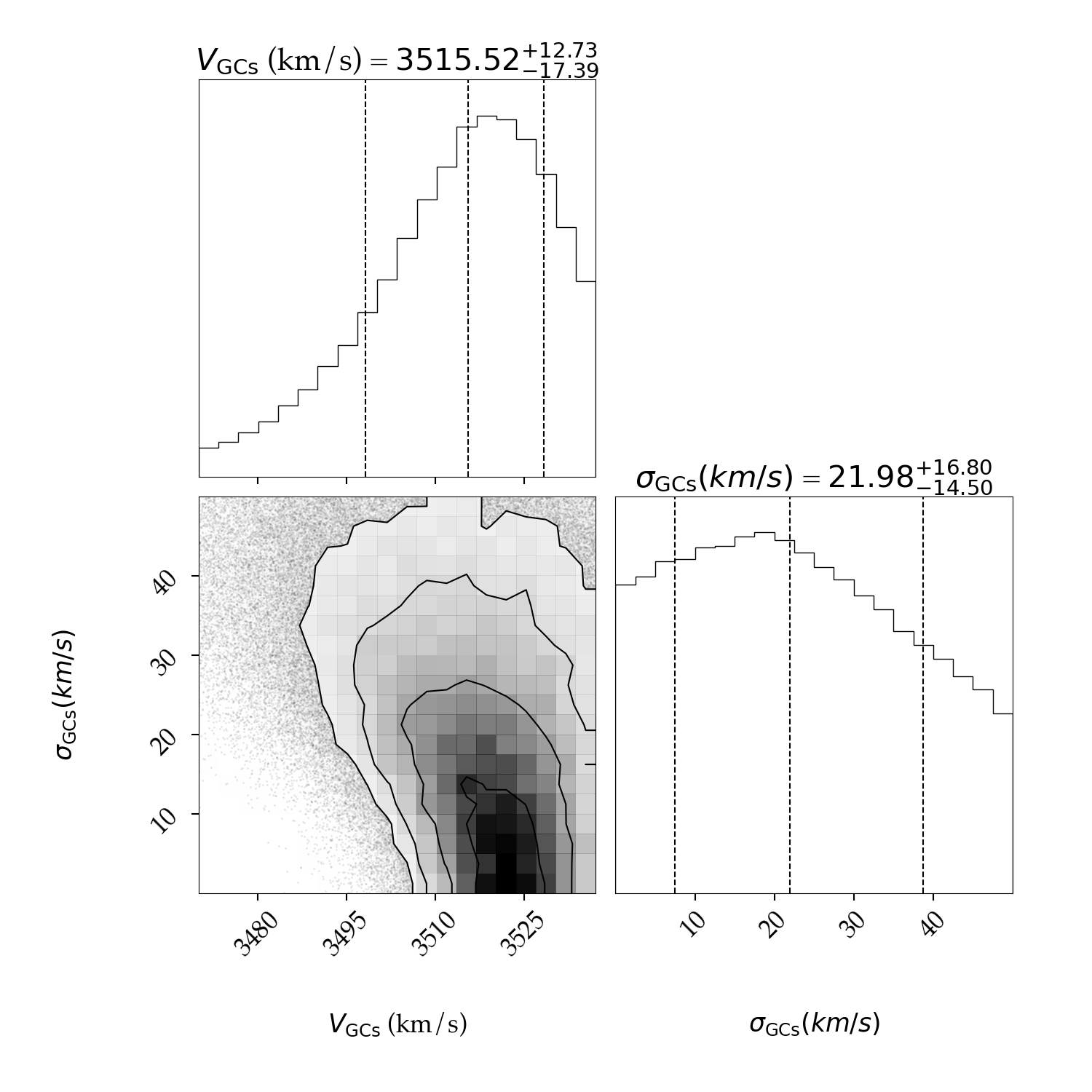}
    \caption{Corner plots showing the results of the MCMC analysis for UDG\,11 GC system with one GC excluded in each panel. The upper left, upper right, lower left, and lower right panels correspond to the exclusion of the first, second, third, and fourth GC from Tab. \ref{tab:spectra_gc}, respectively. Each plot displays the posterior probability distributions of the systemic velocity (top), velocity dispersion (bottom right), and their correlation (bottom left).}
    \label{fig:mcmc_diff}
\end{figure*}

Here, we report the results of the adopted MCMC approach for four different estimates of the GC velocity dispersion and systemic velocity. We obtained four separate estimates using only three GCs of UDG\,11 each time, to verify possible systematic effects due to the misclassification of one of the sources. Figure \ref{fig:mcmc_diff} shows four corner plots, each generated by removing a different single GC from the sample to estimate $\sigma_{\rm GC}$. The upper left panel corresponds to the distribution obtained when removing the first GC listed in Table \ref{tab:spectra_gc}, the upper right panel corresponds to removing the second, the lower left panel to removing the third, and the lower right panel to removing the fourth.
Each cornerplot displays the posterior distributions: the top panel shows the probability distribution of the systemic velocity across the allowed prior range, the bottom right panel shows the distribution of the velocity dispersion, and the bottom left panel illustrates the correlation between the two parameters.

\section{Sample Completeneess}
\label{sec:comple}

\begin{figure*}[h]
    \centering
    \includegraphics[width=\textwidth]{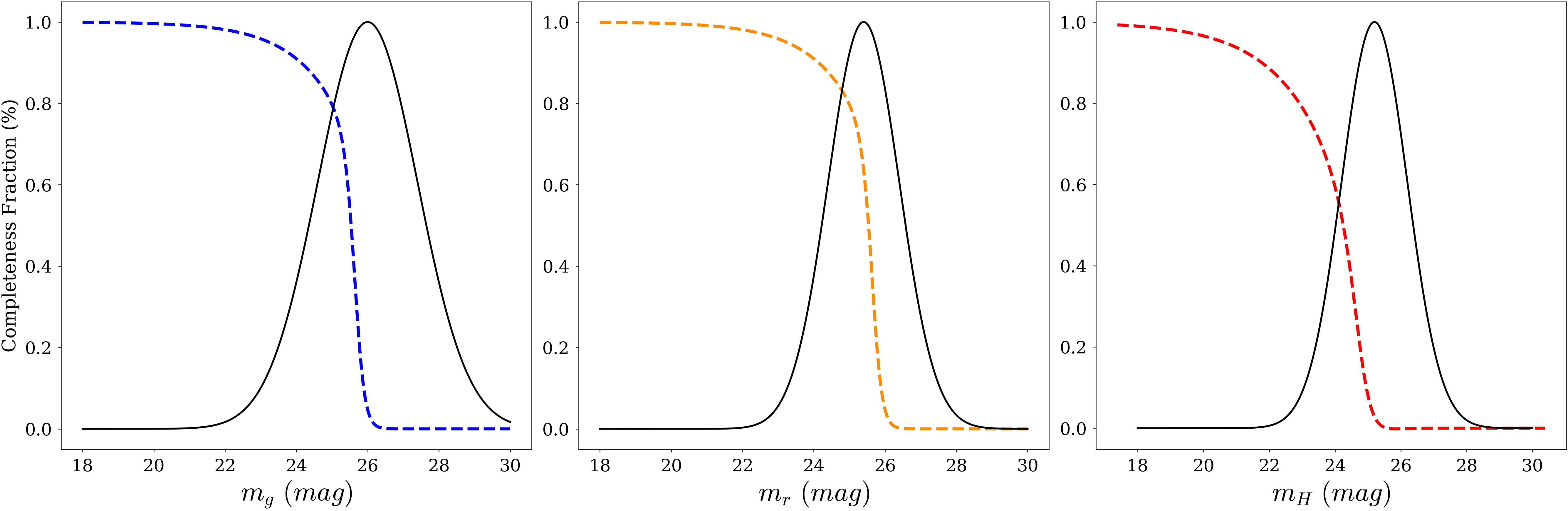}
    \caption{Completeness functions for the $g$ (left panel) and $r$ (middle panel) VST passbands observations of NGC\,3640 \citep{Mirabile24}, and the $H$-band (right panel) observations. The black line shows the expected GCLF.}
    \label{fig:completeness}
\end{figure*}

As discussed in Sect. \ref{sec:tot_n_gc}, we derived the completeness correction using the $H$-band observations. This choice is motivated by the fact that the $H$-band data are shallower than the $g$- and $r$-band data from VST, as well as the MUSE observations. As shown in Fig.~\ref{fig:fields}, the MUSE data are the deepest and sharpest among the available datasets (see also Sect. \ref{sec:dataset}). Here, we demonstrate that the $H$-band observations are the shallowest, meaning they cover only a small fraction of the GCLF. The completeness function for the coadded $H$-band observations was obtained following the procedure described in \citet{Mirabile24}. For the $g$- and $r$-band VST observations, we adopted the completeness function estimated from a different field, specifically the NGC\,3640 field \citep[for more details see][]{Mirabile24}. This choice is justified because the $g$- and $r$-band data for this field were acquired under conditions similar to those of the Hydra I cluster observations (same instrument, exposure time of $\sim$2 h, and comparable image quality). Figure~\ref{fig:completeness} illustrates the completeness functions: in each image, the GCLF is shown as a solid black line; the left panel corresponds to the $g$-band VST completeness, the middle panel to the $r$-band, and the right panel to the $H$-band. Since the $H$-band observations are the shallowest and cover the smallest portion of the GCLF compared to the other passbands, this supports our assumption and shows that the $H$-band drives the overall completeness of our dataset.

\section{$H-$band total magnitude}
\label{sec:hband_tot_mag}

\begin{figure*}[h!]
    \centering
    \includegraphics[width=6.2cm]{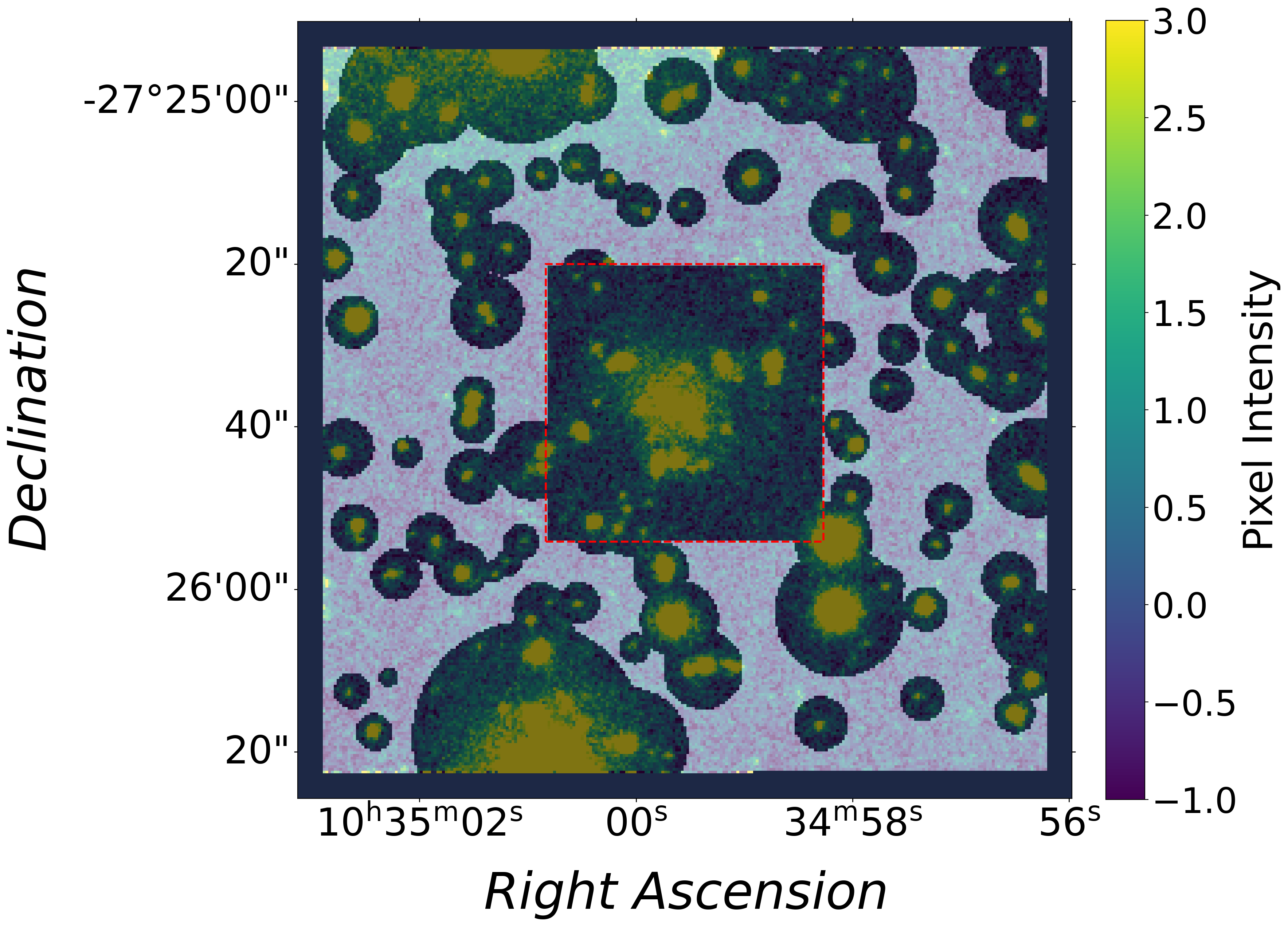}
    \includegraphics[width=5.8cm]{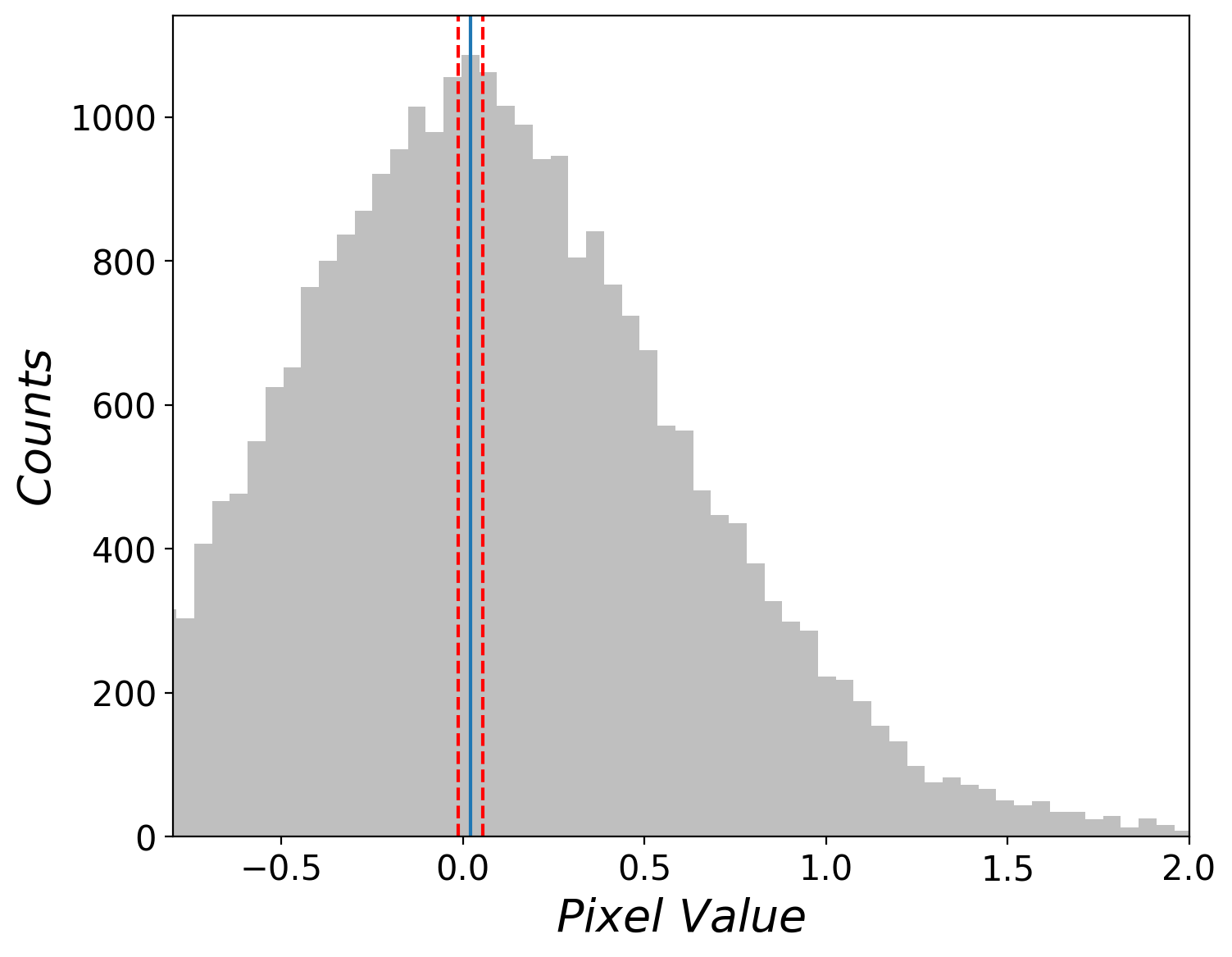}
    \includegraphics[width=5.8cm]{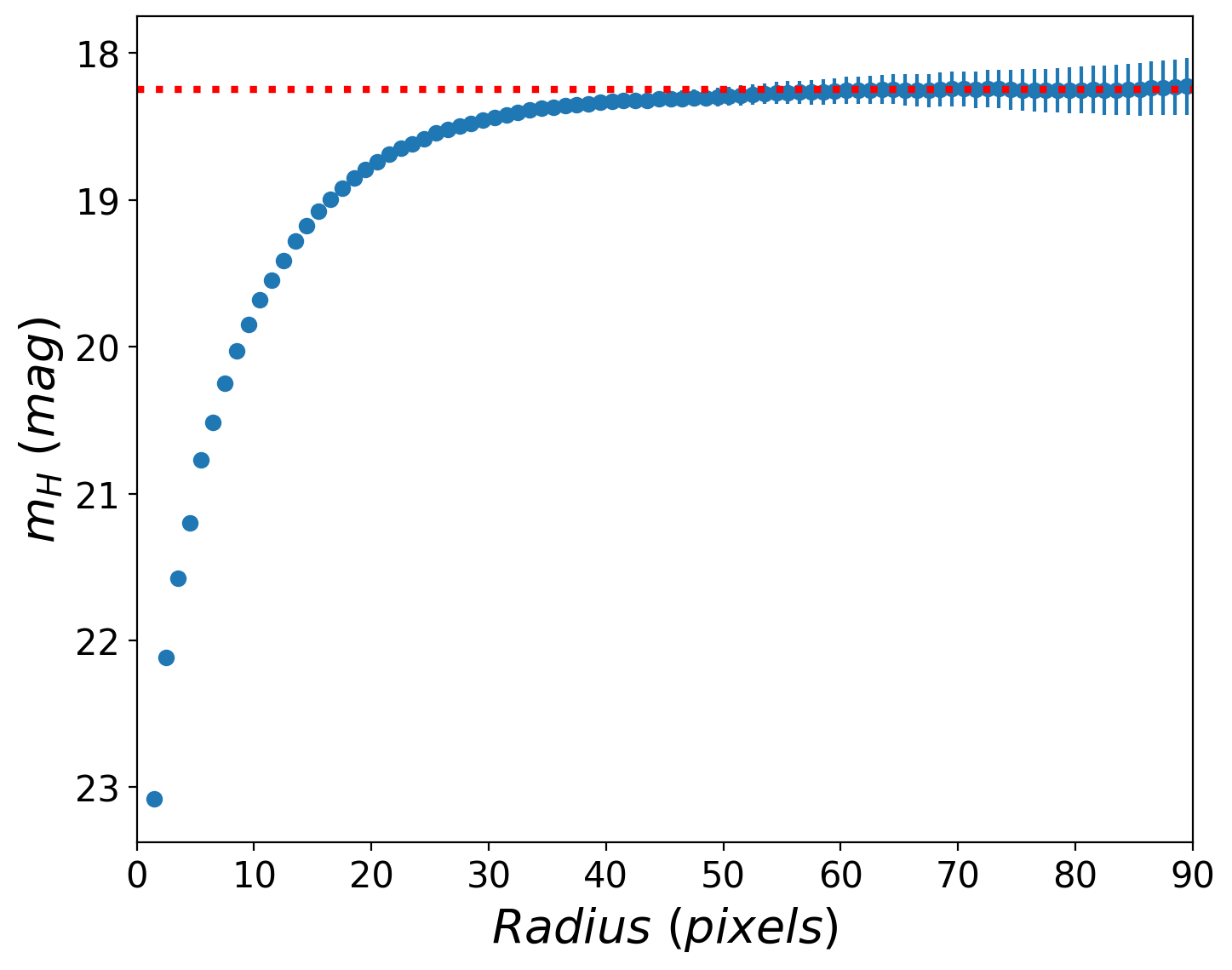}
    \caption{Estimate of the total  $H$-band magnitude for UDG\,11.  Left panel: Darker pixels indicate masked regions. All remaining pixels were used to determine the background level. Central panel: Distribution of the pixels used to estimate the background level. The blue solid line indicates the mode of the distribution, while the red dashed lines show the $\pm1\sigma$ range around the mode. Right panel: Curve of growth obtained from the azimuthally averaged surface brightness profile. The horizontal red dotted line marks the asymptotic value.}
    \label{fig:mag_h_estimate}
\end{figure*}

\begin{figure*}
    \centering
    \includegraphics[width=6cm]{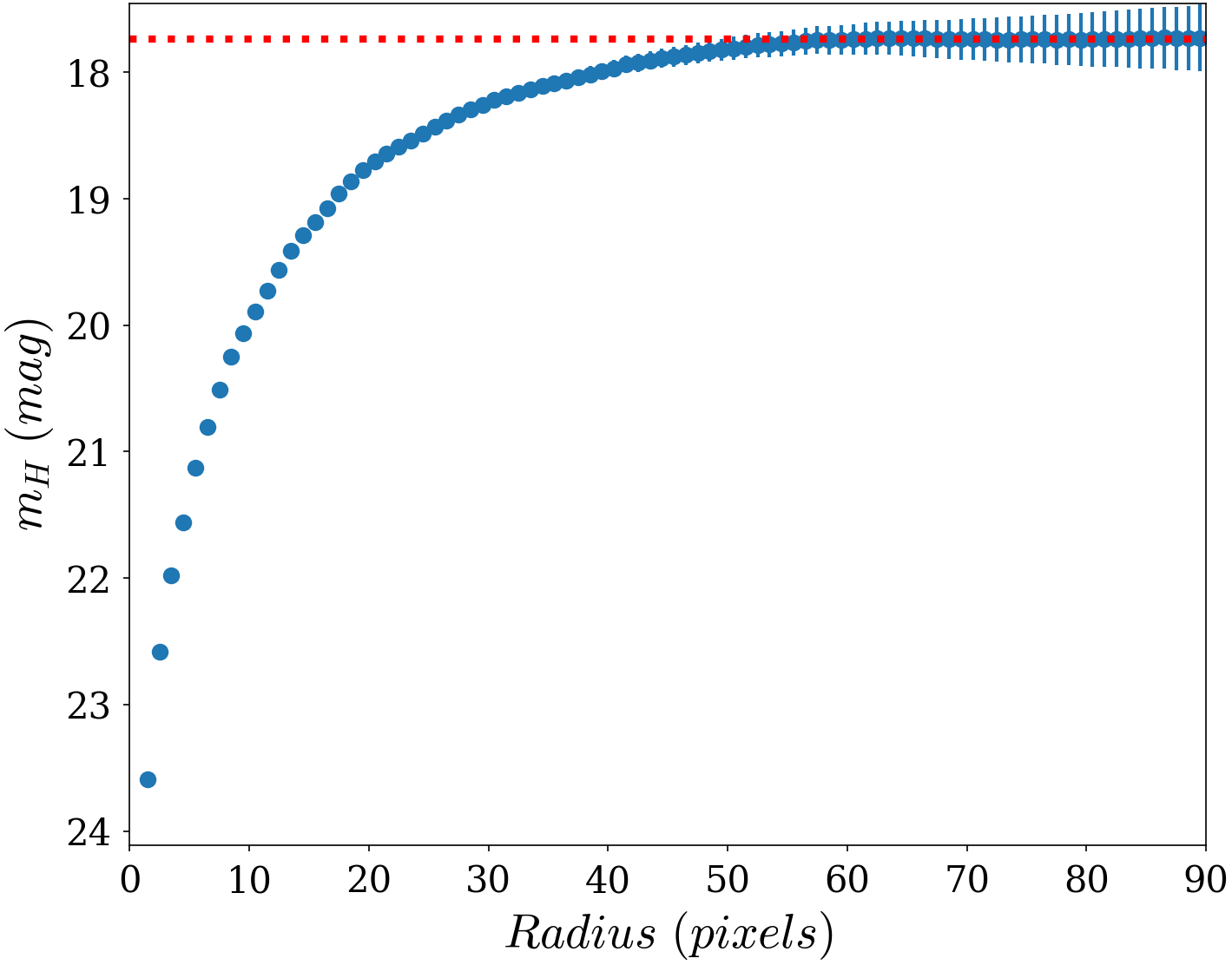}
    \includegraphics[width=6cm]{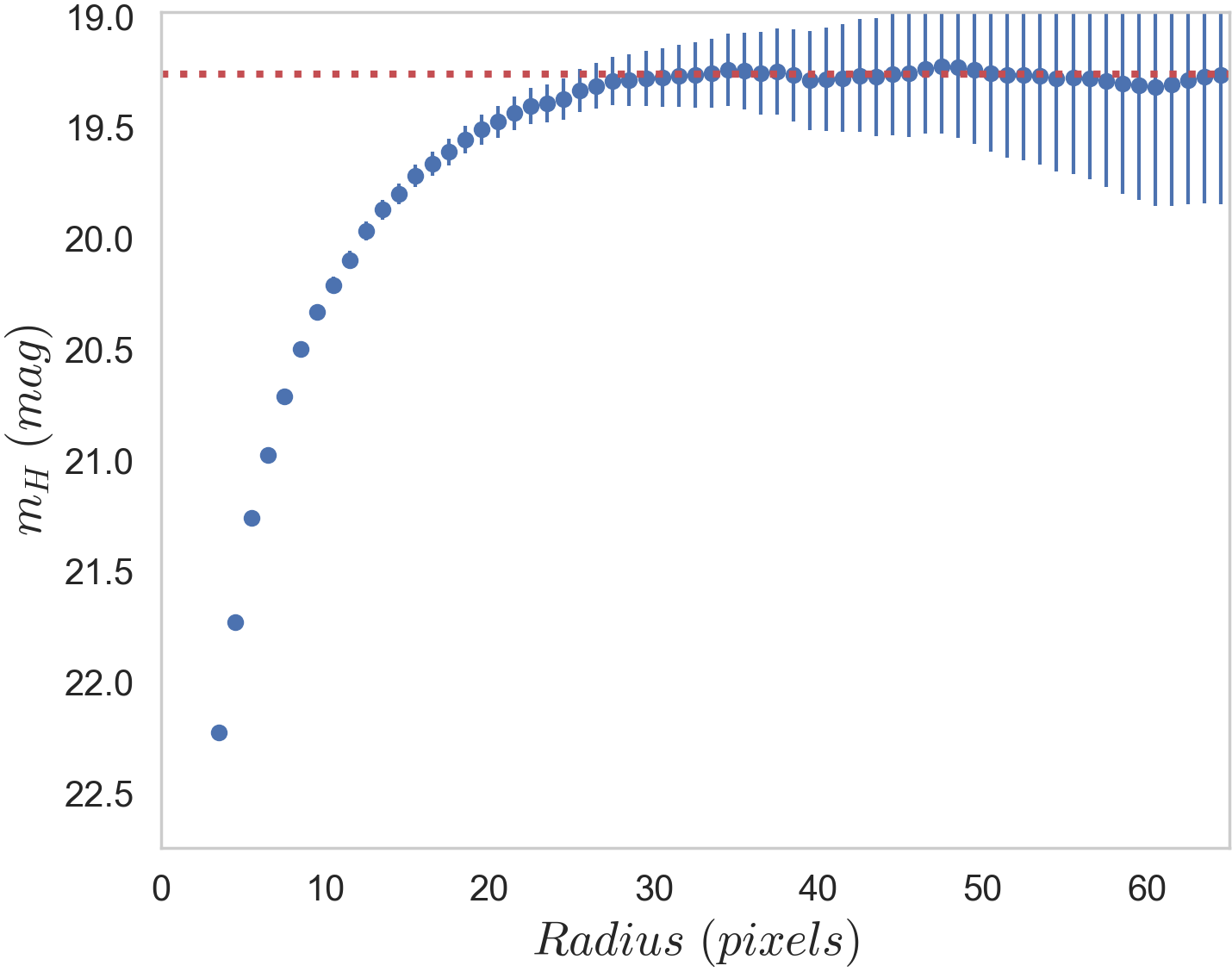}
    \includegraphics[width=6cm]{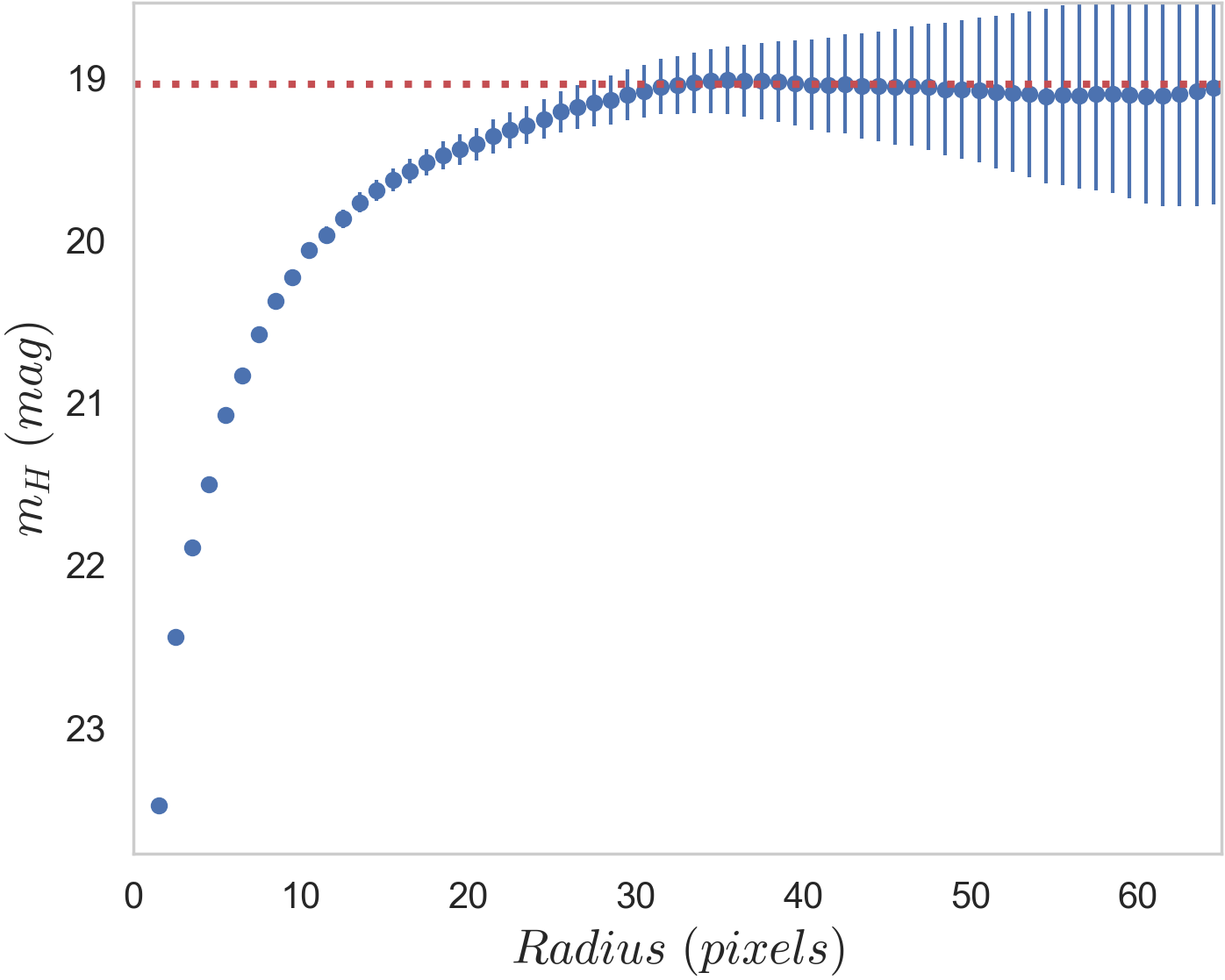}
    \caption{From left to right: curve of growth for UDG\,3, UDG\,7, and UDG\,9, respectively.}
    \label{fig:curve}
\end{figure*}

We developed a procedure to estimate the total $H$-band magnitude using the curve of growth derived from an azimuthally averaged surface brightness profile.
First, the background level for each UDG is estimated as the mode of pixel values. We extracted the pixel distribution from the unmasked pixels beyond a square region of $100\times100$ pixels centered on the galaxy. The first two panels of Fig. \ref{fig:mag_h_estimate} show such a procedure for the UDG\,11. The left panel shows all the masked pixels (darker colors) and the excluded central region (red dashed box). The uncertainty on the background is derived through bootstrap resampling. The azimuthally averaged surface brightness profile is determined by measuring the median flux within concentric annuli centered on the object. Error propagation accounts for both pixel intensity variations within each annulus and uncertainties introduced by background estimation. To construct the curve of growth (right panel in Fig. \ref{fig:mag_h_estimate}), cumulative flux values are calculated by summing flux measurements across radial bins. Uncertainty in cumulative flux is propagated by considering individual errors from each bin and combining them to estimate total uncertainty at each radius.
A weighted average of the asymptotic values in the growth curve is used to obtain a robust estimate of total magnitude. Figure \ref{fig:curve} shows the curve of growth for UDG\,3, UDG\,7 and UDG\,9, respectively. Finally, we inspected the effect introduced by assuming a certain binning of the pixel distribution, by repeating the procedure changing the number of bins and re-estimating background level and asymptotic magnitude from each iteration. The final adopted total magnitudes reported in Tab. \ref{tab:usg_prop} are obtained as median and $rms_{MAD}$ of all the iterations.

\end{document}